\documentclass[manuscript]{aastex}

\slugcomment{Accepted for publication in ApJ, 02/12/14}

\shorttitle{Age gradients in star forming regions}
\shortauthors{Getman et al.}

\begin{document}

\title{Age Gradients in the Stellar Populations of Massive Star Forming Regions Based on a New Stellar Chronometer}

\author{Konstantin V. Getman\altaffilmark{1}, Eric D. Feigelson\altaffilmark{1,2}, Michael A. Kuhn\altaffilmark{1}, Patrick S. Broos\altaffilmark{1}, Leisa K. Townsley\altaffilmark{1}, Tim Naylor\altaffilmark{3}, Matthew S. Povich\altaffilmark{4}, Kevin L. Luhman\altaffilmark{1,2}, Gordon P. Garmire\altaffilmark{5}}

\altaffiltext{1}{Department of Astronomy \& Astrophysics, 525 Davey Laboratory, Pennsylvania State University, University Park, PA 16802, USA}
\altaffiltext{2}{Center for Exoplanets and Habitable Worlds, Pennsylvania State University, University Park, PA 16802, USA}
\altaffiltext{3}{School of Physics and Astronomy, University of Exeter, Stocker Road, Exeter, EX4 4QL, UK}
\altaffiltext{4}{Department of Physics and Astronomy, California State Polytechnic University, 3801 West Temple Ave, Pomona, CA 91768, USA}
\altaffiltext{5}{Huntingdon Institute for X-ray Astronomy, LLC, 10677 Franks Road, Huntingdon, PA 16652, USA}

\begin{abstract}

A major impediment to understanding star formation in massive star forming regions (MSFRs) is the absence of a reliable stellar chronometer to unravel their complex star formation histories.  We present a new estimation of stellar ages using a new method that employs near-infrared (NIR) and X-ray photometry, $Age_{JX}$. Stellar masses are derived from X-ray luminosities using the $L_X - M$ relation from the Taurus cloud. $J$-band luminosities are compared to mass-dependent pre-main-sequence evolutionary models to estimate ages. $Age_{JX}$ is sensitive to a wide range of evolutionary stages, from disk-bearing stars embedded in a cloud to widely dispersed older pre-main sequence stars.  The MYStIX (Massive Young Star-Forming Complex Study in Infrared and X-ray) project characterizes 20 OB-dominated MSFRs using X-ray, mid-infrared, and NIR catalogs.  The $Age_{JX}$ method has been applied to 5525 out of 31,784 MYStIX Probable Complex Members. We provide a homogeneous set of median ages for over a hundred subclusters in 15 MSFRs;  median subcluster ages range between $0.5$~Myr and $5$~Myr. The important science result is the discovery of age gradients across MYStIX regions. The wide MSFR age distribution appears as spatially segregated structures with different ages.  The $Age_{JX}$ ages are youngest in obscured locations in molecular clouds, intermediate in revealed stellar clusters, and oldest in distributed populations. The NIR color index $J-H$, a surrogate measure of extinction, can serve as an approximate age predictor for young embedded clusters.

\end{abstract}


\section{Introduction} \label{introduction_section}

Infrared (IR) and millimeter astronomy, together with astrophysical theory, have made great progress in understanding the formation of single stars in small molecular cloud cores on $\leq 0.1$~pc scales.  However, most stars do not form in such simple environments, arising rather in turbulent giant molecular clouds where different modes of star formation are seen: rich OB-dominated stellar clusters, smaller clusters in dense IR Dark Clouds and filaments, and stellar groups on the peripheries of expanding H\,{\sc{ii}} regions.  It is not clearly known how cloud cores and filaments form, whether clusters form during a single free-fall time or over millions of years, or whether triggering by H\,{\sc{ii}} regions or supernova remnants play minor or major roles.  Combined infrared and X-ray surveys, such as the {\it Chandra Carina Complex Project} \citep[CCCP;][]{Townsley2011},  have the advantage of locating both recently formed disk-bearing (proto)stars and older pre-main sequence (PMS) stars.  These studies reveal spatially distinct clusters that appear to have different ages, and often a population of older widely distributed stars dispersed from earlier generations of star formation.   Our group is now engaged in the Massive Young Star-Forming Complex Study in Infrared and X-ray \citep[MYStIX;][]{Feigelson2013} that examines 20 star forming complexes.  MYStIX fields include a range of small and large cloud structures producing sparse and rich stellar clusters. 

A major impediment to understanding star formation in massive star forming regions (MSFRs) has been the absence of a reliable stellar chronometer so that the star formation history in a region can be reconstructed.  While embedded protostars with Class 0 and Class I infrared spectral energy distributions (IR SEDs) clearly represent the most recent episodes of star formation activity, the sequence of past activity is difficult to unravel from IR SEDs alone because dusty protoplanetary disks have a wide range of longevities.  Approximately 10\% of stars lose their disks in $\leq 0.5$~Myr, 10\% retain their disks for $\geq 6$~Myr, and the remaining 80\% have disk longevities smoothly distributed between 0.5 and 6~Myr \citep{Hernandez2008, Mamajek2009}.  The disk fraction commonly used to estimate the age of a young stellar sample \citep[][ and citations thereof]{Haisch2001}, roughly the ratio  \#(Class~I + Class~II)/\#(Class~I + Class~II + Class~III), therefore can only give a crude measure of stellar ages.  Other stellar properties once proposed as chronometers, such as the depletion of photospheric lithium or the evolution of stellar rotation, do not show simple scalings with stellar age during the PMS phase \citep{Piau02, Eggenberger12, Bouvier13}.    

The best hope for an accurate stellar age estimator during the PMS evolutionary phase has been location on the Hertzsprung-Russell diagram (HRD) or photometric color-magnitude diagram (CMD), where stars of a fixed mass are predicted to progress along distinct tracks defined by convective interior models.  Unfortunately, many problems are now recognized to affect a star's location in HRDs:  photometric variability from accretion and magnetic activity, different accretional histories, binarity, extinction uncertainty, veiling from accretion, scattering and absorption by disks, stellar interiors model uncertainty, and distance uncertainty \citep[e.g.,][]{Baraffe2009, Naylor2009, Soderblom2010, Jeffries2011, Jeffries2012}.  These issues have recently been reviewed by \citet{Preibisch2012} who emphasizes that `severe misinterpretations, gross overestimates of the age spread, and ill-based conclusions about the star formation history' can result from insufficient treatment of these effects.  Preibisch also emphasizes the importance of disentangling the distinct subgroups that are often present in a star forming region in order to establish the `temporal sequence of several discrete star formation episodes'.

In \S\S \ref{concept_section} and \ref{understanding_method_section}, we describe a new PMS stellar age estimator, $Age_{JX}$,  that depends on near-IR stellar photospheric emission and on hard X-ray luminosity which arises mostly from enhanced magnetic flaring of PMS stars.  Near-IR and X-ray photometric measurements are provided in surveys such as CCCP and MYStIX where they are used in the selection of MSFR members.   $Age_{JX}$ is applied only to carefully chosen subsamples of members with appropriate near-IR and X-ray data, and is based on an empirical X-ray/mass relation calibrated to well-studied Taurus PMS stars \citep{Telleschi2007} and to theoretical evolutionary tracks calculated by \citet{Siess2000}.  This age estimate mitigates some (although not all) of the above problems encountered in HRD- and CMD-based chronometry (\S\ref{some_advantage_subsection}). Individual star $Age_{JX}$ values for the MYStIX MSFRs range from $<1$ to $\sim 5$~Myr (Table~\ref{tbl_individual_ages}).

As with other methods, we find a range of PMS $Age_{JX}$ values in large-scale MSFRs that could be considered to be a general wide spread in ages (\S\ref{reddening_effect_section}-\ref{subregions_subclusters_section}).  But, in accord with the advice of \citet{Preibisch2012}, we recompute ages averaged over spatially distinct subclusters (\S\ref{subregions_subclusters_section}) that were recently obtained by \citet{Kuhn2014c} for MYStIX MSFRs.  In that study, a statistical `finite mixture model' of isothermal ellipsoids was fit to the sky distribution of probable cloud members using maximum likelihood techniques.  Each star can then be assigned as a member of a specific subcluster or as a member of a distributed unclustered population.  

The result of computing $Age_{JX}$ averaged over these subclusters is often remarkable: the full MSFR age distribution spanning several million years now appears as spatially segregated structures with narrowly constrained ages.   Furthermore, we often see that the subclusters form spatial-absorption gradients consistent with astronomically reasonable patterns of star formation histories, with older lightly absorbed structures on one side, younger heavily absorbed structures on the other side, and intermediate-age structures in the middle (\S\ref{age_gradients_section}).  Widely distributed populations are nearly always older.   The progression of star formation within a MSFR can be traced, and patterns of star formation histories can be compared between MSFRs.  We provide a homogeneous set of averaged $Age_{JX}$ for a hundred subclusters and regions of interest in MYStIX MSFRs (Table~\ref{tbl_cluster_ages}).

\section{The $Age_{JX}$ Method} \label{concept_section}

\subsection{Concept}

The method developed here has three steps. 
\begin{enumerate}

\item The stellar mass ($M$) is derived from the intrinsic X-ray luminosity measurement ($L_X$) of a star using the well-established empirical $L_X - M$ relationship for PMS stars of \citet{Telleschi2007} (\S \ref{lx_mass_pms_section}). 

\item The sample is truncated to include only low-mass stars and the absolute $J$-band magnitude ($M_J$) is derived by correcting the apparent $J$ magnitude for a photometrically estimated absorption $A_J$ and distance. Notice that, unlike optical magnitudes (or IR magnitudes at longer wavelengths), $J$ is less prone to the effects of absorption, accretion, and diffuse nebular background (or disks and nebular background) (\S \ref{some_advantage_subsection}). $J$ is more prone to the effects of absorption than $H$ and $K$ near-infrared magnitudes, but is less affected by thermal disk emission.  

\item We estimate stellar ages $Age_{JX}$ using obtained mass estimate $M$ and the absolute $J$-band magnitude $M_J$ by comparison to the PMS isochrones of \citet{Siess2000}.  This $M_J-M$ diagram is a surrogate for the traditional HRD (Figure \ref{fig_mj_vs_mass_ajfromccd_SMOOTH_MEDIANS_V2}).  

\end{enumerate}
The observational inputs to the calculation are:  the distance to the MSFR\footnote{To calculate both $L_X$ and $M_J$ we use distances to each MSFR adopted for the MYStIX project \citep[Table 1 of][]{Feigelson2013}.}, the absorption-corrected X-ray luminosity in the $(0.5-8)$~keV energy band, and the apparent $JHK_s$ magnitudes as well as the $L_X-M$ and interstellar reddening calibration relations and  theoretical evolutionary tracks.

\subsection{Sample selection} \label{sample_section}

The 31,784 MYStIX Probable Complex Members (MPCMs) constitute the initial stellar sample for this study \citep{Broos2013}.  The MPCM sample is extracted from source catalogs constructed by special analysis of data from the {\it Chandra X-ray Observatory}, {\it Spitzer Space Telescope}, United Kingdom Infra-Red Telescope, and 2MASS telescopes.  The MPCMs are a union of statistically classified X-ray selected stars, photometric infrared excess stars, and published OB stars.  The multistage construction of the MPCM sample is described by \citet{Feigelson2013} and its accompanying papers.  \citet{Kuhn2013a} and \citet{Townsley2013} present X-ray sources and properties,  \citet{King2013} present the near-IR sources and properties, \citet{Kuhn2013b} present the mid-IR sources and properties, \citet{Naylor2013} describe the X-ray/infrared source matching procedure, \citet{Povich2013} identify infrared-excess stars, and \citet{Broos2013} statistically derive  the integrated sample of MPCMs in the 20 MYStIX MSFRs. 

Intrinsic stellar X-ray luminosities $L_X$ (in the $0.5-8$~keV band) are determined in a uniform way using the non-parametric method $XPHOT$ \citep{Getman2010}. $XPHOT$ is applicable to X-ray sources with greater than several net counts, and allows an approximate recovery of the soft ($\la 1$~keV) X-ray plasma component, which is often missed in the X-ray data of weak and/or highly absorbed sources when using traditional methods of parametric model fitting, such as XSPEC. We therefore distill, from the full MPCM catalog, sources with XPHOT $L_X$  estimates.

Individual stellar age estimates $Age_{JX}$ are restricted to 5525 MPCM members.  First, the full MPCM sample was restricted to stars with sufficient X-ray counts to have an $XPHOT$ estimate of $L_X$. All MYStIX Infrared Excess \citep[MIRES;][]{Povich2013} sources without X-ray counterparts, and very faint X-ray sources, are omitted.   Second, three MYStIX regions (W~3, W~4, and NGC~3576), which were not subject to spatial clustering analysis by \citet{Kuhn2014c}, and two MYStIX regions (RCW~38 and Trifid Nebula) with poor $Age_{JX}$ sampling are ignored from our detailed age analysis.  Third, the X-ray source sample is truncated to retain sources with reliable NIR photometry, retaining sources with uncertainties on individual $JHK_s$ measurements of $<0.1$~mag or uncertainties on $J-H$ and $H-K_s$ colors of $<0.1$~mag.  This permits reasonably accurate absorption correction. 

\begin{figure}
\centering
\includegraphics[angle=0.,width=5.5in]{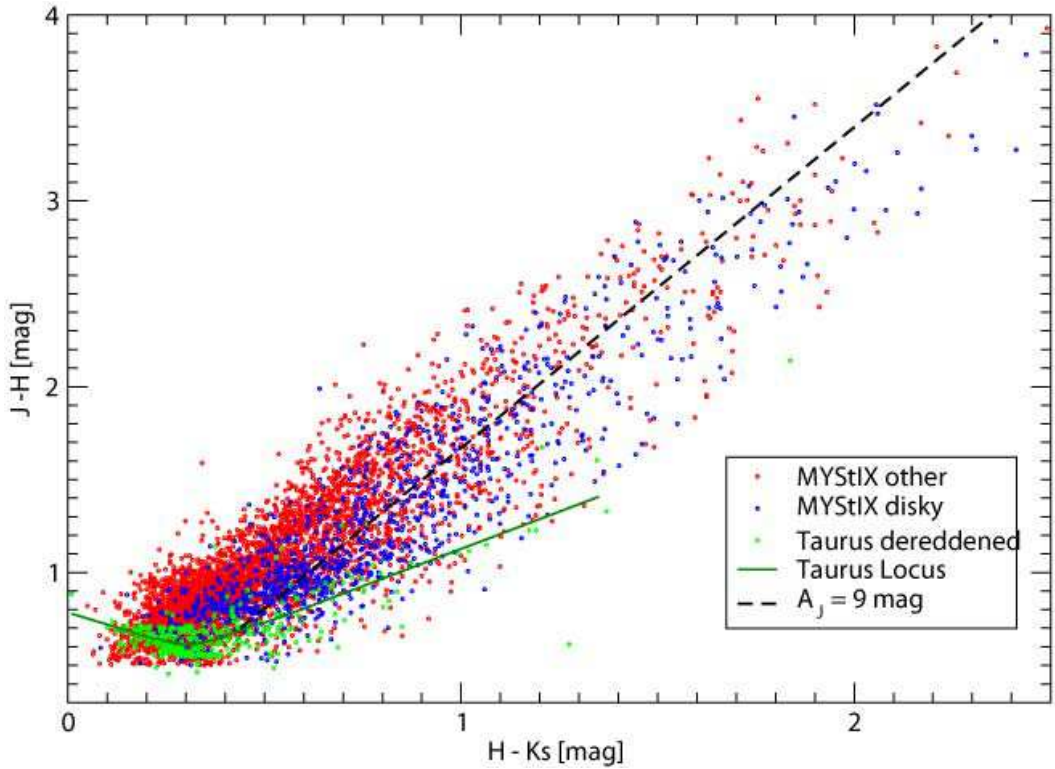}
\caption{Near-IR color-color diagram for 5525 MYStIX young stars that are selected for the age analysis. Disk-bearing stars are in blue, and the remaining MYStIX stars are in red.  The dereddened Taurus stars are marked by green points; Friedman's super smoother regression fit \citep{Friedman1984} to these data is shown by the green line. Dashed black line is a reddening vector of $A_J = 9$~mag originating at $\sim 0.1$~M$_{\odot}$\citep[for the age of 1 Myr;][]{Siess2000}.  \label{fig_jhk_ccd}}
\end{figure}

Fourth, the sample is culled of objects with large $K_s$-excesses for which there is no clear NIR dereddening procedures.  Objects with $H-K_s$ colors redder than those of the well studied disk-bearing stars in the Taurus star forming region are removed from the sample (Figure \ref{fig_jhk_ccd}). Many of these objects with extreme $K$-excess emission are likely protostars \citep{Lada2000}. Future MYStIX papers will focus on properties of these objects. Finally, we apply two cuts to the sample to restrict consideration to lower-mass stars that follow convective Hayashi tracks in the HRD.  First, stars with $L_X > 3 \times 10^{30}$~erg s$^{-1}$ are removed.  From the $L_X-M$ calibration curve of \citet{Telleschi2007}, this deletes stars with $M \ga 1.2$~M$_{\odot}$.   Second, stars with $J-H > 0.5$~mag are removed; this excludes lightly absorbed hotter early-type stars.

It is important to note that for the stars in the final MYStIX $Age_{JX}$ sample, the typical number of X-ray net counts per source is $\sim 20$ counts and the typical Poisson noise and modeling uncertinties on X-ray luminosity inferred by the XPHOT method are $\sigma/L_X \sim 40-50$\% each.

Thus from the original 31,784 MPCM stars, we are left with a subsample of 5525 well-characterized relatively X-ray-bright low-mass stars for $Age_{JX}$ determination.

\subsection{Dereddening \label{dereddening_subsection}} 

The extinctions in the $J$-band ($A_J$) are derived from dereddening to the intrinsic color locus of the Taurus low-mass stars in the $J-H$ vs. $H-K_s$ color-color diagram (Figure \ref{fig_jhk_ccd}). The Taurus locus is produced using the data of \citet{Luhman2010} (and K. Luhman, private communication); the reddening laws $A_J/A_{K_s} = 2.5$ and $A_H/A_{K_s} = 1.55$ are used \citep{Indebetouw2005}. The absolute magnitudes $M_J$ are then determined from $J$, $A_J$, and the MYStIX region distance estimates compiled by \citet{Feigelson2013}. Notice that the XPHOT $L_X$ estimates are totally independent of NIR reddening; $L_X$'s were corrected for absorption using X-ray median energies as surrogates for X-ray absorption column densities, as described by \citet{Getman2010}.

\subsection{Mass and age estimation}

Stellar mass estimates are derived from the $L_X$ values using the clear (but poorly understood; \S \ref{lx_mass_pms_section}) regression relationship $\log(L_X) = 1.69 \times \log(M/M_{\odot}) + 30.33$ found for PMS stars in the Taurus region \citep{Telleschi2007}. The minor effect of the weak temporal evolution of X-ray luminosity in PMS stars \citep{PreibischFeigelson2005}, if real, is ignored in our analysis (\S \ref{lx_mass_pms_section}). 

$Age_{JX}$ estimates are then produced by minimizing $\chi^2$ between the observed absorption-corrected $J$-band magnitudes and X-ray-based mass estimates ($M_J$, $\sigma_{M_J}$, $M$, $\sigma_{M}$) and the grid of evolutionary tracks associated with the PMS stellar interior models of \citet{Siess2000}. The age step in this grid is 0.1~Myr. Since smaller $M_J$ separations among older isochrones on the $M_J-M$ diagram could result in unrealistically high individual age estimates, the evolutionary models used here stop at 5~Myr. The resulting $Age_{JX}$ estimates for 5525 MYStIX stars are given in Table \ref{tbl_individual_ages} along with the MYStIX designation, sky coordinates, NIR photometry, absorption-corrected X-ray luminosity, and cluster membership from \citet{Kuhn2014c}.

\section{Understanding the $Age_{JX}$ Method \label{understanding_method_section}}

Before presenting the results from the $Age_{JX}$ method on the ages of the MYStIX regions and subclusters (\S \ref{subregions_subclusters_section}), in the current section, we discuss various aspects of the $Age_{JX}$ estimates of PMS ages: the underlying relationships between stellar properties, sources of uncertainty and scatter, and comparison with other methods. 

\subsection{Relationship between X-ray Luminosity and Mass \label{lx_mass_pms_section}}

The $Age_{JX}$ method relies on an assumption of a universal $L_X - M$ relationship for $\la 5$~Myr old PMS stars. The relationship was originally found in $ROSAT$ studies of nearby star forming regions \citep{Feigelson93}, and has been described in major studies of the Taurus \citep[XEST;][]{Telleschi2007} and Orion Nebula Cluster \citep[COUP;][]{Preibisch2005} populations.  The correlation in Taurus stars is tighter than in Orion stars, probably due to their more reliable stellar mass estimates. $L_X - M$ distributions similar to that seen in Taurus and the ONC are seen in other regions; for instance, in the $1-3$~Myr old NGC~2264 region \citep{Flaccomio2006}, the $2-3$~Myr old Cep~OB3b region \citep{Getman2006,Getman2009}, the $\la 1$~Myr old W~40 region \citep{Kuhn2010}, and the $2-3$~Myr old IC~348 region \citep{Stelzer2012}. The sources of scatter in the $L_X-M$ relationship will be statistical noise, X-ray variability from magnetic activity, binarity, intrinsic imperfection of the relationship, modeling uncertainties on $L_X$ and mass, and suppression of time-integrated X-ray emission in accreting versus non-accreting stars.
  
While both accreting `classical' T Tauri stars and non-accreting `weak-line' T Tauri stars have broad X-ray luminosity functions ranging from $\log L_X \simeq 28-31$ erg s$^{-1}$, it was found with $ROSAT$ \citep{Neuhauser95} that the X-ray luminosity function of accreting stars is somewhat lower than that of non-accreting stars. This is seen despite the frequent presence of a soft X-ray component associated with the accretion spot on the stellar surface \citep{GudelTelleschi2007}. However, the effect could be mass dependent: for the PMS population in the nearby Taurus region, \citet{Telleschi2007} report that accreting stars are less X-ray active than non-accreting stars for a wide range of masses ($>0.3$~M$_{\odot}$); meanwhile, for the richer PMS populations in the more distant Orion Nebula Cluster and IC~348 region, \citet{Preibisch2005} and \citet{Stelzer2012} report no differences in X-ray activity between accreting and non-accreting stars in the mass range  $\sim 0.5-1.2$~M$_{\odot}$. The latter is a typical mass range for the MYStIX-$Age_{JX}$ stars (Figure \ref{fig_mj_vs_mass_ajfromccd_SMOOTH_MEDIANS_V2}).

On the other hand, PMS stars lie coincident with the activity-saturation locus of main-sequence (MS) stars on the $L_X/L_{bol}$-Rossby (or $L_X/L_{bol}$-Rotation) diagram; albeit, the scatter in $L_X/L_{bol}$ is large. Examples of the activity-Rossby relationship for the PMS stars in ONC, Taurus, NGC~2264, and IC~348 can be found in \citet{Preibisch2005,Briggs2007,Dahm2007} and \citet{AlexanderPreibisch2012}. Since the PMS stars are in the magnetically saturated regime of late-type stars, one may reasonably expect their $L_X/L_{bol}$ ratio to remain constant in time and $L_X-M$ to be simply a consequence of the mass-bolometric luminosity relation ($L_{bol} \propto M^{\alpha}$) coupled with the effect of the X-ray saturation. However, the $L_{bol}-M$ relationship for PMS stars is a strong function of time as the stars descend the convective Hayashi tracks \citep{Siess2000}.

\begin{figure}
\centering
\includegraphics[angle=0.,width=4.5in]{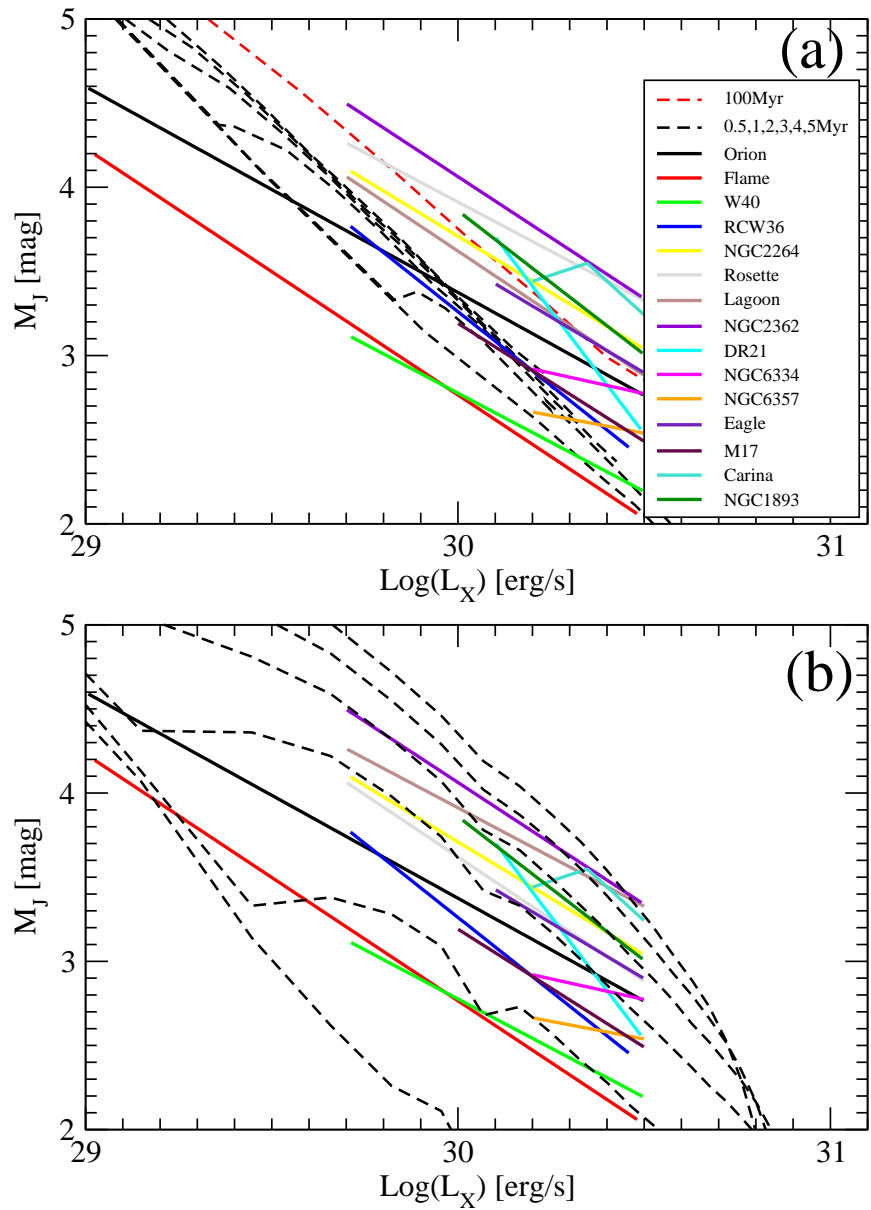}
\caption{The MYStIX data in the form of the B-spline 50\% quartile regression lines \citep[colored solid lines;][]{HeNg1999} are compared to the theoretical PMS isochrones of \citet{Siess2000} transformed onto $M_J - \log(L_X)$ plane, as described in the text (dashed lines). (a) The Seiss et al. isochrones are given for the two scenarios: (a) constant in time $L_X/L_{bol}$ ratio, and (b) constant in time $L_X - M$ relation. \label{fig_mj_vs_lx_two_views}}
\end{figure}

Having the largest and most comprehensive IR/X-ray dataset of PMS stars at a wide range of ages allows us to consider if either of these two relationships ($L_X - L_{bol}$ or $L_X - M$) may remain unvarying (or slowly varying) with time during the early evolutionary phase of PMS stars. We transform the $M_J$, $L_{bol}$, and $M$ outcome of the theoretical evolutionary PMS models of \citet{Siess2000} onto an $M_J - \log(L_X)$ plane based on two different assumptions: $L_X-L_{bol}$ is fixed for all stars and evolutionary stages; and similarly $L_X-M$ is fixed for all stars and evolutionary stages. In the former case, the $L_{bol}$ outcome of Siess et al. is substituted with $L_X$ using the respective transformation from \citet{Telleschi2007}, $\log(L_X) = 1.05 \times \log(L_{bol}/L_{\odot}) + 30.00$. In the latter case, the $M$ outcome is substituted with $L_X$ using the respective transformation from \citet{Telleschi2007}, $\log(L_X) = 1.69 \times \log(M/M_{\odot}) + 30.33$.  Figure \ref{fig_mj_vs_lx_two_views} compares these two treatments with the \citet{Siess2000} isochrones.  The top panel shows inconsistencies when constant $L_X-L_{bol}$ is assumed.  Various MYStIX regions appear either younger than 0.5~Myr, older than 100~Myr, or have $M_J-\log L_X$ distributions that cross isochrones.  The lower panel, where $L_X-M$ is assumed constant, shows much better consistency between models and data.  Each MYStIX region distribution in the $M_J- \log L_X$ diagram roughly follows a reasonable theoretical PMS isochrone; for instance, the embedded Flame Nebula and W~40 regions are associated with the youngest isochrones ($t \simeq 0.5-1$~Myr; \S \ref{age_flame_subsection} and \ref{age_w40_subsection}) while the isolated NGC~2362 cluster is associated with the oldest isochrones ($t \simeq 3-5$~Myr; \S \ref{age_ngc2362_subsection}).

Notice that \citet{PreibischFeigelson2005} report the effect of the weak temporal evolution of X-ray luminosity during the first several Myr of PMS evolution, $L_X \propto t^{-1/3}$. We tested the impact of this possible effect on our inferred age estimates by modifying the shape of the \citet{Siess2000} model isochrones shown in Figure~\ref{fig_mj_vs_lx_two_views}b with the following formula: $\log(L_X) = 30.33 + 1.69 \times \log(M/M_{\odot}) - 0.3 \times \log(t/t_0)$, where the fiducial $t_0$, as a typical age for both the Taurus and Orion regions, was varied between $[1.6-2]$~Myr. This effect is tiny for the young $\la 2$~Myr regions, but produces systematically higher ages for older $\ga 3$~Myr regions. However, for the oldest of the MYStIX regions, NGC~2362, these new age estimates become too high (over $5$~Myr on the time scale of \citet{Siess2000}, over 6~Myr on the time scale of \citet{Baraffe1998}) to be consistent with the published optical ages for NGC~2362 (\S \ref{age_ngc2362_subsection}). The effect is thus ignored from our further analyses. 

The $L_X-L_{bol}$ and $L_X - M$ relationships are astrophysically poorly understood. One possibility consistent with the saturation limit is that X-ray luminosity scales with surface area because larger stars would allowi more flaring magnetic loops. Another possibility is that the X-ray activity scales with stellar volume allowing more opportunity for a convective dynamo.  In both cases, the $L_X-M$ relation arises indirectly from the scaling of mass with radius.  Finally, $L_X-M$ might reflect the early trapping of higher magnetic flux in more massive stars, although it seems unlikely that fully convective stars would be dominated by `fossil' magnetic fields rather than by dynamo-generated fields.  In any case, these astrophysical explanations could be too simplistic to reflect reality. 

But even without clear scientific understanding for a universal $L_X-M$ for pre-main sequence stars, this relationship, when calibrated to the Taurus sample of \citet{Telleschi2007}, has considerable predictive power.  The use of poorly understood empirical relationships for calibrating other properties has a long tradition in astronomy: the Cepheid P-L relation of \citet{LeavittPickering1912}, the $L \sim M^{3.5}$ relation for main sequence stars of \citet{Eddington1924}, the relations involving kinematics and luminosities in galaxies of \citet{FaberJackson1976} and \citet{TullyFisher1977}, and the SN Ia light curve relation to $L_{bol}$ widely used today \citep{Hamuy1996}.

\subsection{Relationship between $Age_{JX}$ and Absorption \label{reddening_effect_section}}

Absorption plays an important role in $Age_{JX}$ estimation.  Both the $J$ magnitude and the full-band X-ray luminosity are corrected for absorption for individual stars prior to their placement on the $L_X-M$ and $M_J-M$ diagrams.  Note that these corrections may not be self-consistent.  The $J$ reddening correction is based on the NIR color-color diagram and physically arises from scattering and absorption by interstellar dust.  The $L_X$ absorption correction is based on the median energy of observed X-ray photons as a surrogate for soft X-ray absorption by a column density of interstellar gas \citep{Getman2010}.   We have chosen this divided approach, rather than a unified approach, to avoid assuming that the gas-to-dust ratio in interstellar clouds is fixed and known. 

But absorption should also play an astronomical role that can help validate the median $Age_{JX}$ estimates for MYStIX subclusters. We expect that older PMS stars will exhibit lower interstellar absorption than the youngest stars, which must reside in their natal dense molecular cores and filaments.  As star formation progresses, older stars may kinematically drift from these obscured regions, cloud material can be depleted through conversion of gas to stars, and cloud material can be photoevaporated and ablated by the radiation and winds of newborn massive stars. A strict relationship between age and absorption is not expected due to variations in line-of-sight geometries and cloud star formation histories.  Since the $J-H$ color index is a reddening indicator that is relatively insensitive to stellar spectral type or protoplanetary disk thermal emission, we expect stellar ages $Age_{JX}$ to be anti-correlated with $J-H$.  

\begin{figure}
\centering
\includegraphics[angle=0.,width=3.7in]{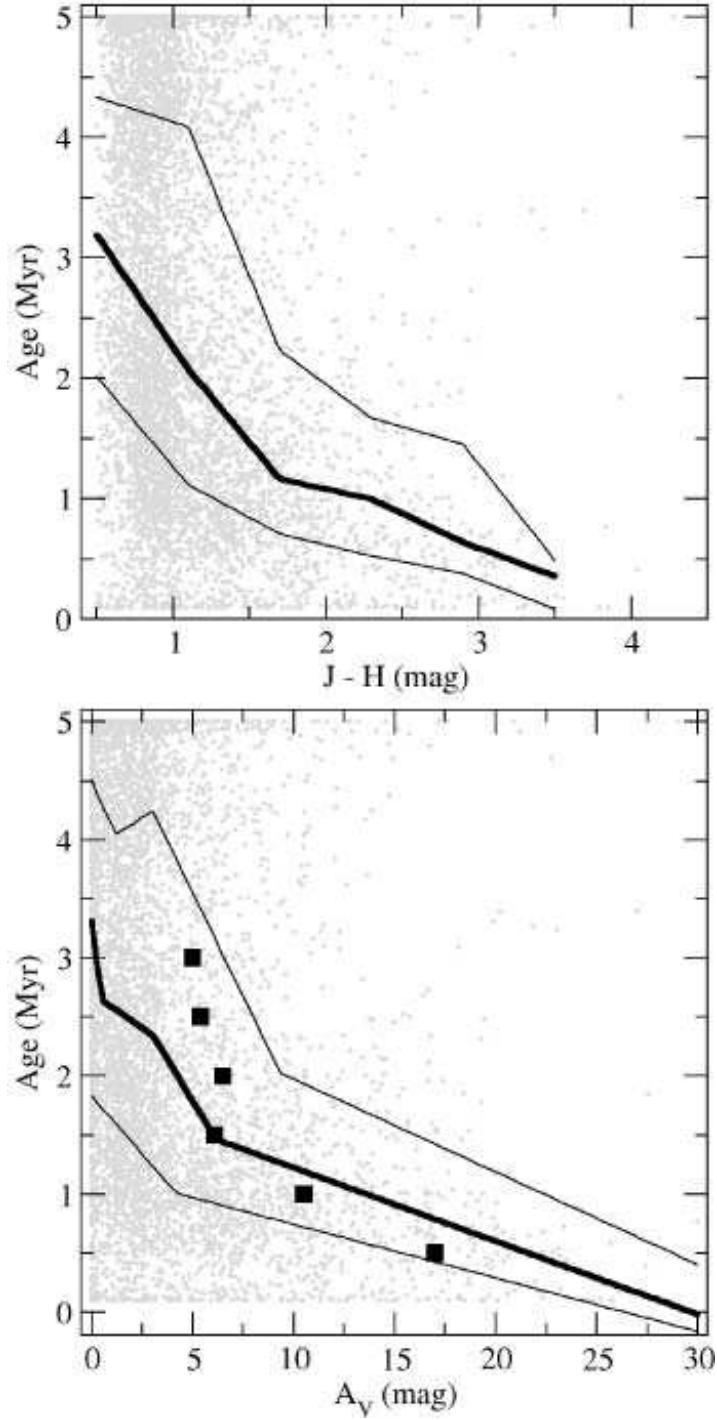}
\caption{The relationship between $Age_{JX}$ and absorption for 5525 MYStIX stars (grey $\circ$) where absorption is measured by (a) $J-H$ and (b) $A_V$.  The black thick (thin) lines show the relationship of the median ($25$\% and $75$\% quartiles) of $Age_{JX}$ and absorption obtained from spline regression (see text for details).  Values obtained by \citet{Ybarra2013} for Rosette region clusters are shown as black $\sq$. \label{fig_AgeJX_vs_Reddening}}
\end{figure}

We first consider the integrated sample of 5525 MPCM stars averaged over the full fields of MYStIX MSFRs.  These fields are determined by the outline of single or mosaicked exposures of the {\it Chandra X-ray Observatory} and typically include a well-known rich OB-dominated stellar cluster plus subclusters, embedded young stars,  and widely distributed young stars.  Most regions have heterogeneous structures with stars ranging from lightly to heavily absorbed.   Figure \ref{fig_AgeJX_vs_Reddening} shows the $Age_{JX}$~vs.~$J-H$ scatter diagram. Note that in the range of $(J-H) \la 1.2$~mag, the spread in the J-H color is caused by the effect of absorption and the variation in the intrinsic colors of low-mass diskless stars,  $(J-H)_{0} \sim 0.6-0.7$~mag, and disk-bearing stars,  $(J-H)_0 \sim 0.6 - 1.2$~mag. The bottom panel shows the similar scatter diagram using $A_V = A_J/0.28$ \citep{Cardelli89}.  Due to the wide spread in ages at a chosen absorption, a standard regression line does not give an effective measure of this relationship.  We apply instead the B-spline quantile regression described by \citet{HeNg1999} and implemented within the {\bf R} statistical software system in CRAN package {\it CABS} \citep{NgMaechler2011}.  The spline fits for the 25\%, 50\% (median), and 75\% quartiles to the $Age_{JX}$ values as a function of the $J-H$ absorption indicator are shown in Figure \ref{fig_AgeJX_vs_Reddening}.  No monotonicity or concavity constraint is imposed, and `spline knot' selection is based on the likelihood-based Akaike Information Criterion.   

The median $Age_{JX}$ regression relationship with $J-H$ is well approximated by two lines: 
\begin{eqnarray}
Age_{JX} &=& 4.00 - 1.71 \times (J-H) ~{\rm Myr} ~~~{\rm for}\; 0.5 < J-H < 1.7  \\
                &=& 1.77 - 0.40 \times (J-H) ~{\rm Myr} ~~~{\rm for}\; 1.7 < J-H < 3.5 
\end{eqnarray}

\noindent These equations can be used for estimating cluster ages from $J-H$, provided that the bulk of the reddening is local to the cluster, i.e., local molecular cloud and circumstellar environments.  A star with $J-H = 1.0$, for example, has an estimated age of $Age_{JX} = 2.3$~Myr.  From the upper and lower quartile curves, we estimate that half of an ensemble of such stars would lie in the range $1.2-4.2$~Myr. A heavily absorbed star with $J-H=3.0$ has estimated $Age_{JX}=0.6$~Myr with interquartile range $0.4-1.4$~Myr.  Clearly, the prediction for any single star is very uncertain.  

The $Age_{JX}$ $vs.$ $A_V$ relationship in Figure~\ref{fig_AgeJX_vs_Reddening}(b) favorably compares to a similar relationship recently obtained by \citet{Ybarra2013} from a disk fraction measurement calibrated to a disk evolution timescale and averaged over several clusters in the Rosette Molecular Cloud.  These two age estimates agree well above $A_V > 5$~mag, but Ybarra gets older ages for $A_V \leq 5$~mag. The agreement at high absorption is gratifying given the very different approaches taken in the two studies.  The difference at low absorption might be attributed to different methods for obtaining $A_V$; we estimate extinction by photometric dereddening as shown in Figure~\ref{fig_jhk_ccd}, while Ybarra et al.\ estimate $A_V$ using a reduction in background NIR star surface density.  In any case, we note that the Ybarra et al. values lie within our $25$\%-to-$75$\% interquartile MYStIX range, and thus the two estimates are in formal agreement even at low absorption.
  
As the scatter around the median regression line in Figure~\ref{fig_AgeJX_vs_Reddening} is smaller, and the agreement with \citet{Ybarra2013} is better, for $J-H>1.5$~mag, we use the regression lines at high absorption (equations 1 and 2 above) as a calibration for estimating ages of young, highly reddened subclusters. This is especially useful in cases of subclusters with poor $Age_{JX}$ sampling. The calibration curve is thus employed as an alternative to the median $Age_{JX}$ estimates for highly reddened MYStIX subclusters; these new estimates are called $Age_{JH}$  in \S\ref{subregions_subclusters_section} and Table~\ref{tbl_cluster_ages}.

\subsection{Uncertainties of Individual Stellar Ages\label{error_propagation_section}}

There are many sources of uncertainties in estimating ages of individual PMS stars.  Here we ignore `systematic' uncertainties that can globally shift age scales, such as errors in the distance to a MSFR or errors in the theoretical evolutionary tracks. The effects of distance uncertainties on the MYStIX age gradients resulted from our age analysis (\S \ref{results_subsection}) are further discussed in \S \ref{some_advantage_subsection}.  We concentrate on possible sources of error in $Age_{JX}$ that can produce scatter among age estimates of samples that are truly coeval.  

\begin{figure}
\centering
\includegraphics[angle=0.,width=5.5in]{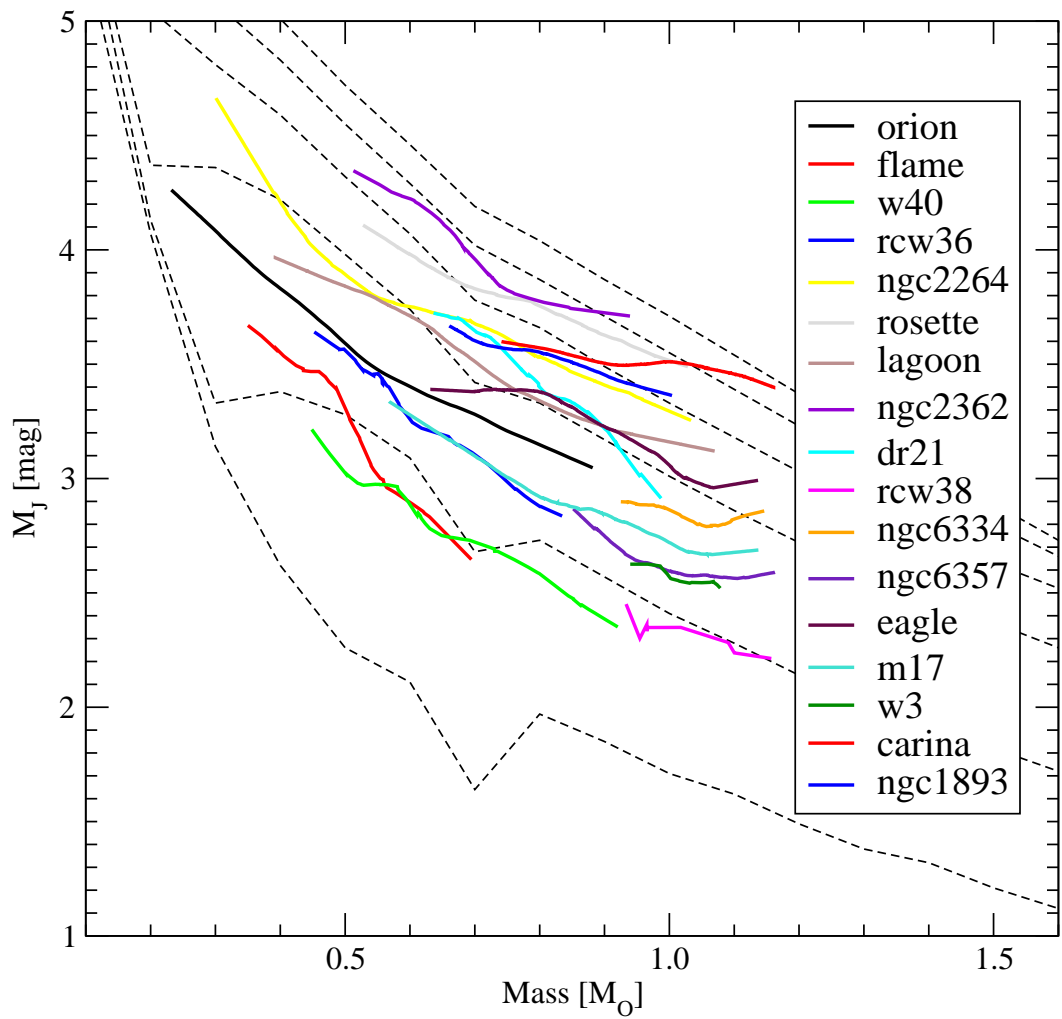}
\caption{Friedman's super smoother regression fits \citep{Friedman1984} derived for 17  MYStIX massive star forming regions from $\sim 5500$ MYStIX Probable Complex Members (MPCM) in a diagram of absolute $J$-band magnitude $vs.$ stellar mass. The dashed lines show, from bottom to top,  0.5, 1, 2, 3, 4 and 5~Myr PMS isochrones from \citet{Siess2000}. \label{fig_mj_vs_mass_ajfromccd_SMOOTH_MEDIANS_V2}}
\end{figure}

For MYStIX stars, typical photometric errors on $J$-band magnitudes (mostly from UKIRT observations) are $\simeq 0.015$~mag. From a multi-epoch NIR study of nearby star forming regions, \citet{Scholz2012} finds that in the $J$-band `half the objects show variations of 5-20 per cent [0.05-0.2 mag]' and  `the other half is not variable'.  Based on this work, we assume a typical scatter of $\Delta J \sim 0.1$~mag due to stellar variability. From the NIR color-color diagram (\S \ref{dereddening_subsection}), the typical scatter of the $J-H$ color around the Taurus Locus for the Taurus (above and below the locus) and the MYStIX (below the locus) stars is $\sim 0.15$~mag. This scatter is likely due to measurement errors, intrinsic imperfection of the locus, and to a lesser extent to abnormal individual extinction for some stars. This scatter of $\Delta (J-H) \sim 0.15$~mag is equivalent to the accuracy of source extinction of $\Delta A_V \sim 1.5$~mag. Thus, the reddening errors probably exceed or are comparable to the variability errors, which in turn exceed the photometric errors.  For the majority of the stars treated here, this reddening uncertainty translates into uncertainties $\Delta Age_{JX} \leq 1$~Myr (Figure~\ref{fig_mj_vs_mass_ajfromccd_SMOOTH_MEDIANS_V2}).

The Poisson noise and modeling uncertainties on $L_X$ for the MYStIX $Age_{JX}$ stars are about 0.2~dex each for typical MYStIX stars treated here \citep{Broos2013}.  But the standard deviation of the residuals on the calibration $L_X - M$ diagram of \citet{Telleschi2007} is much larger with $\sigma_{\log(L_X)} = 0.4$~dex. Recall that multiple factors could contribute to this spread, including X-ray variability, binarity, possible suppression of X-rays by accretion, and intrinsic imperfection of a power law relationship. This scatter in $L_X - M$ corresponds to about 50\% uncertainty in the mass estimation ($\sigma_M = 0.5 \times M$) which in turn propagates into $\sigma_{t} \sim 2$~Myr uncertainty on individual stellar ages (for a typical MYStIX star with mass around $0.8$~M$_{\odot}$ and age around $2$~Myr;  Figure~\ref{fig_mj_vs_mass_ajfromccd_SMOOTH_MEDIANS_V2}). 

Thus, the individual age estimates are expected to be very uncertain, and wide scatter in individual $Age_{JX}$ estimates should be present.  Application of the $L_X-M$ transformation is likely the major source of the uncertainty.  The age uncertainties are higher for more massive and/or older stars, but are lower for less massive and/or younger stars.  For example, the stars we treat in the Orion Nebula Cluster (ONC) have median $Age_{JX}$  (mass) of $1.6$~Myr ($0.6$~M$_{\odot}$), and we expect $\sigma_{t} \sim 1$~Myr uncertainty on individual stellar ages. Formally, this is similar to the uncertainty on individual stellar ages that are reported for {\it Hubble Space Telescope} photometry and ground-based optical spectroscopy and photometry of stars in ONC \citep{Reggiani2011}: $\sigma_{log(t)} \sim 0.15$~dex for typical age values of $2.2$~Myr.

{\it It is critical to restate our objectives here. We are not principally interested in accurately evaluating individual stellar ages, or interested in the spread of the age distribution for an entire MSFR.  Our principal interest is a `typical' age value for spatially distinct subclusters within a MSFR.}  Rather than using the mean as an estimate of `typical' age, we use the median value.  The median has many advantages: it is robust to outliers and non-Gaussian scatter, invulnerable to the choice of logarithmic or other variable transformation, and is unaffected by a truncation at old ages seen in some regions\footnote{The evolutionary models used in our analysis stop at 5~Myr. Note that for any MYStIX subcluster, the number of 5~Myr old members is aways much lower than the number of $<5$~Myr members, so the `age saturation' effect at $5$~Myr does not affect the estimate of the median age.}.  

We estimate  an uncertainty for each subcluster median age using nonparametric `bootstrap case resampling' \citep[e.g., Wikipedia\footnote{\url{http://en.wikipedia.org/wiki/Bootstrapping\_(statistics)}};][]{DavisonHinkley1997}. This technique generates numerous sets of synthetic stellar ages by resampling with replacement from the observed stellar ages in a MYStIX subcluster and computing the median of each synthetic data set.  The standard deviation of the ensemble of synthetic medians is then used as an estimate of the uncertainty of the median age. If the stellar ages in a subcluster are normally distributed, then this standard deviation represents a 68\% confidence interval on the sample median.

A different well established technique for characterizing uncertainty on the median statistic, based on integrals of the binomial distribution \citep[e.g.,][]{Conover1980}, can calculate a confidence level for any confidence interval whose end points lie at values found in the data set.  Using this binomial technique, we made a second estimate for the uncertainty of our sample median  by searching among the allowed confidence intervals for the one with the smallest confidence level that exceeds 68\%.  For most MYStIX subclusters, the two error estimates differ by $<35$\%.

We report the bootstrap errors as the uncertainties on subcluster ages in Table~\ref{tbl_cluster_ages}.  These uncertainties are $\sim 0.5$~Myr for typical MYStIX subclusters ($N \sim 20$ stars with $Age_{JX}$ estimates), which is often sufficiently accurate to discriminate age differences and gradients among subclusters.  Richer subclusters naturally have smaller median age uncertainties.

\subsection{Comparing $Age_{JX}$ to Ages from Optical Data \label{individual_ages_section}}

The $Age_{JX}$ estimates can be compared to age estimates independently derived from optical HRDs/CMDs. For this analysis, we choose the nearby and best studied MYStIX star forming regions, the Orion Nebula ($d = 414$~pc) and NGC~2264 ($d = 913$~pc), two of the most lightly absorbed MSFRs in the MYStIX survey.   

For the Orion Nebula, optical ages are obtained from \citet{DaRio2010} where they were computed by transforming measurements onto the HRD and applying the  \citet{Siess2000} PMS tracks. $T_{eff}$ and $L_{bol}$ values were derived from ground-based optical $UBVI$ and TiO-band photometry and spectroscopy. Their detailed analysis includes construction of synthetic spectra for 2~Myr Siess models to estimate bolometric corrections, derivation of some spectral types from TiO indices, and corrections for reddening and accretion luminosity.   Da Rio et al. also use a distance of 414~pc to the region.  The Da Rio et al. analysis results in a broad range of ages with most Orion stars lying between 0.3 and 10~Myr with a peak at $2-3$~Myr.  The age dispersion is present for all stellar masses. We compare  $Age_{JX}$ and the Da Rio et al. ages for 263 stars in common between the two samples.

For NGC~2264, the optical data are obtained from \citet{Mayne2007} who compute ages from the $V~vs.~V-I$ CMD. Their sample is a union of PMS members obtained from proper motion, X-ray, H$\alpha$, periodic variability, and radial velocity surveys. Their sample is then  truncated to a lightly absorbed stellar sub-sample with $J-H<1$~mag.  Mayne et al. compare their observed CMD to the model predictions of \citet{Siess2000} and obtain a collective age $\simeq 3$~Myr with some age spread. Since Mayne et al. do not report individual ages and their assumed distance to the region of 832~pc differs from our assumed distance of 913~pc, here we recalculate optical ages using their optical $VI_C$ photometry data.  The optical photometric ages are computed assuming the uniform source extinction for the lightly-absorbed stellar sample of $A_V = 0.22$~mag \citep{Dahm2005} and the reddening law of $A_I/A_V = 0.6$ \citep{BessellBrett1988}. The comparison sample with optical and $Age_{JX}$ age estimates in NGC~2264 has 152 stars.

\begin{figure}
\centering
\includegraphics[angle=0.,width=5.5in]{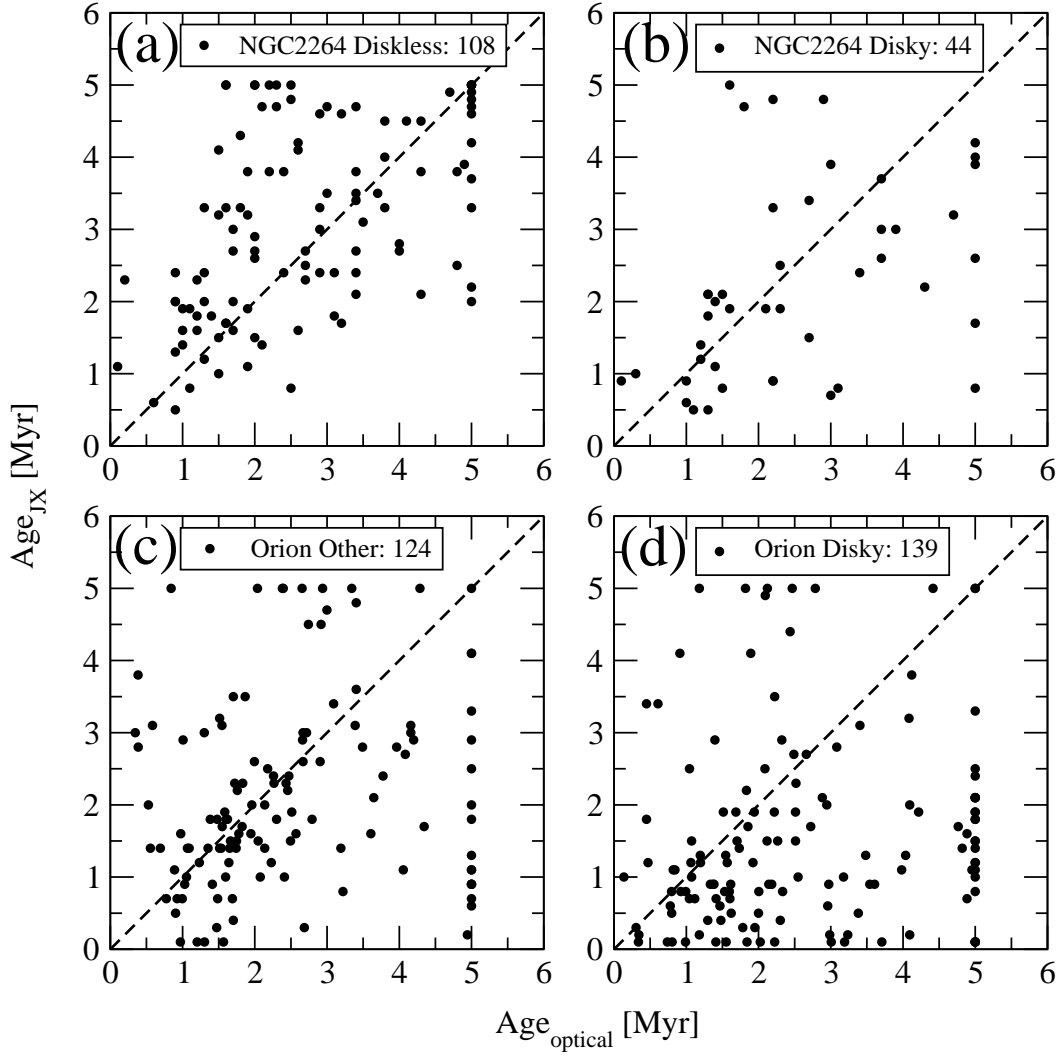}
\caption{Comparison of stellar ages derived here from infrared and X-ray data ($Age_{JX}$) with ages derived from optical Hertzsprung-Russell or color magnitude diagrams $Age_{opt}$.  Left panels show mainly disk-free stars and right panels show disk-bearing stars.  Top panels show NGC~2264 stars and bottom panels show Orion Nebula stars.  Black dashed lines indicate expected locus if the two methods give identical ages. Stars shown at 5~Myr (0.1~Myr) have inferred age estimates $\ga 5$~Myr ($\la 0.1$~Myr). \label{fig_validation_ngc2264_orion}}
\end{figure}

Figure \ref{fig_validation_ngc2264_orion} plots the $Age_{JX}$ against the ages derived from the optical HRD (Orion) or CMD (NGC~2264);  agreement would distribute the stars along the dashed lines.  The discrimination between diskless and disk-bearing stars is based on the analysis of \citet{Povich2013} for NGC~2264 and \citet{Megeath2012} for Orion.  In all four panels, the correlation between the $Age_{JX}$ and optical ages is statistically significant based on Kendall's $\tau$ nonparametric measure of bivariate correlation with $\tau \simeq 0.35$ for NGC~2264 and $\tau \simeq 0.17$ for the Orion Nebula.   

But the agreement between the two age estimates for many individual stars is generally poor with dispersion at any chosen age only slightly smaller than that present in the full sample. These large spreads are likely due to the poor age estimates for individual stars from optical \citep[e.g.,][]{Reggiani2011} and MYStIX (\S \ref{error_propagation_section}) data. The greater scatter at older ages is due to the smaller $M_J$ ($\log(L_{bol})$) separation of older isochrones and thus the higher age uncertainties for older stars. Only small offsets in the medians of stellar ages are seen for the sample of all NGC~2264 stars and the sample of mainly diskless Orion stars. We have checked to ensure that different assumed extinction laws, the $A_J/A_{K_s} = 2.65$ choice of \citet{Megeath2012} versus our choice of $A_J/A_{K_s} = 2.5$ \citep{Indebetouw2005}, have a negligible effect on the dispersion or bias in the scatter diagram for the disk-bearing Orion stars.

The comparison between $Age_{JX}$ and optical ages changes with different methods, assumptions for source extinction, and catalogs (published versus new unpublished optical catalogs).  There are ways to improve the optically-derived ages (N. Da Rio, private communication). No clear scientific conclusion can be inferred from the poor association between our new age estimator and traditional HRD- and CMD-based age estimators. If there was confidence that $Age_{opt}$ values were accurate, then we would infer that $Age_{JX}$ values were not useful.  But, as discussed in \S \ref{error_propagation_section}, both $Age_{JX}$ and $Age_{opt}$ values for individual stars are inaccurate (see also \S\ref{some_advantage_subsection}). The only conclusion we can draw from Figure \ref{fig_validation_ngc2264_orion} is that inferences based on stellar ages, such as age spreads within clusters and age gradients across star formation complexes, may differ when $Age_{JX}$ rather than $Age_{opt}$ values are used. However, most MYStIX MSFRs suffer too much obscuration for optical photometry and spectroscopy, so $Age_{opt}$ estimates can not be readily obtained for our sample.

\section{Ages of MYStIX Regions and subclusters} \label{subregions_subclusters_section}

\subsection{Overview}

\begin{figure}
\centering
\includegraphics[height=0.7\textheight]{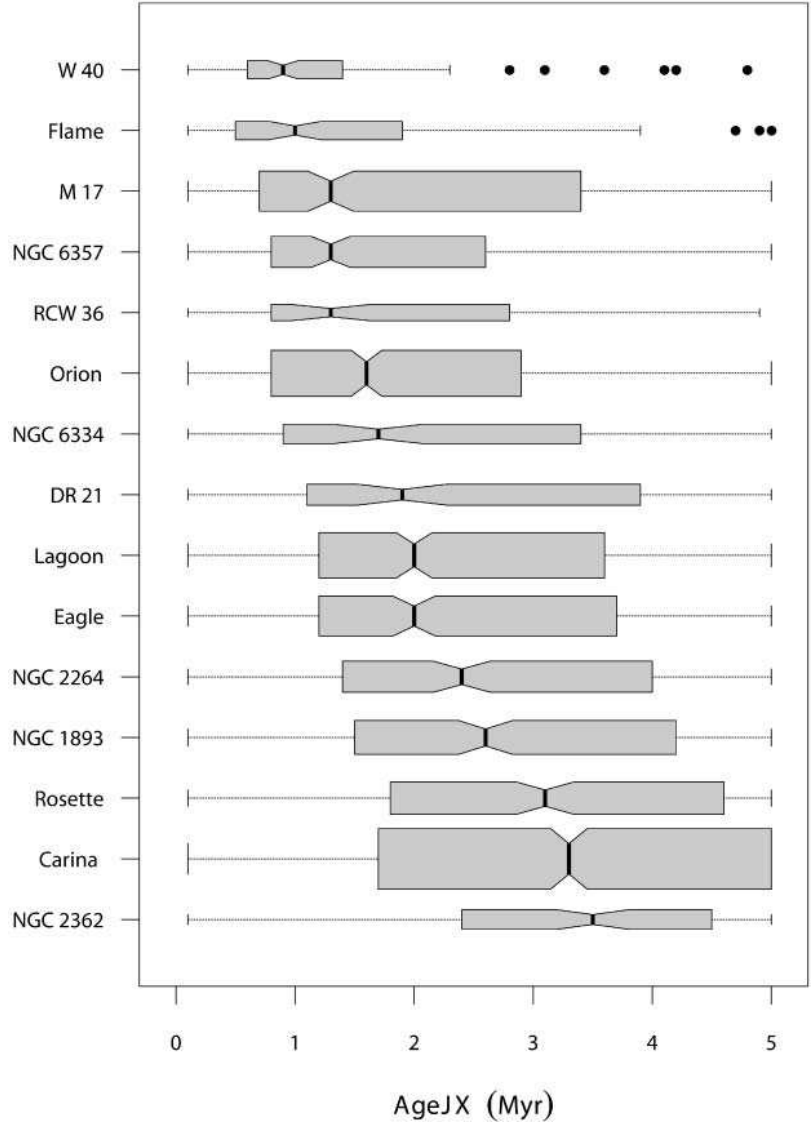}
\caption{Boxplots of the $Age_{JX}$ distributions for the MYStIX massive star forming regions shown in order of youngest (top) to oldest (bottom) median ages. The median age is shown as a dark bar.  See text for details.  
\label{fig_agejx_full_boxplot}}
\end{figure}

Figure~\ref{fig_agejx_full_boxplot} shows boxplots of the $Age_{JX}$ distributions for the 15 MYStIX regions under consideration. The `box' indicates the 25\%, 50\% (median), and 75\% quartiles with `whiskers' set at 1.5 times the interquartile range.  Points lying beyond these values are plotted individually as outliers; a Gaussian distribution would have few outliers.  The notches around the median approximate the confidence intervals of the median.  The box width is scaled to the square root of the sample size.  Recall that older ages are all set to be 5~Myr.   

All MYStIX regions show wide age distributions.  Some distinction between regions can be discerned:  W~40 and Flame Nebula are youngest (medians $<1.0$~Myr) while Rosette Nebula, Carina Nebula, and NGC~2362 are oldest (medians $>3.0$~Myr).  But the interquartile ranges of even these extreme cases greatly overlap.  Much of this dispersion is undoubtedly due to the diverse environments in these fields. The Rosette Nebula fields, for example, have embedded clusters as well as the main revealed cluster \citep{Wang2010}.  The Carina Nebula has both embedded groups in the South Pillars and the older cluster Tr~15 and thousands of older dispersed stars \citep{Townsley2011}. Figure~\ref{fig_AgeJX_vs_Reddening} shows that stars drawn from a range of absorptions $-$ revealed and embedded clusters $-$ will exhibit a wide range of ages.  In our view, the study of ages averaged over the full MYStIX regions (as shown in Figure~\ref{fig_agejx_full_boxplot}) is not a fruitful enterprise. 

We instead follow the lead of \citet{Kuhn2014c} who have divided the MPCM population into spatially distinct clusters using an objective statistical algorithm.  A `finite mixture model' composed of multiple isothermal ellipsoidal clusters and an unclustered population is fit to the two-dimensional positions of MPCM stars.  Each subcluster component has six parameters (right ascension, declination, core radius, central stellar density, ellipticity, and ellipse orientation angle) and the fit is found by maximizing the likelihood of the parameters. The number of clusters in a MYStIX region is chosen by maximizing the penalized likelihood Akaike Information Criterion assisted, in some cases, with minimizing extrema in smoothed residual maps.  The mixture model is then used as a `soft classifier' that, with additional decision rules, allows individual MPCM stars to be assigned to the clusters and to the unclustered component.

Table~\ref{tbl_cluster_ages} presents some of the cluster properties for over a hundred subclusters (denoted "A", "B", ...) and  unclustered or ambiguous components (denoted "U") found by \citet{Kuhn2014c} in 15 MYStIX MSFRs\footnote{Five regions are omitted due to the poor $Age_{JX}$ sampling of MYStIX subclusters: RCW~38, W~3, W~4, Trifid Nebula, and NGC~3576.}. Each subcluster component is assigned an age (column 11), defined as the median $Age_{JX}$ of the stellar members of the component. The uncertainties shown for the median ages are symmetric 68\% confidence intervals estimated by bootstrap resampling (\S\ref{error_propagation_section}). $Age_{JX}$ values in column (11) are oomitted when the subcluster is very sparse with $<3-5$ members that satisfy the criteria for $Age_{JX}$ analysis (\S\ref{sample_section}).  For 40 highly reddened subclusters with $(J-H)_{tot} > 1.5$~mag, column (12) lists the $Age_{JH}$ estimated from the spline fit to the $Age$ $vs.$ $(J-H)$ data given by equations (1)-(2) in \S~\ref{reddening_effect_section}. These $Age_{JH}$ estimates are fairly consistent with the $Age_{JX}$ estimates obtained from the median of individual star members of the subcluster shown in column (11).  For 21 sub-clusters with both age estimates, there is only a small bias $(Age_{JX} - Age_{JH}) = 0.2$~Myr and a dispersion of only 0.3~Myr.

For the Orion Nebula, Flame Nebula, and M~17 MSFRs, we also investigate age estimates for additional groups of stars defined by combining or subdividing \citet{Kuhn2014c} clusters. The respective results on these special age estimates are given in the current paper for M~17 (\S \ref{age_m17_subsection}) and in a companion paper for Orion and Flame \citep{Getman2013ApJL}.

\subsection{Principal Results \label{results_subsection}}

\begin{figure}
\centering
\begin{minipage}[t]{1.0\textwidth}
  \centering
  \includegraphics[angle=0.,width=5.5in]{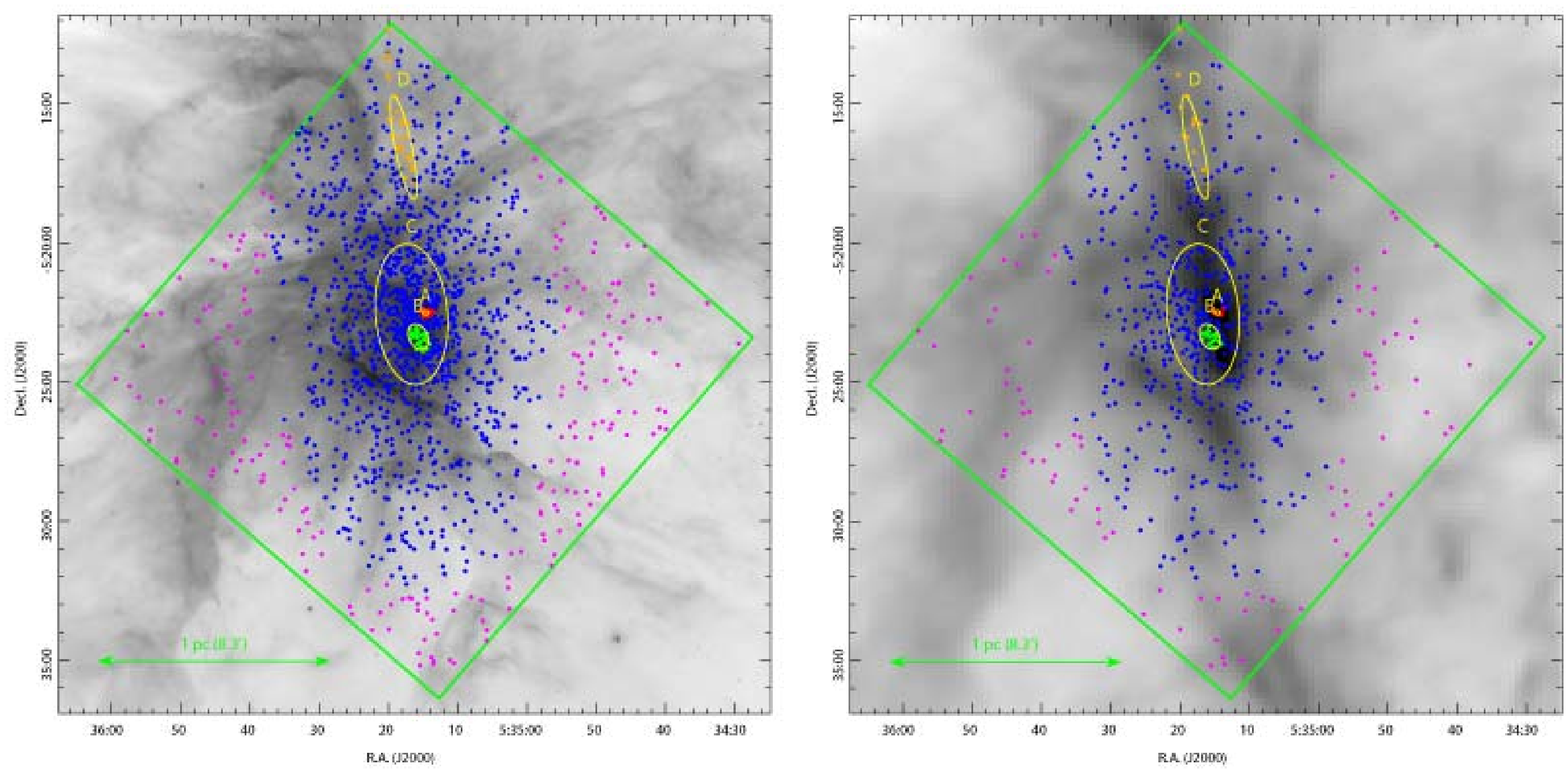} \hspace{0.0in}
  \includegraphics[angle=0.,scale=0.7]{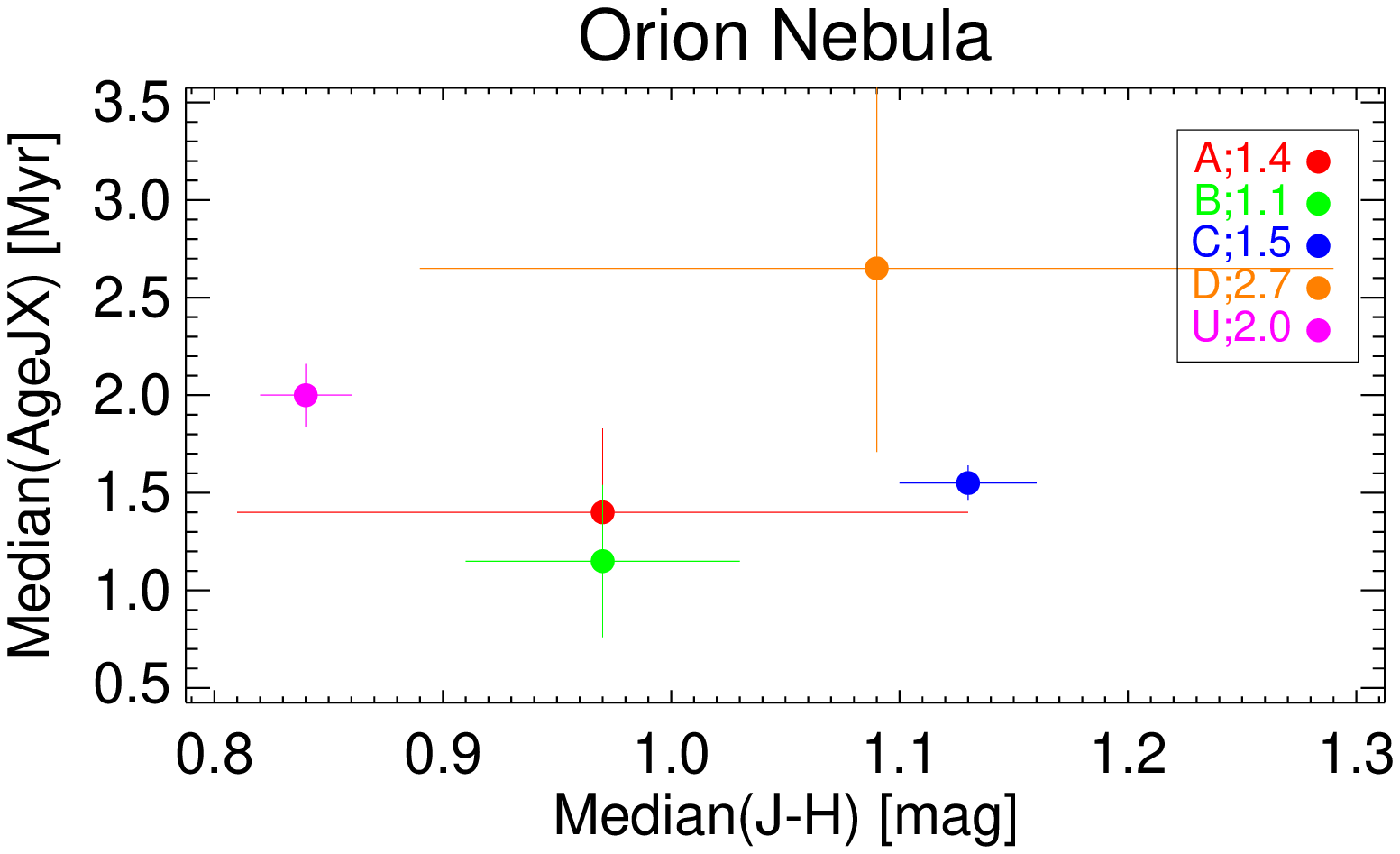} \hspace{0.0in}
\caption{\small $Age_{JX}$ analysis of the Orion Nebula. {\it Upper left:} The full MPCM sample with subcluster membership coded by color and the elliptical contours (yellow) showing the core radii of the isothermal structures from \citet{Kuhn2014c}.  These symbols are  superimposed on the 8.0~$\mu$m $Spitzer$-IRAC image. In this and following figures, the unclustered/ambiguous stars are shown in magenta color, and the {\it Chandra} field of view is outlined by the green polygon.  {\it Upper right:} The subset of MPCM stars available for $Age_{JX}$ analysis and the elliptical contours superposed on the 500~$\mu$m $Herschel$-SPIRE image.   {\it Lower:} The median $Age_{JX}$ estimates for the four subclusters and the unclustered component plotted against their respective median $(J-H)$ color indices.  The legend identifies each point with a subcluster and states the median age in Myr.  These ages are listed in Table~\ref{tbl_cluster_ages}.  \label{fig_age_gradients_orion}}
\end{minipage}
\end{figure}

\begin{figure}
\centering
\begin{minipage}[t]{1.0\textwidth}
  \centering
  \includegraphics[angle=0.,width=5.5in]{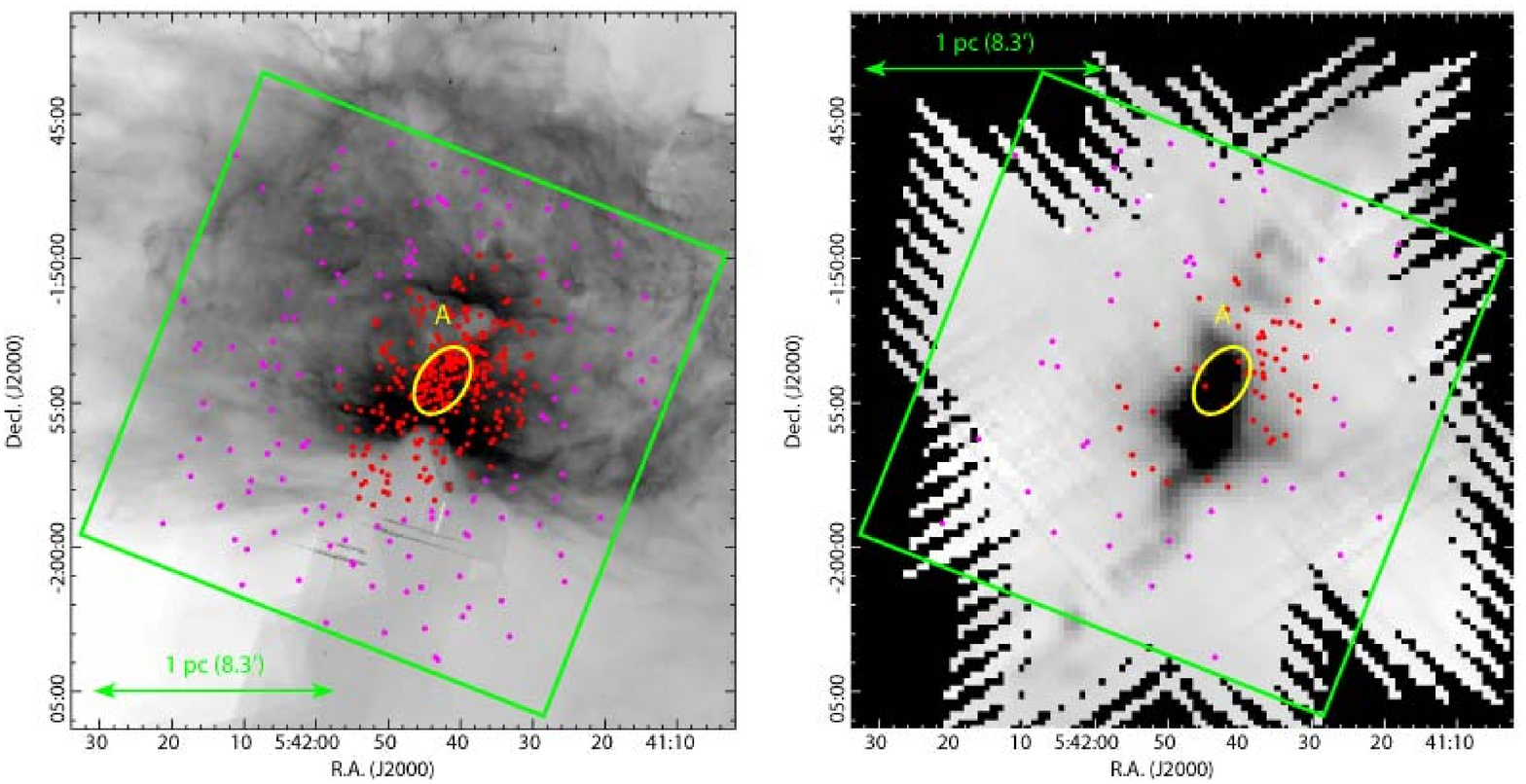} \hspace{0.0in}
  \includegraphics[angle=0.,scale=0.9]{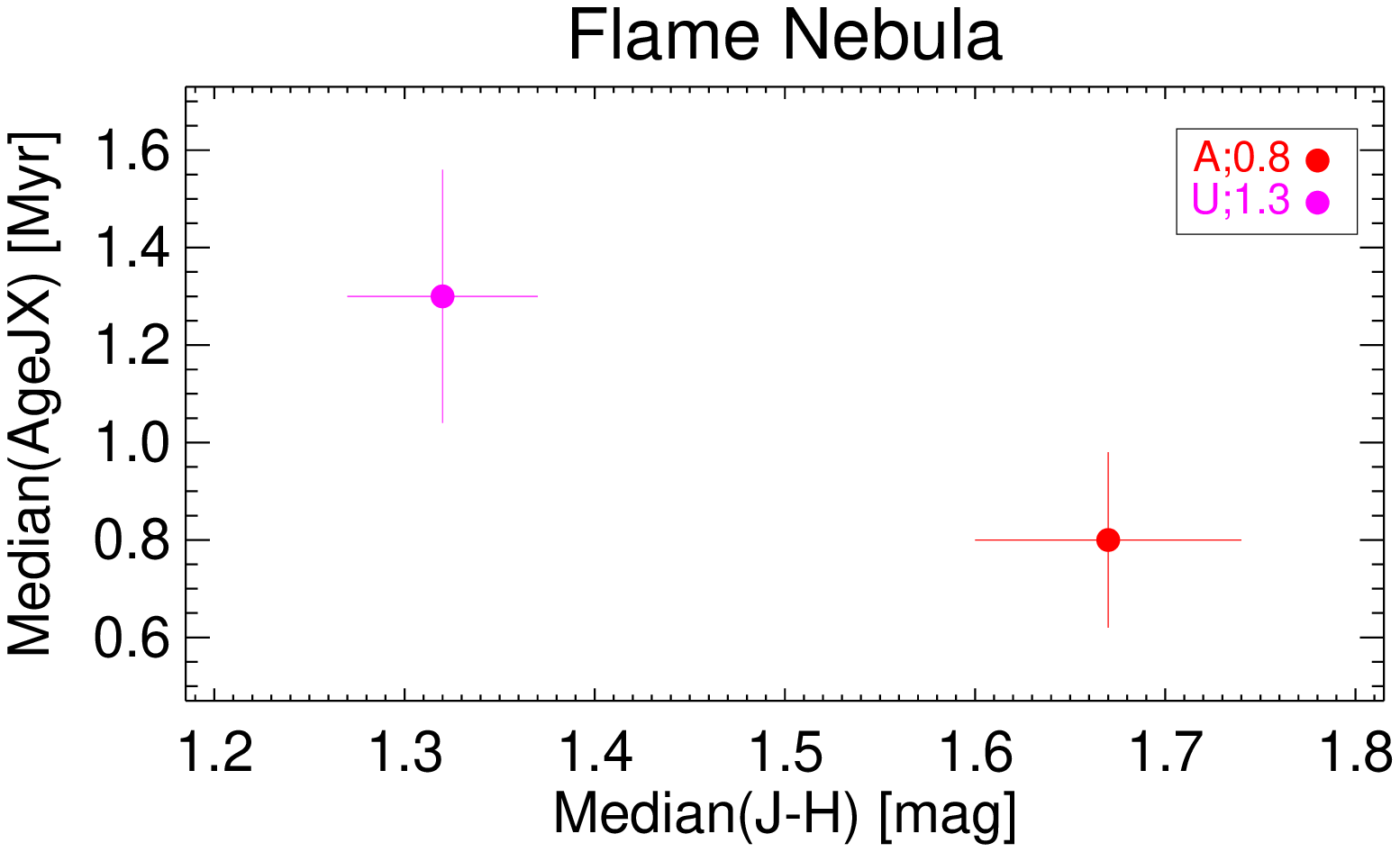} \hspace{0.0in}
\caption{$Age_{JX}$ analysis of Flame Nebula. See Figure~\ref{fig_age_gradients_orion} for description.  \label{fig_age_gradients_flame}}
\end{minipage}
\end{figure}

\begin{figure}
\centering
\begin{minipage}[t]{1.0\textwidth}
  \centering
  \includegraphics[angle=0.,width=5.5in]{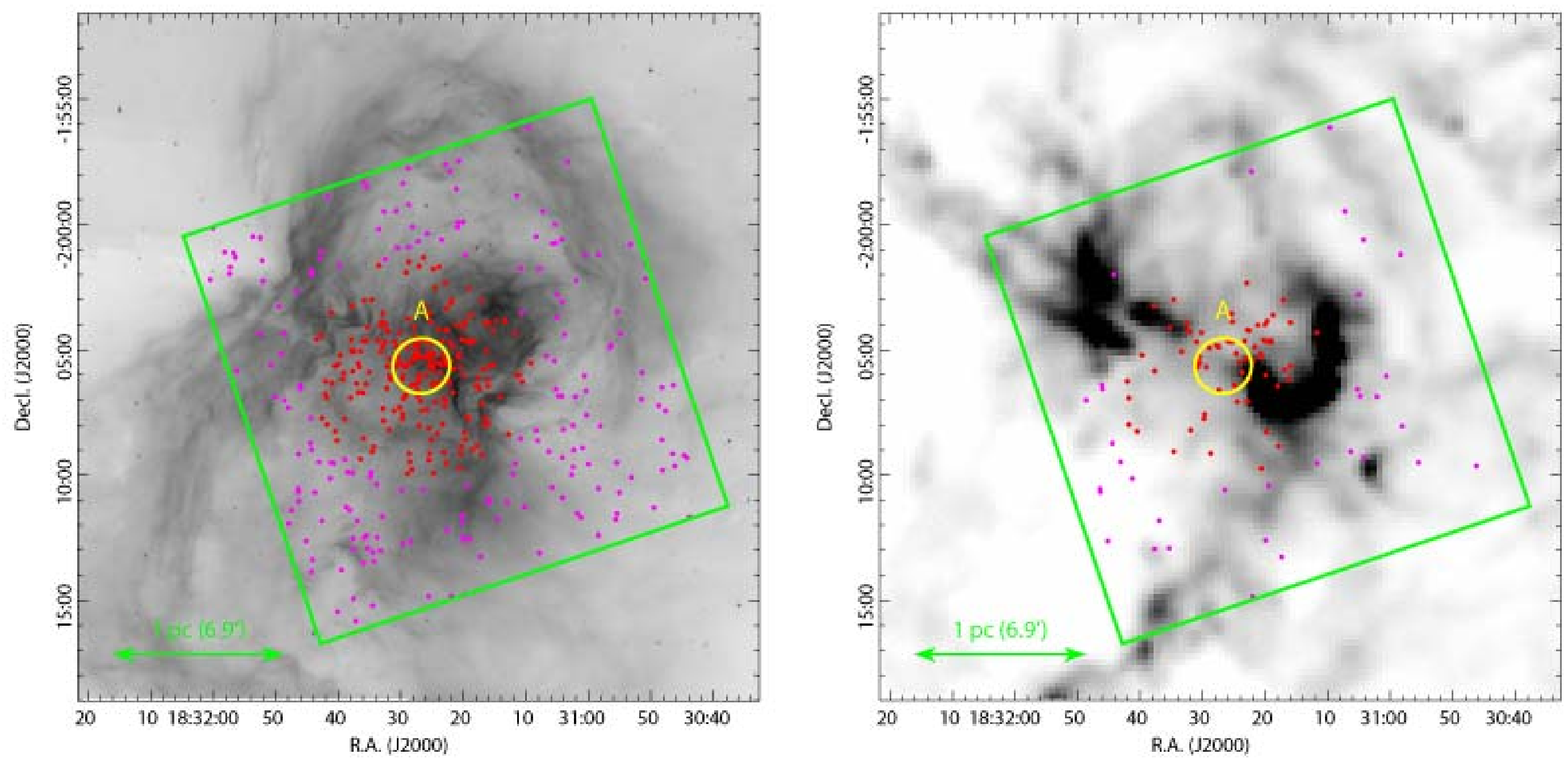} \hspace{0.0in}
  \includegraphics[angle=0.,scale=0.9]{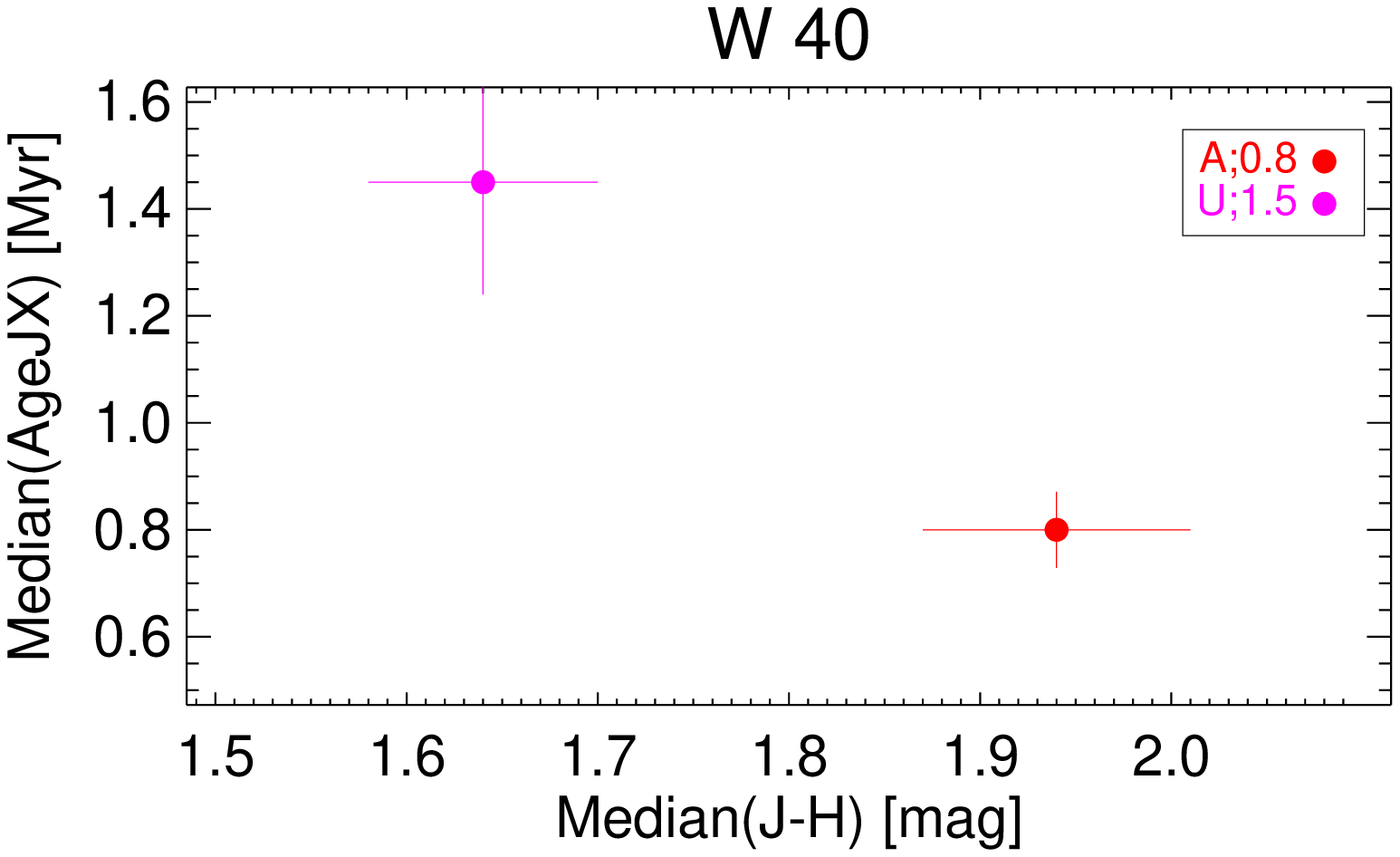} \hspace{0.0in}
\caption{$Age_{JX}$ analysis of W~40. See Figure~\ref{fig_age_gradients_orion} for description.   \label{fig_age_gradients_w40}}
\end{minipage}
\end{figure}

\begin{figure}
\centering
\begin{minipage}[t]{1.0\textwidth}
  \centering
  \includegraphics[angle=0.,width=5.5in]{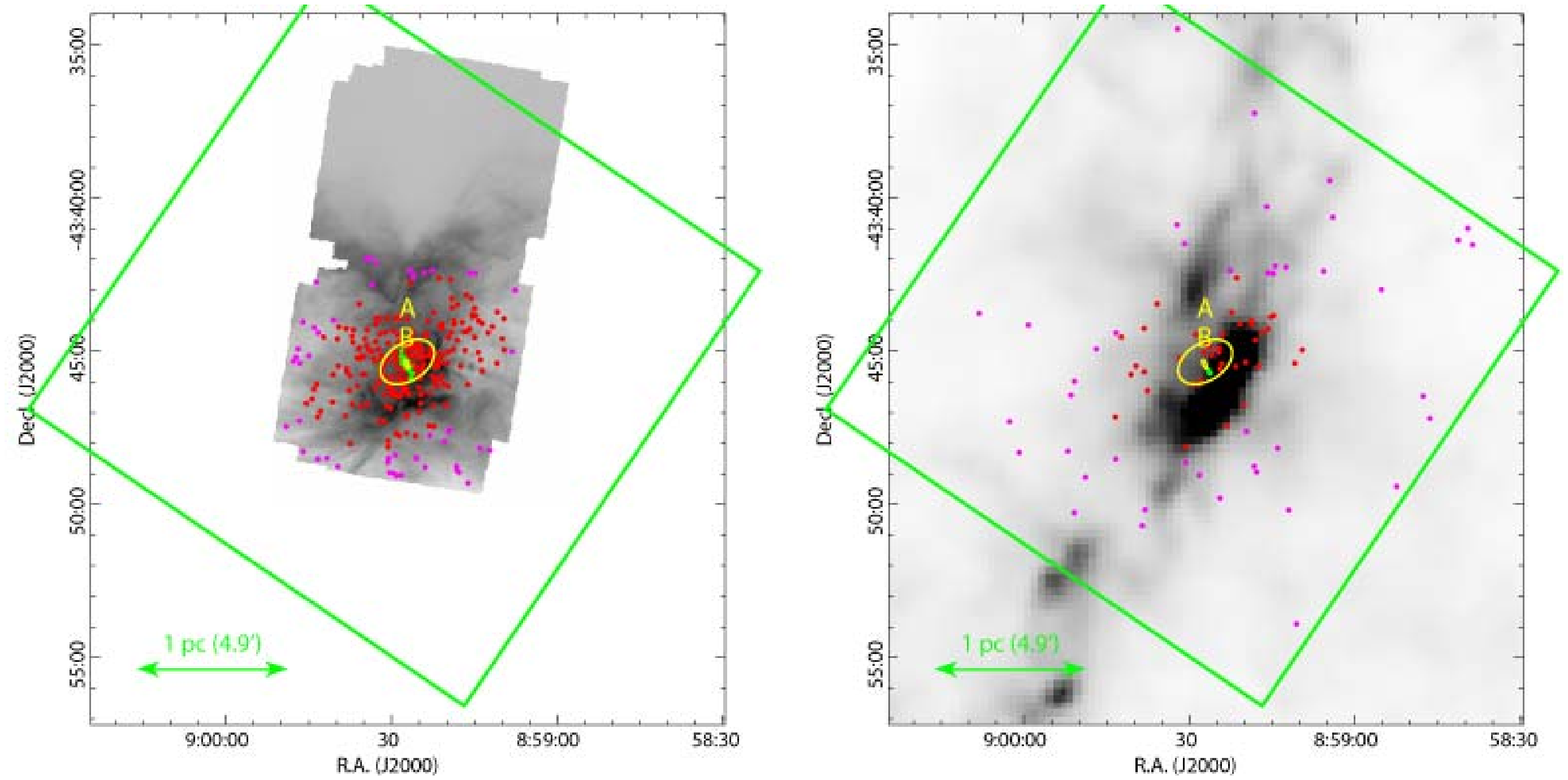} \hspace{0.0in}
  \includegraphics[angle=0.,scale=0.9]{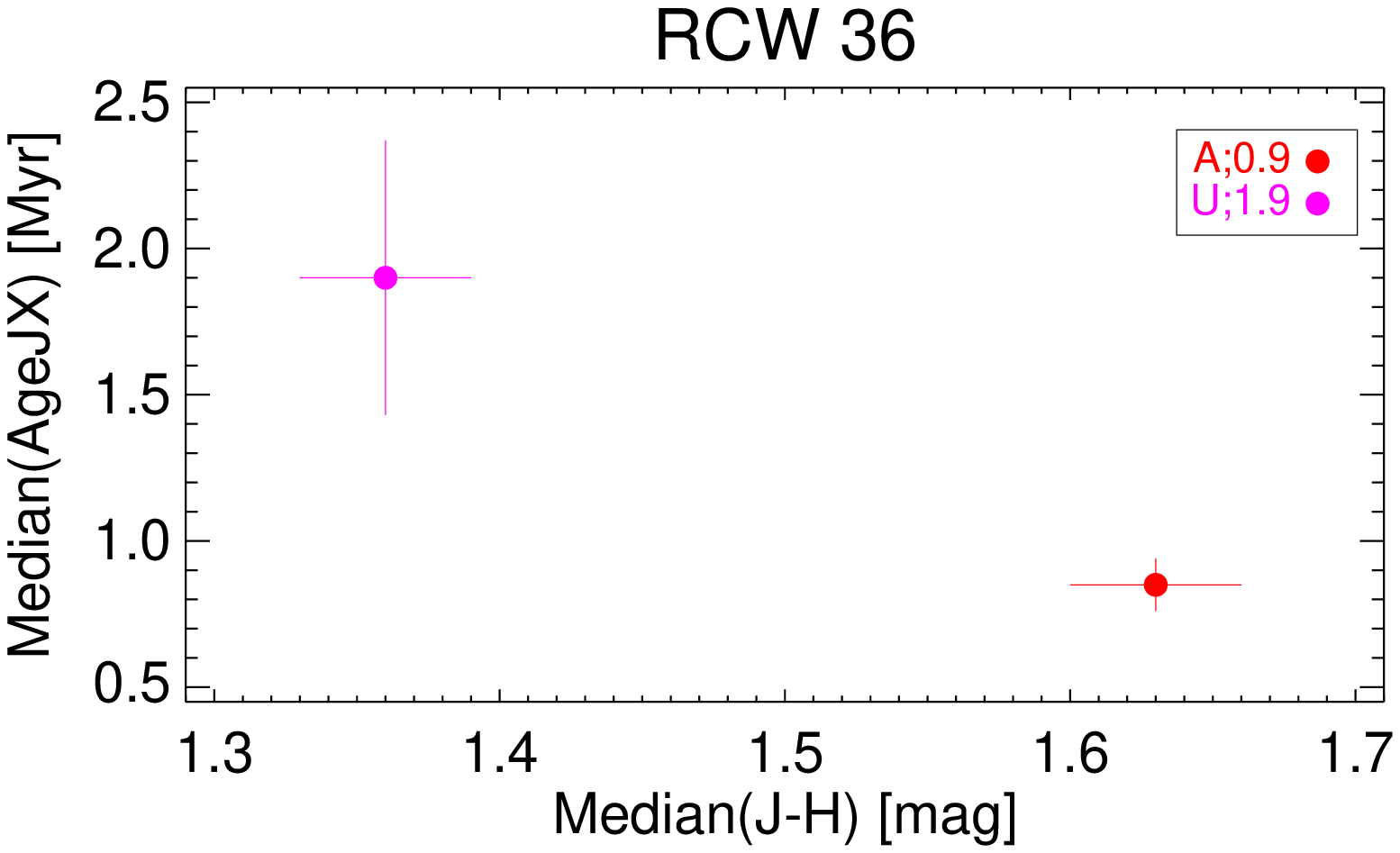} \hspace{0.0in}
\caption{$Age_{JX}$ analysis of RCW~36. See Figure~\ref{fig_age_gradients_orion} for description.   \label{fig_age_gradients_rcw36}}
\end{minipage}
\end{figure}

\begin{figure}
\centering
\begin{minipage}[t]{1.0\textwidth}
  \centering
  \includegraphics[angle=0.,width=5.5in]{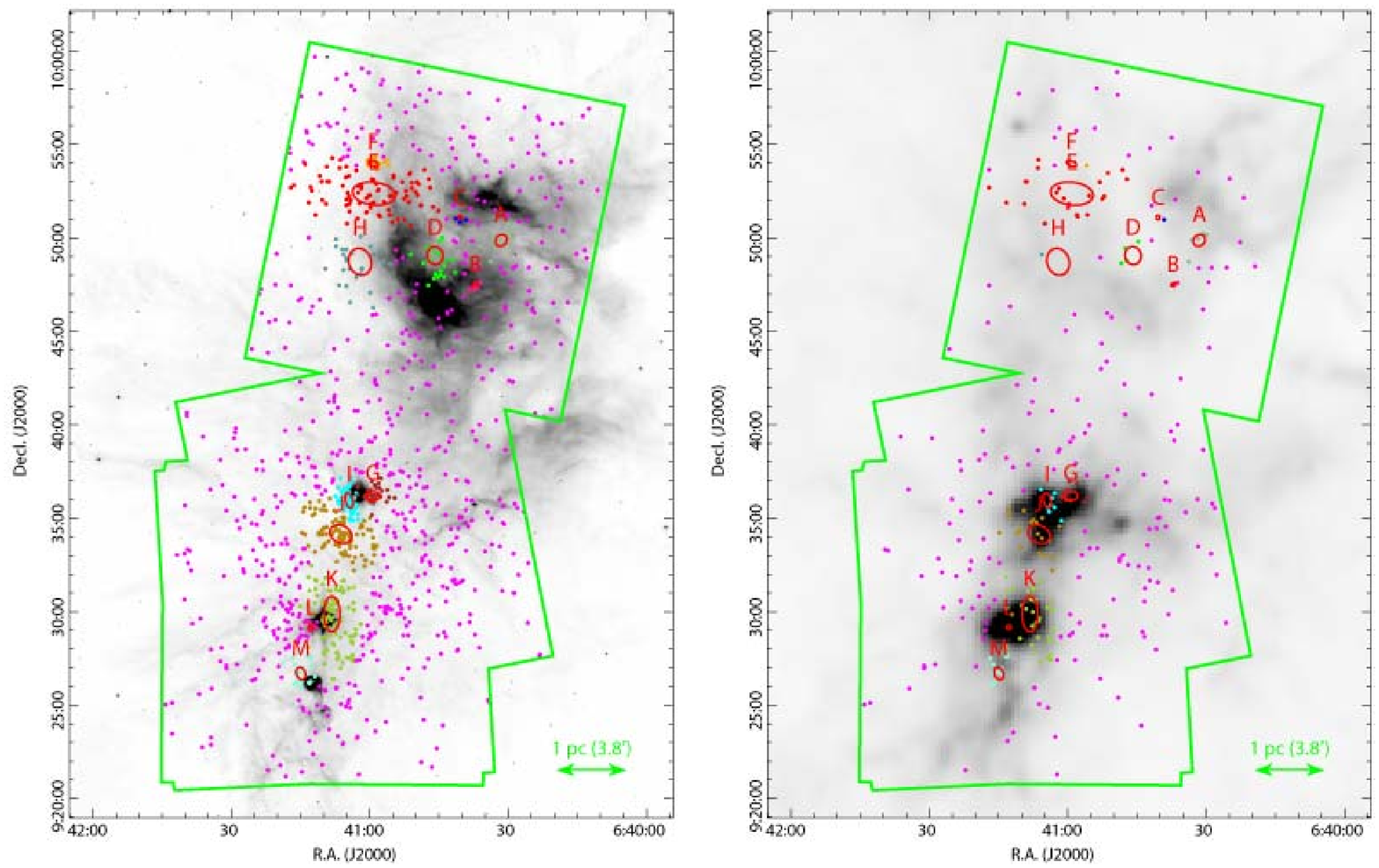} \hspace{0.0in}
  \includegraphics[angle=0.,scale=0.9]{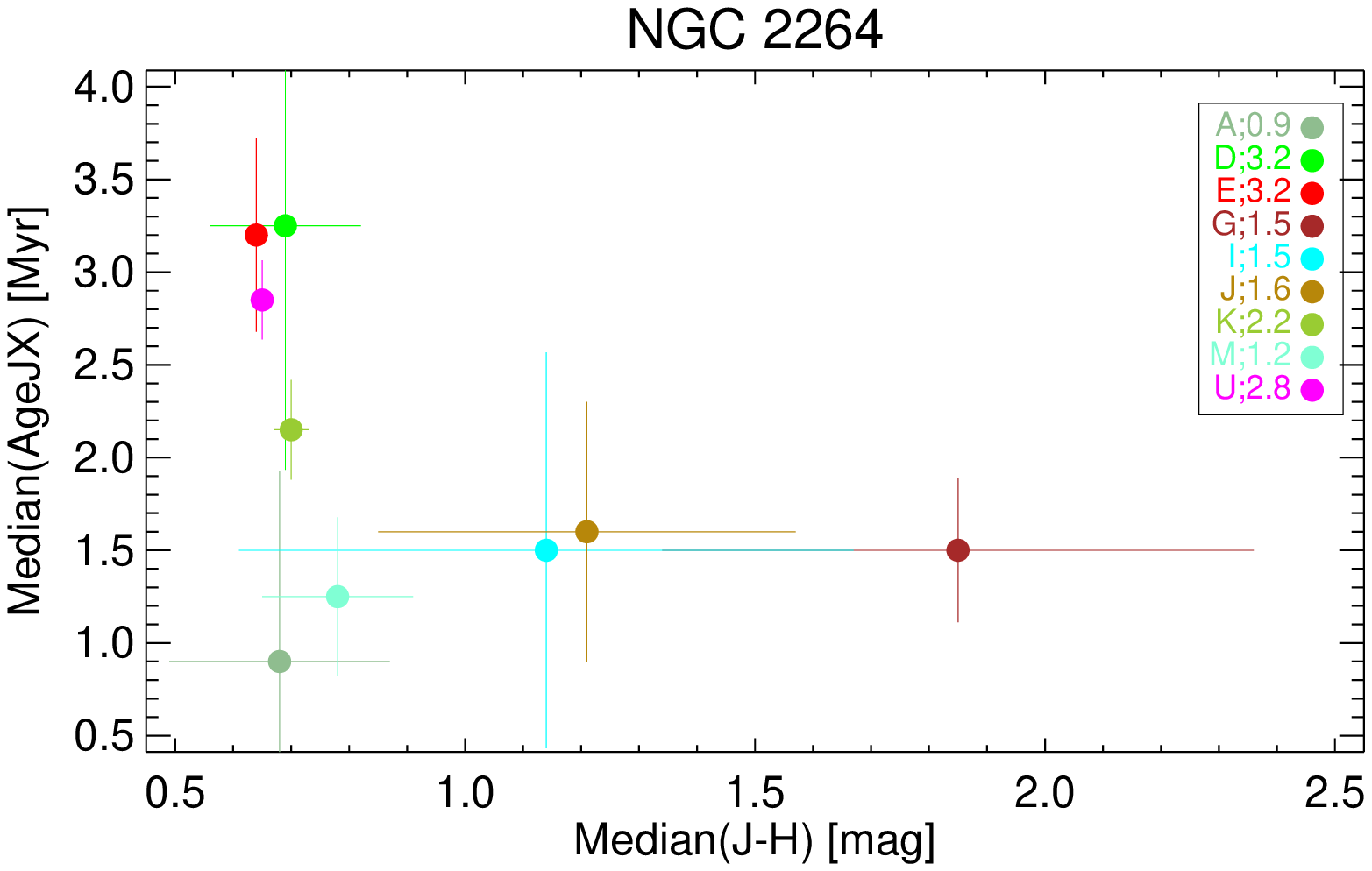} \hspace{0.0in}
\caption{$Age_{JX}$ analysis of NGC~2264. See Figure~\ref{fig_age_gradients_orion} for description. \label{fig_age_gradients_ngc2264}}
\end{minipage}
\end{figure}

\begin{figure}
\centering
\begin{minipage}[t]{1.0\textwidth}
  \centering
  \includegraphics[angle=0.,width=5.5in]{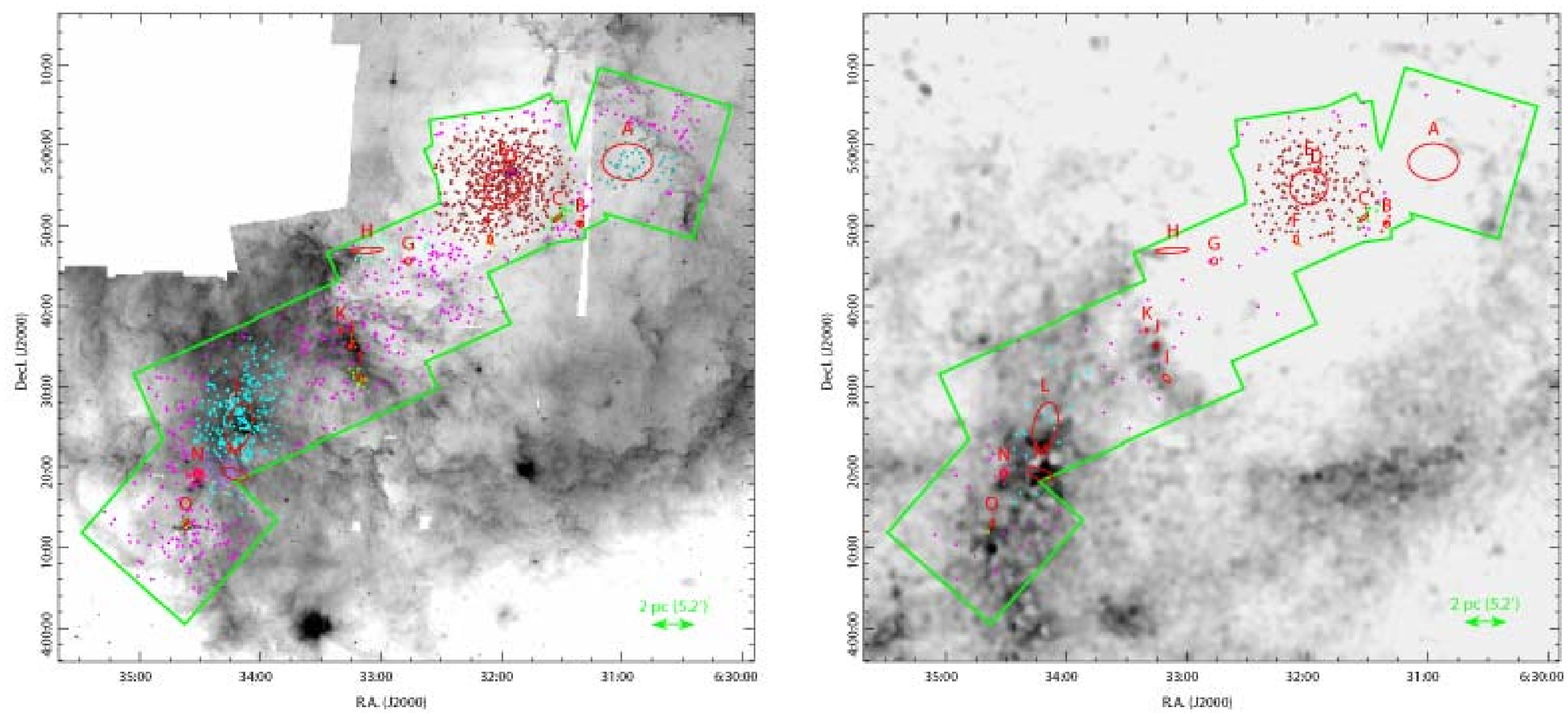} \hspace{0.0in}
  \includegraphics[angle=0.,scale=0.9]{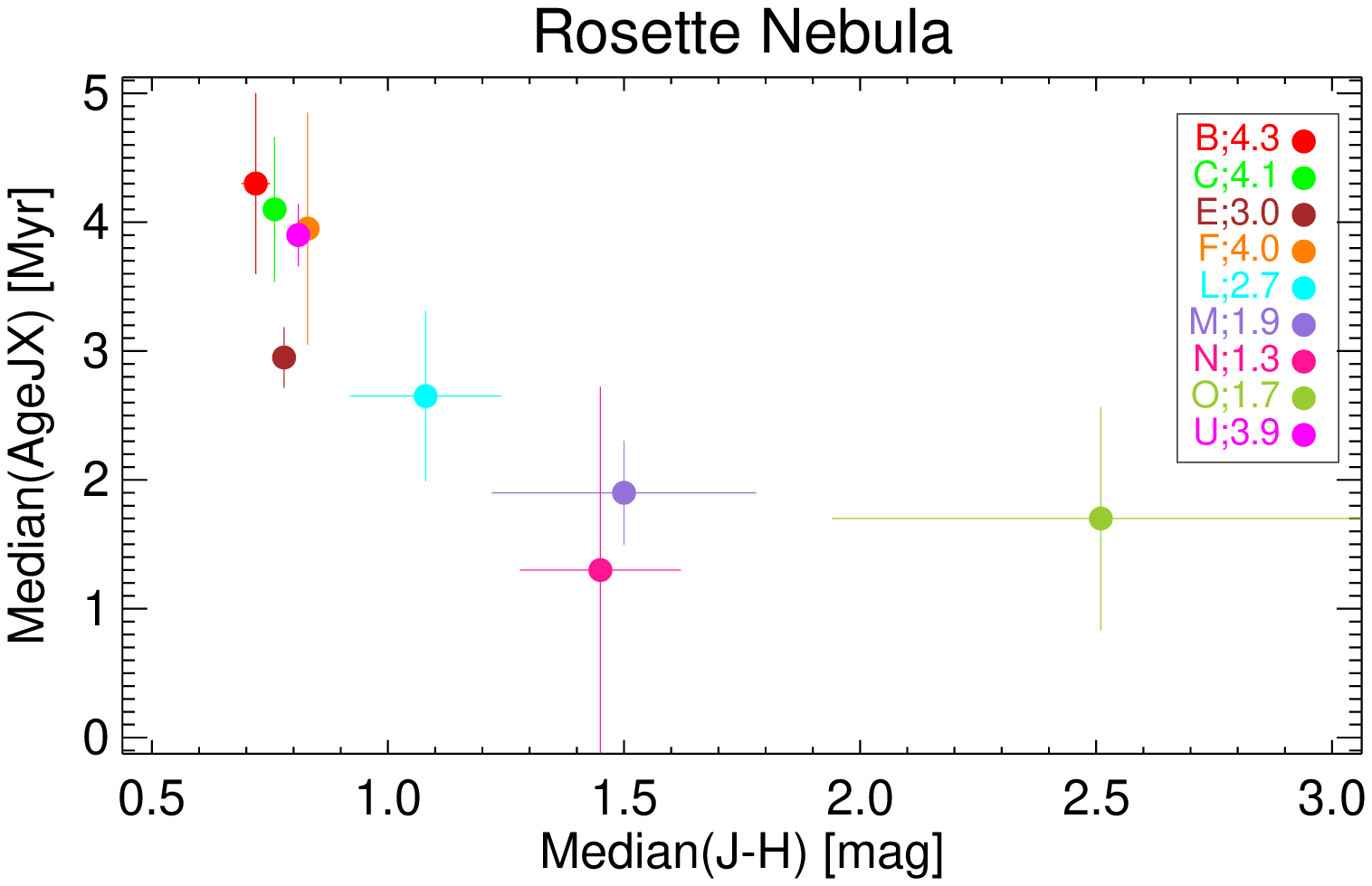} \hspace{0.0in}
\caption{$Age_{JX}$ analysis of the Rosette Nebula. See Figure~\ref{fig_age_gradients_orion} for description. {\it Upper right:} The subset of MPCM stars available for $Age_{JX}$ analysis and the elliptical contours superposed on the $A_V$ map from \citet{Broos2013}.    \label{fig_age_gradients_rosette}}
\end{minipage}
\end{figure}

\begin{figure}
\centering
\begin{minipage}[t]{1.0\textwidth}
  \centering
  \includegraphics[angle=0.,width=5.5in]{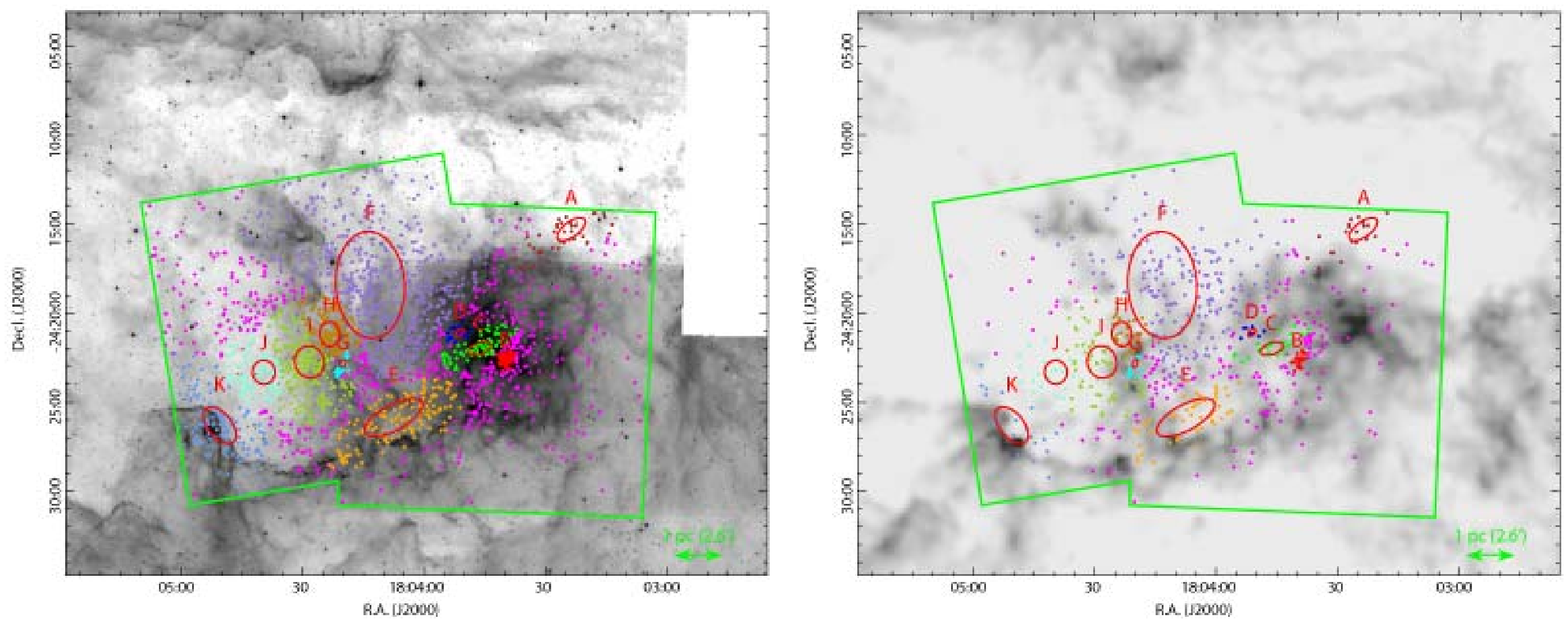} \hspace{0.0in}
  \includegraphics[angle=0.,scale=0.9]{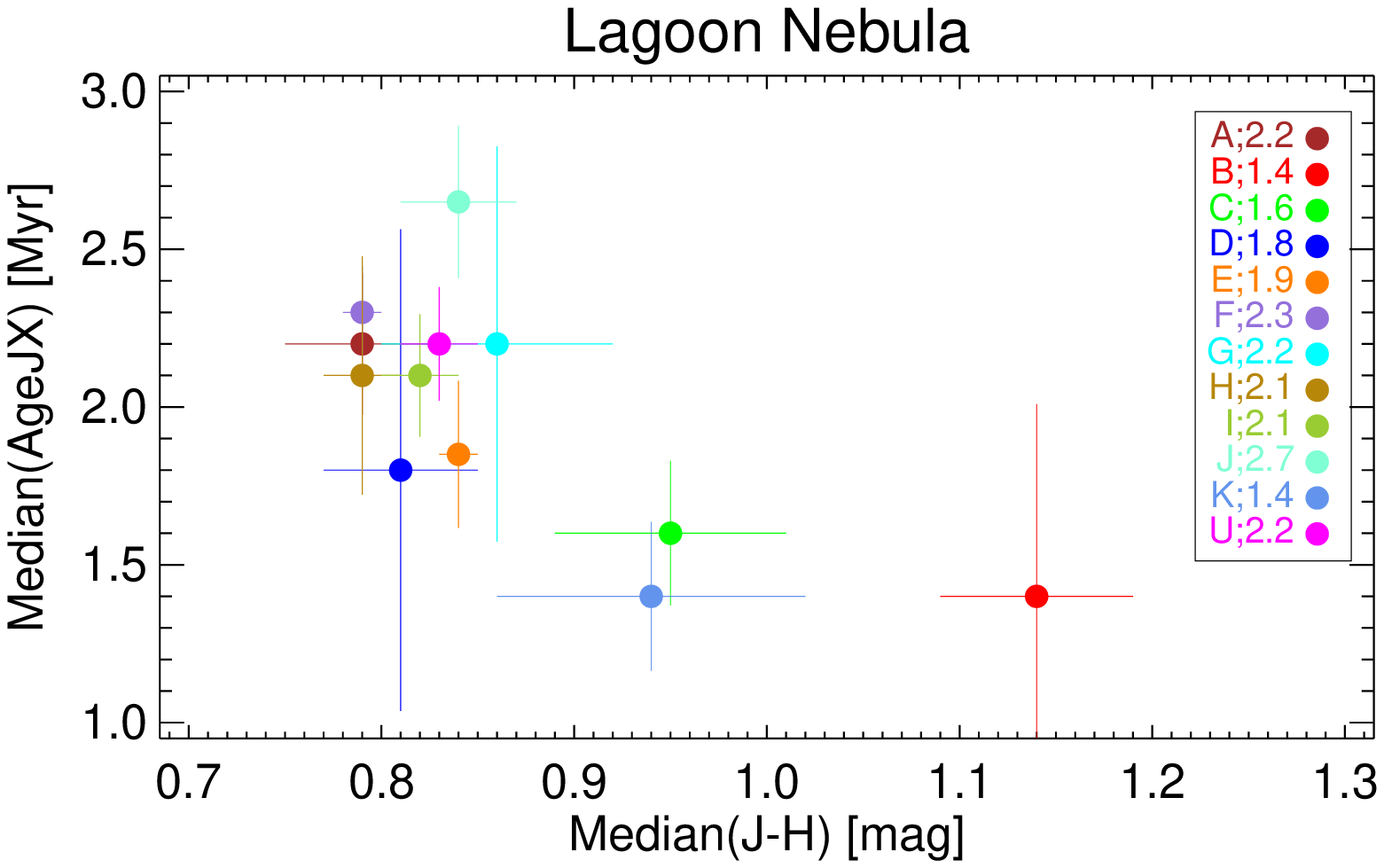} \hspace{0.0in}
\caption{$Age_{JX}$ analysis of the Lagoon Nebula. See Figure~\ref{fig_age_gradients_orion} for description. {\it Upper right:} The subset of MPCM stars available for $Age_{JX}$ analysis and the elliptical contours superposed on the $A_V$ map from \citet{Broos2013}. \label{fig_age_gradients_lagoon}}
\end{minipage}
\end{figure}

\begin{figure}
\centering
\begin{minipage}[t]{1.0\textwidth}
  \centering
  \includegraphics[angle=0.,width=5.5in]{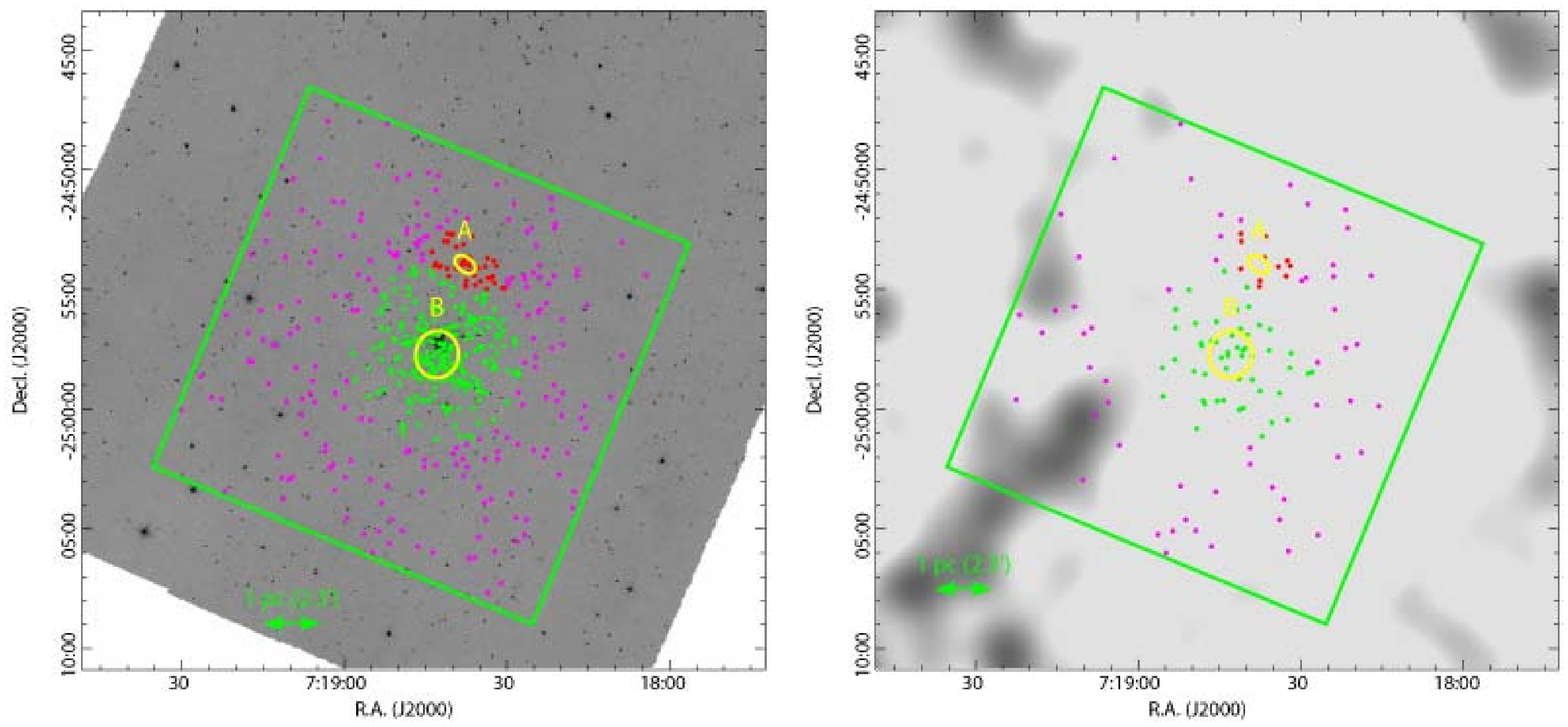} \hspace{0.0in}
  \includegraphics[angle=0.,scale=0.9]{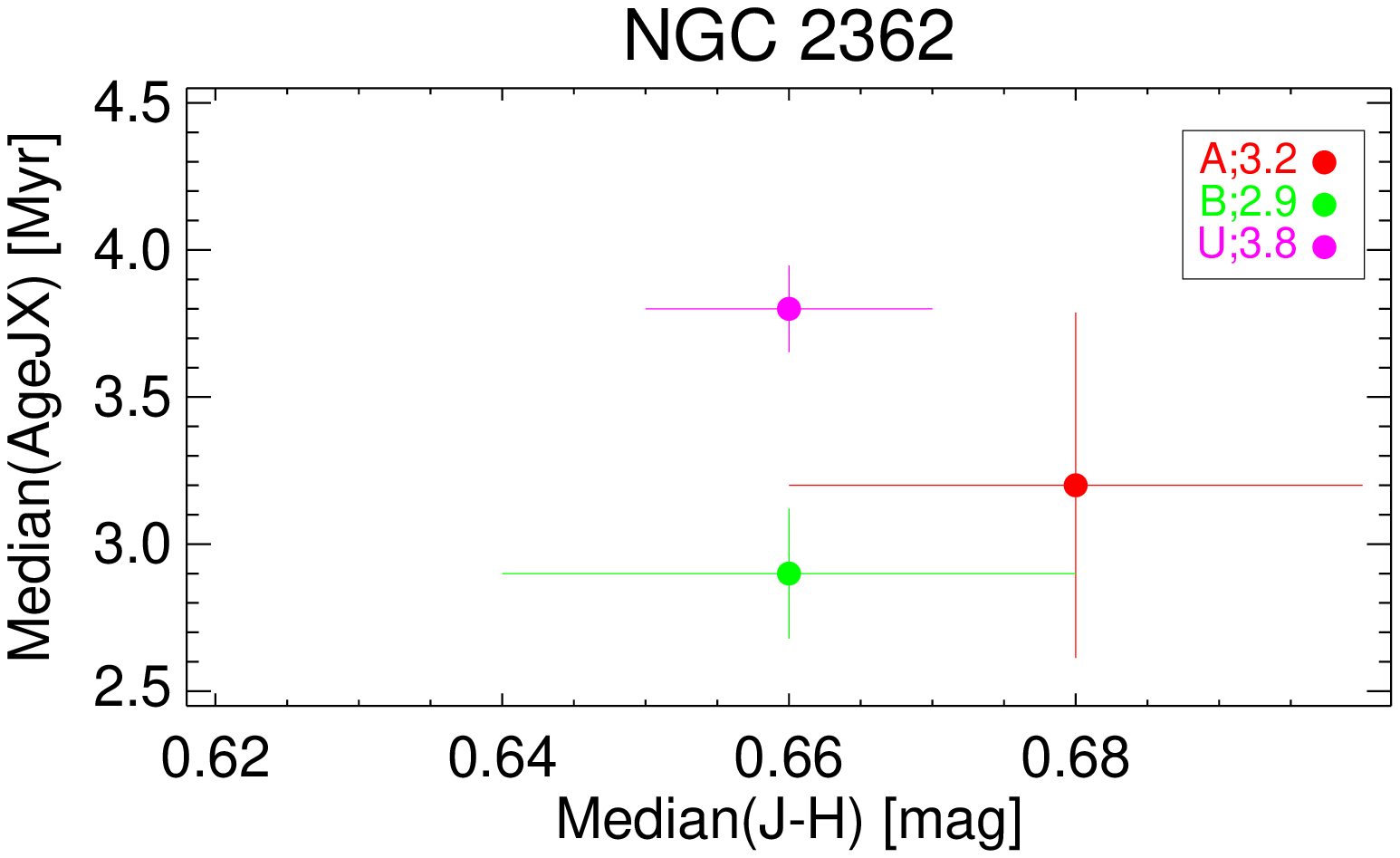} \hspace{0.0in}
\caption{$Age_{JX}$ analysis of NGC~2362. See Figure~\ref{fig_age_gradients_orion} for description. {\it Upper right:} The subset of MPCM stars available for $Age_{JX}$ analysis and the elliptical contours superposed on the $A_V$ map from \citet{Broos2013}. \label{fig_age_gradients_ngc2362}}
\end{minipage}
\end{figure}

\begin{figure}
\centering
\begin{minipage}[t]{1.0\textwidth}
  \centering
  \includegraphics[angle=0.,width=5.5in]{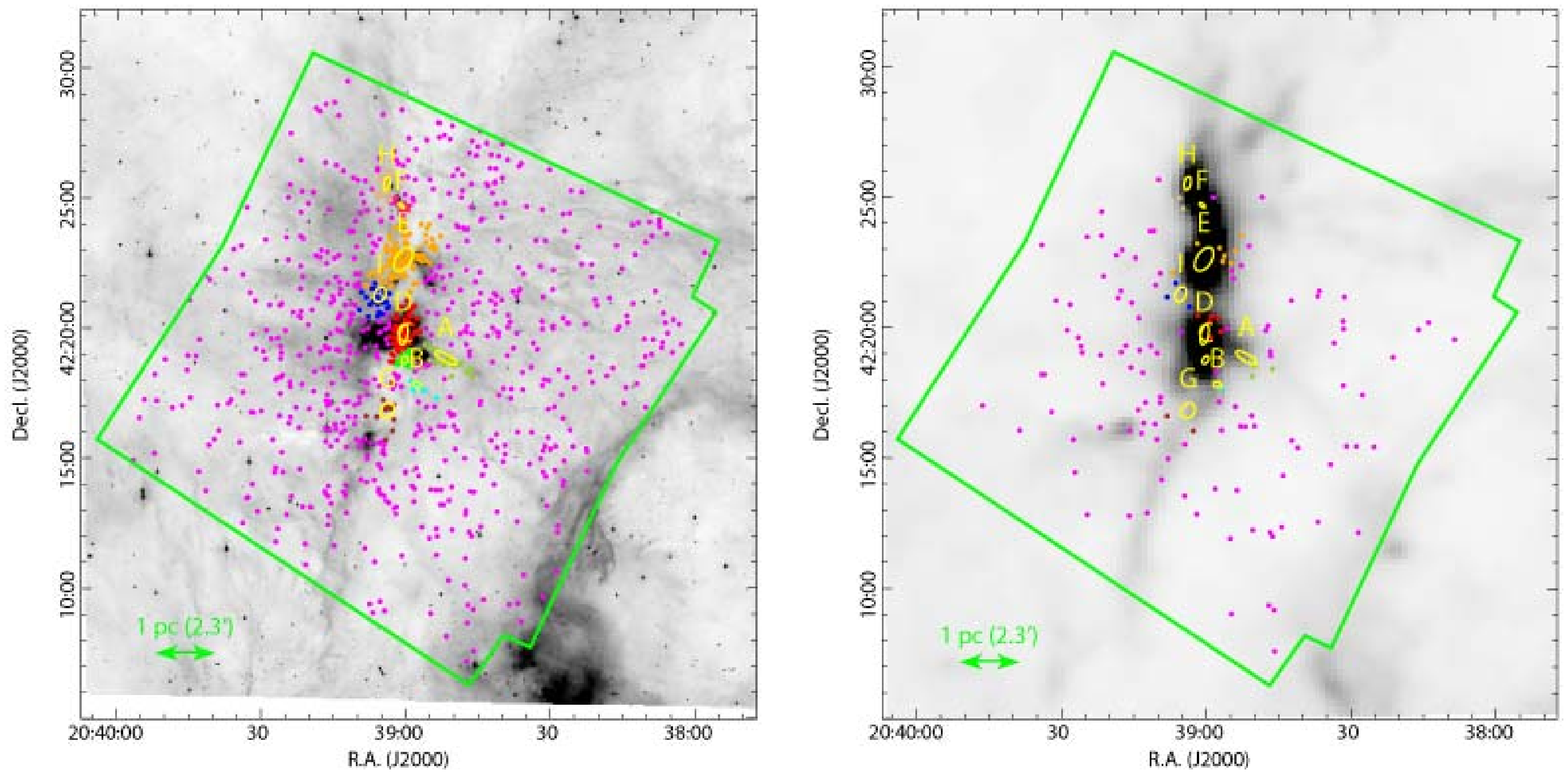} \hspace{0.0in}
  \includegraphics[angle=0.,scale=0.9]{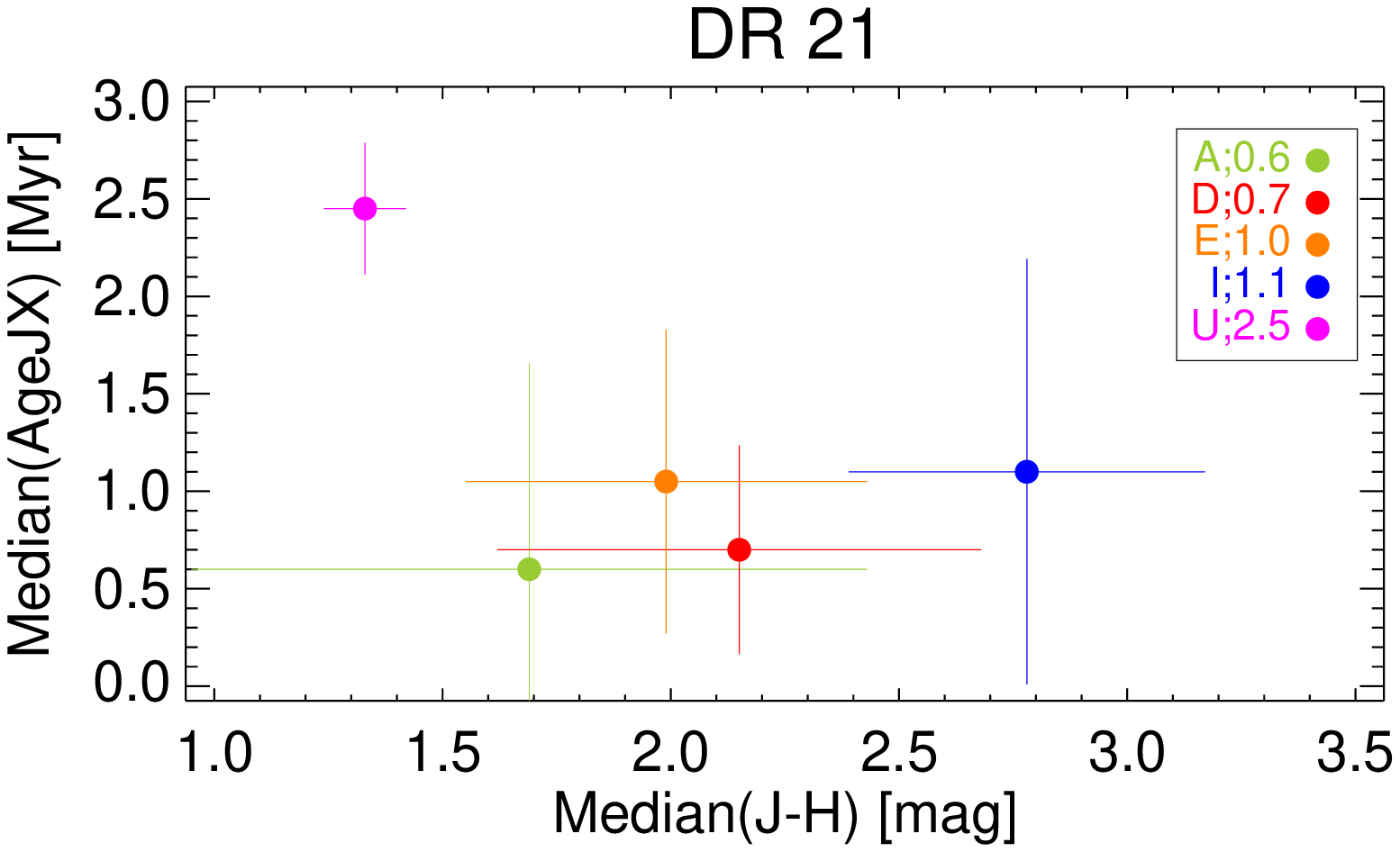} \hspace{0.0in}
\caption{$Age_{JX}$ analysis of DR~21. See Figure~\ref{fig_age_gradients_orion} for description. \label{fig_age_gradients_dr21}}
\end{minipage}
\end{figure}

\begin{figure}
\centering
\begin{minipage}[t]{1.0\textwidth}
  \centering
  \includegraphics[angle=0.,width=5.5in]{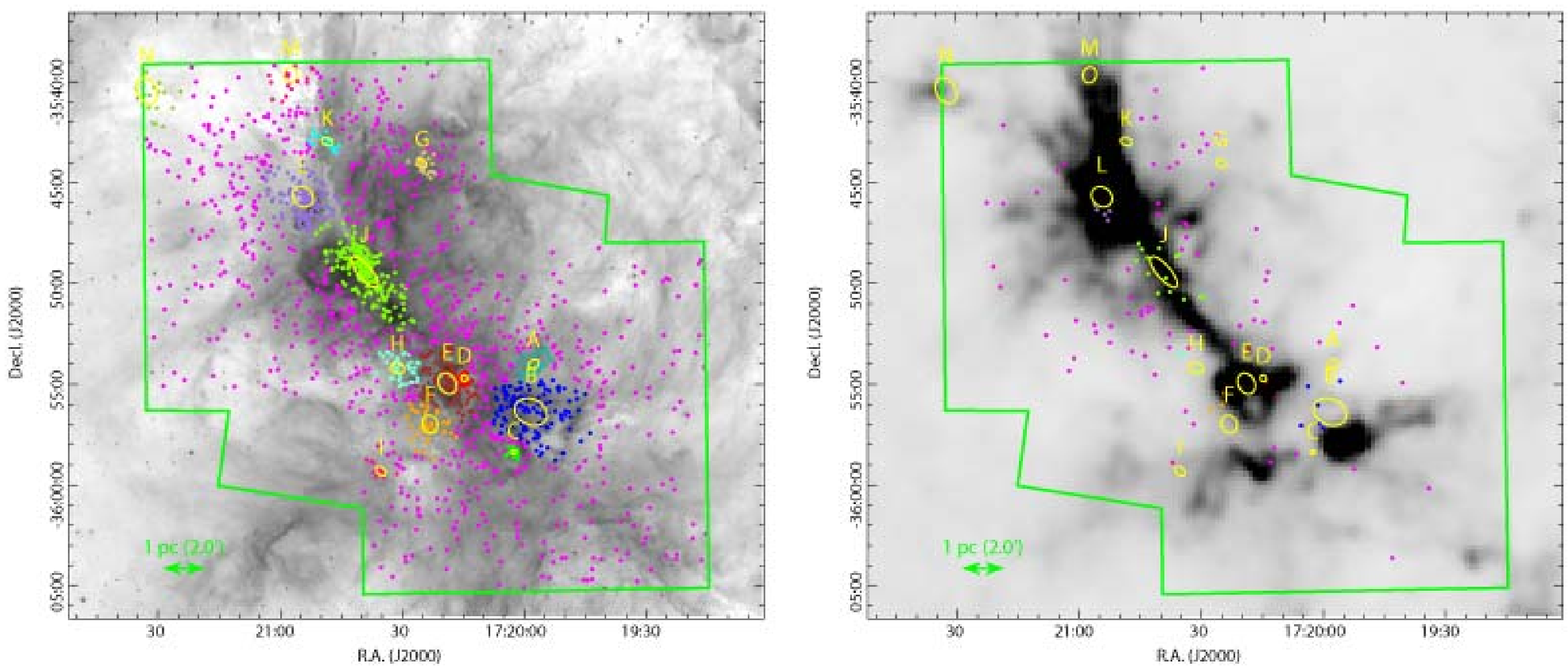} \hspace{0.0in}
  \includegraphics[angle=0.,scale=0.9]{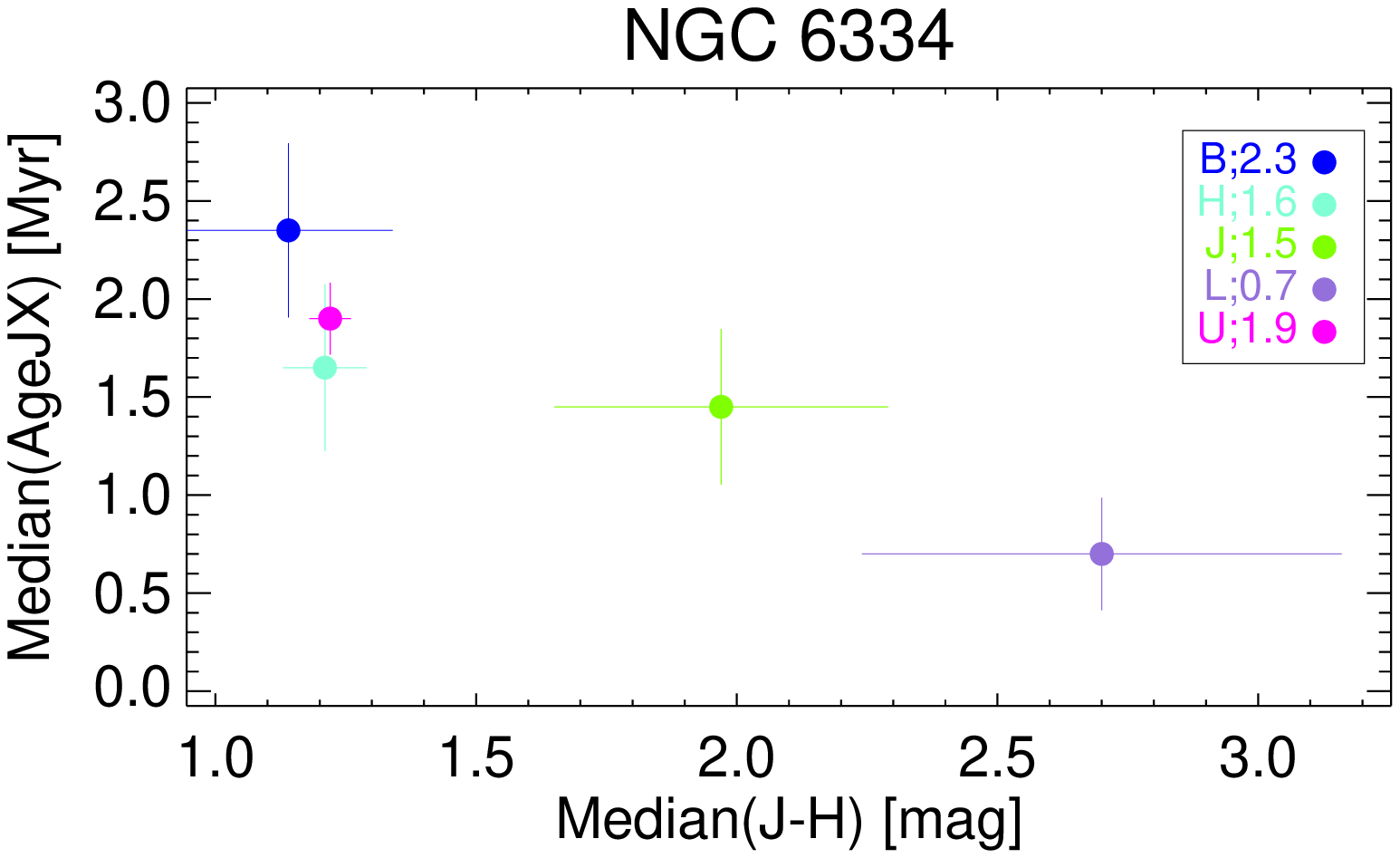} \hspace{0.0in}
\caption{$Age_{JX}$ analysis of NGC~6334. See Figure~\ref{fig_age_gradients_orion} for description. \label{fig_age_gradients_ngc6334}}
\end{minipage}
\end{figure}

\begin{figure}
\centering
\begin{minipage}[t]{1.0\textwidth}
  \centering
  \includegraphics[angle=0.,width=5.5in]{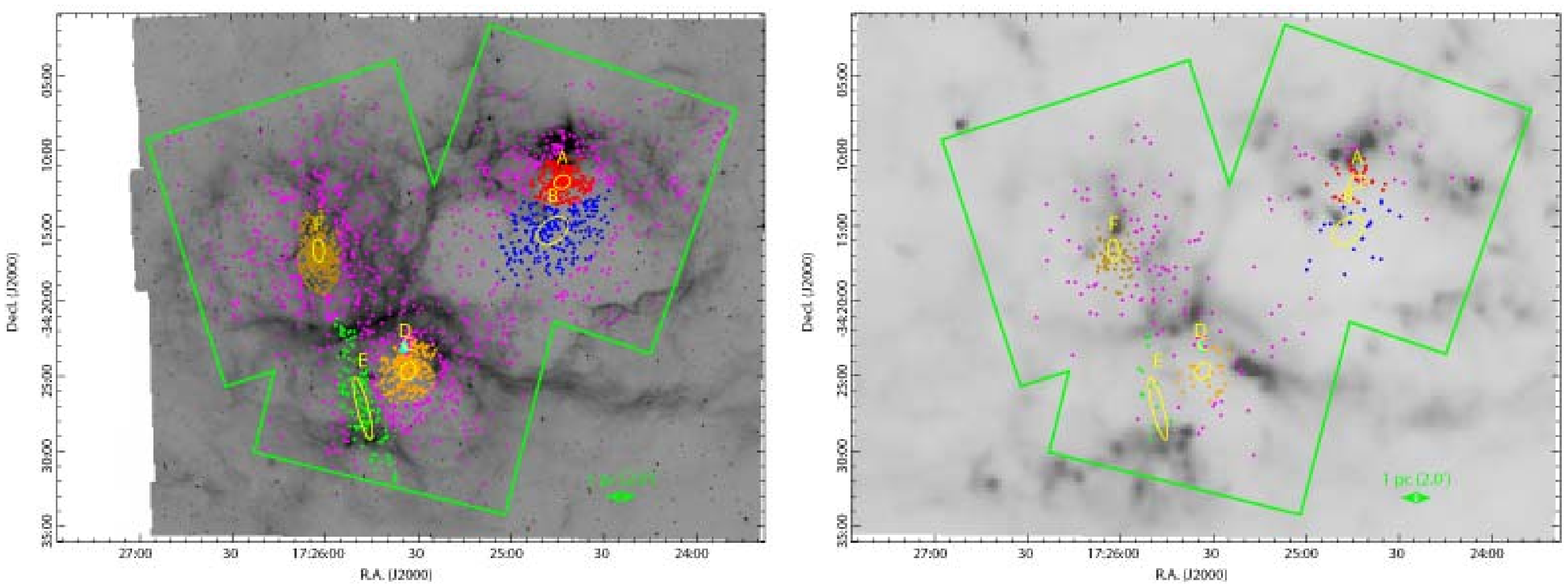} \hspace{0.0in}
  \includegraphics[angle=0.,scale=0.9]{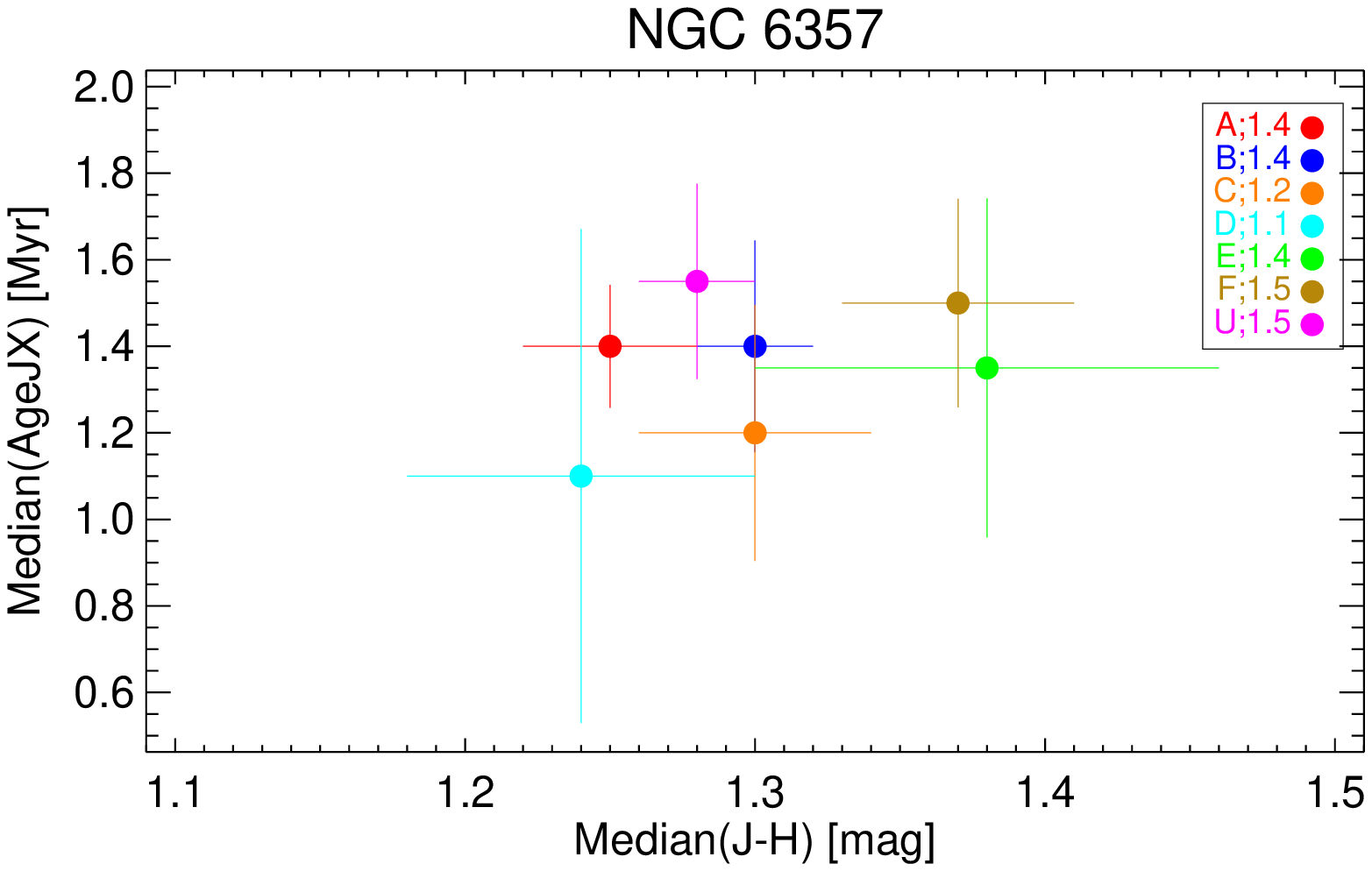} \hspace{0.0in}
\caption{$Age_{JX}$ analysis of NGC~6357. See Figure~\ref{fig_age_gradients_orion} for description. \label{fig_age_gradients_ngc6357}}
\end{minipage}
\end{figure}

\begin{figure}
\centering
\begin{minipage}[t]{1.0\textwidth}
  \centering
  \includegraphics[angle=0.,width=5.5in]{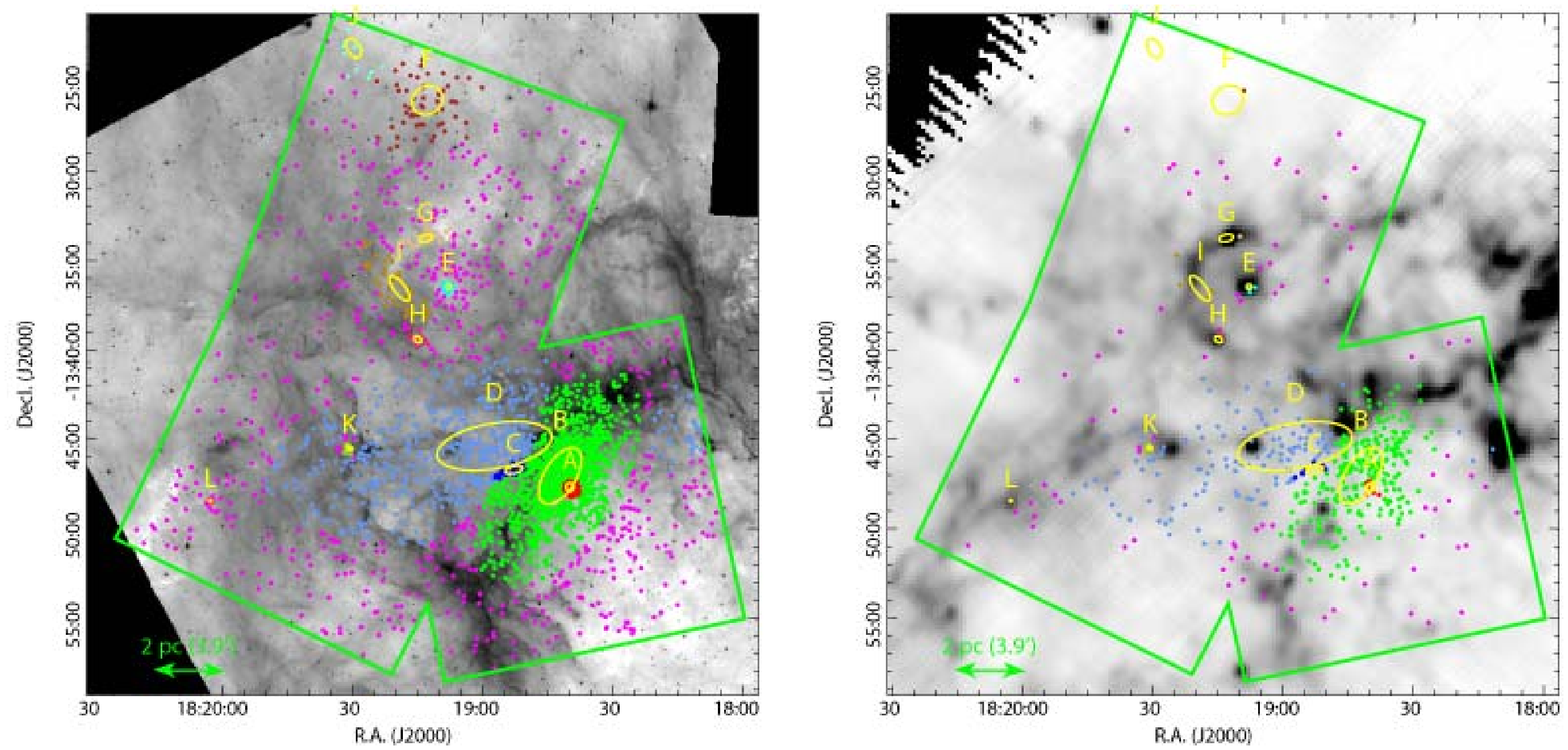} \hspace{0.0in}
  \includegraphics[angle=0.,scale=0.9]{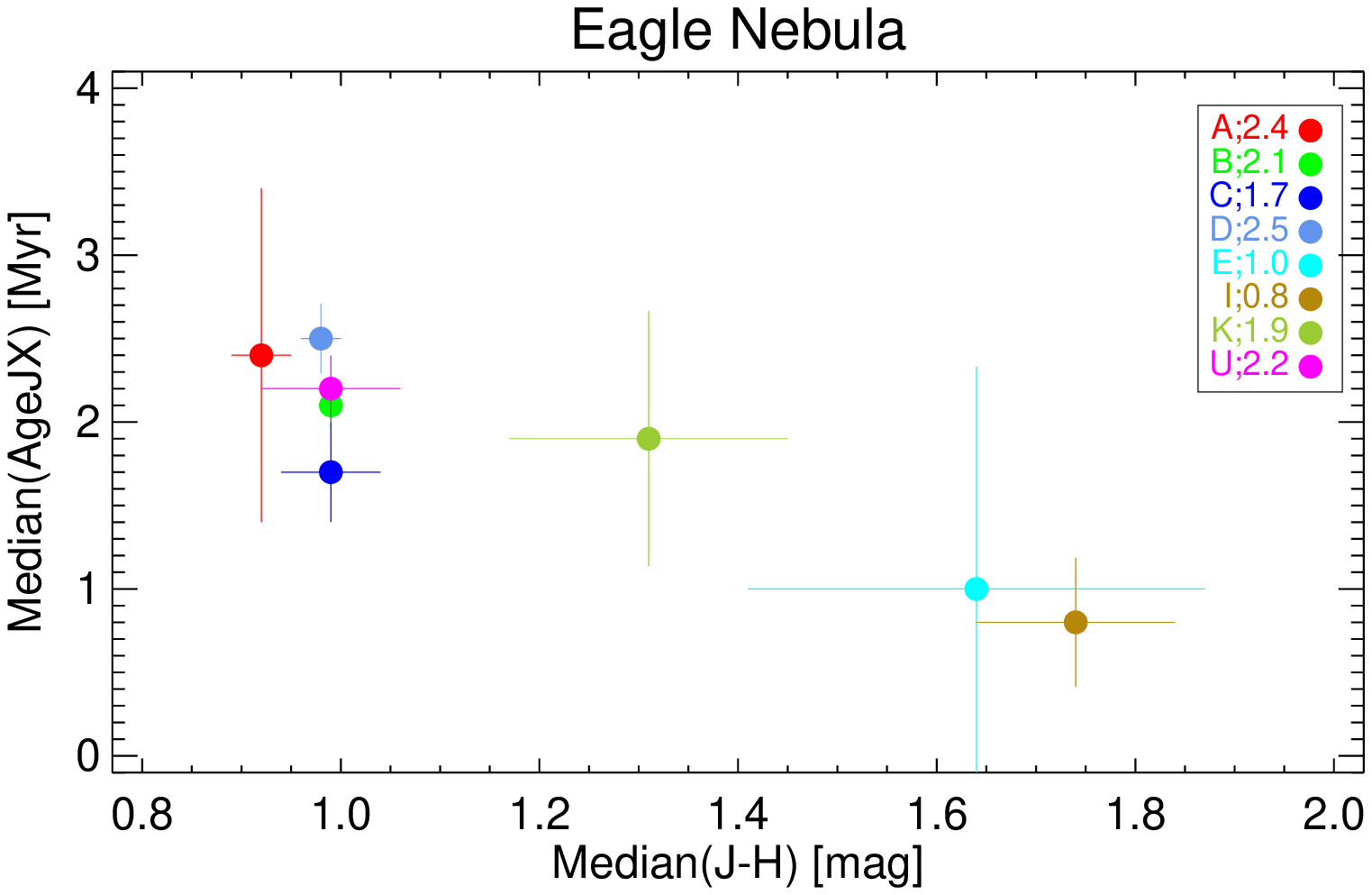} \hspace{0.0in}
\caption{$Age_{JX}$ analysis of the Eagle Nebula. See Figure~\ref{fig_age_gradients_orion} for description. \label{fig_age_gradients_eagle}}
\end{minipage}
\end{figure}

\begin{figure}
\centering
\begin{minipage}[t]{1.0\textwidth}
  \centering
  \includegraphics[angle=0.,width=5.5in]{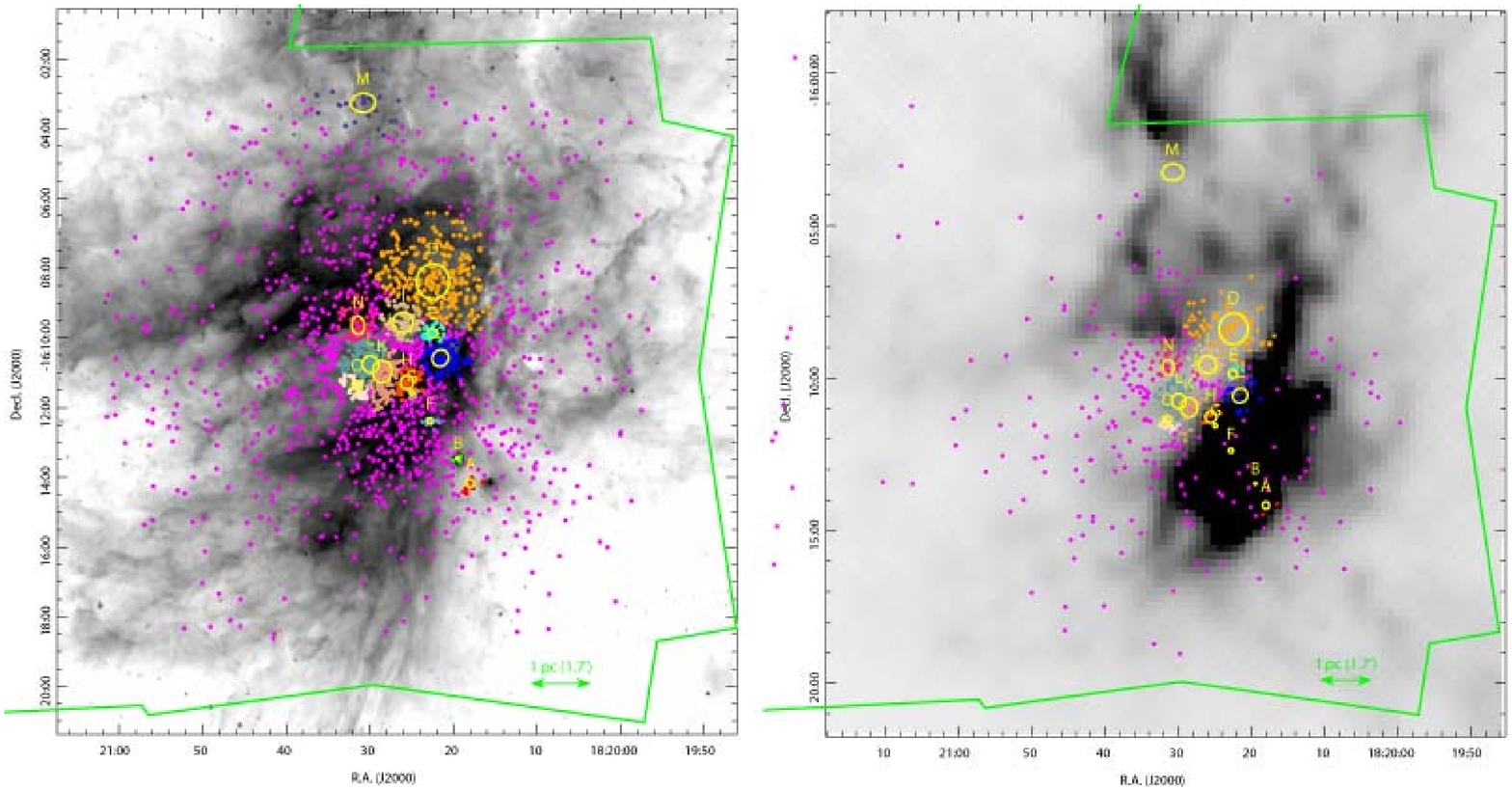} \hspace{0.0in}
  \includegraphics[angle=0.,scale=0.9]{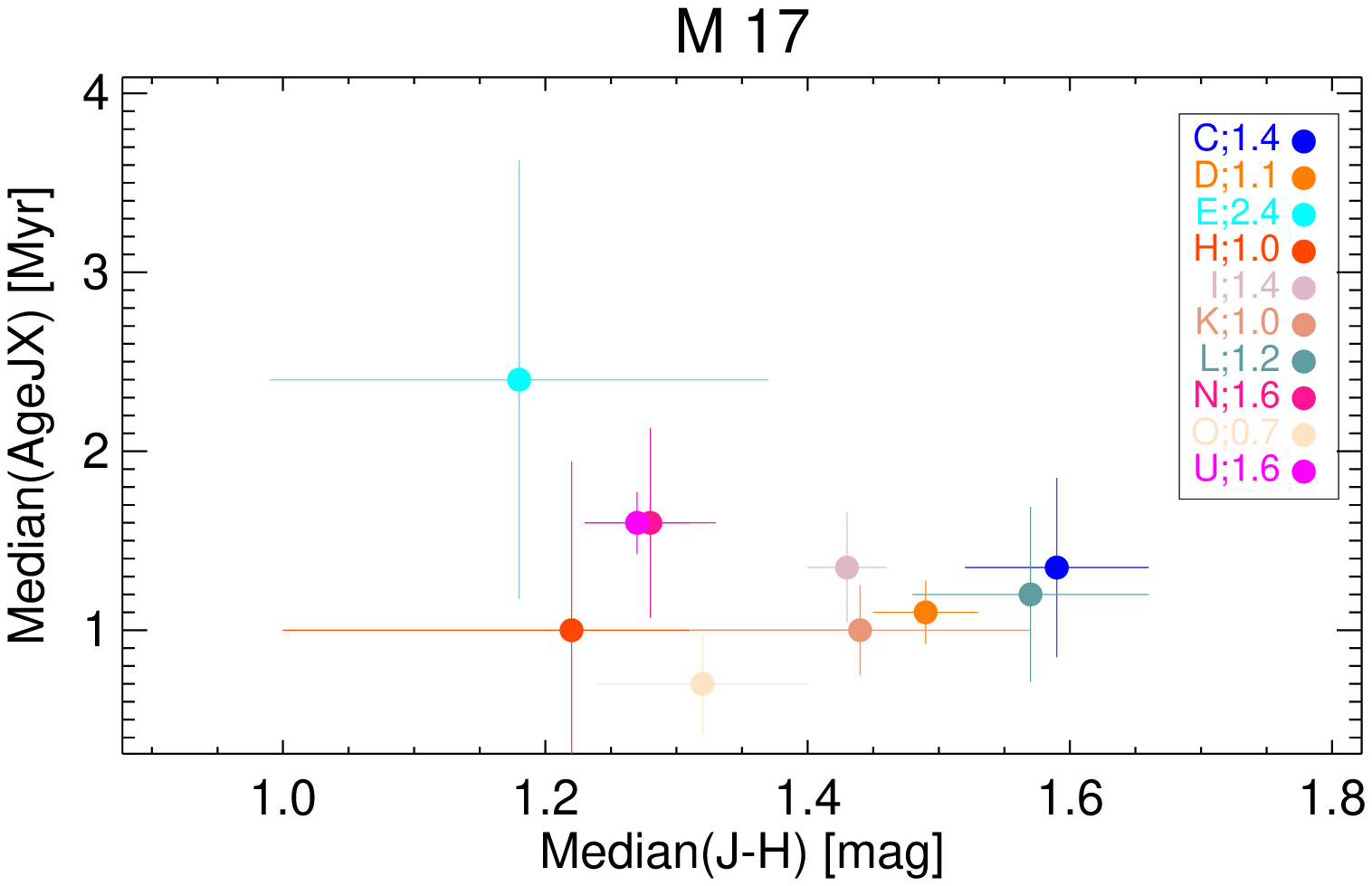} \hspace{0.0in}
\caption{$Age_{JX}$ analysis of M~17. See Figure~\ref{fig_age_gradients_orion} for description. \label{fig_age_gradients_m17}}
\end{minipage}
\end{figure}

\begin{figure}
\centering
\begin{minipage}[t]{1.0\textwidth}
  \centering
  \includegraphics[angle=0.,width=3.5in]{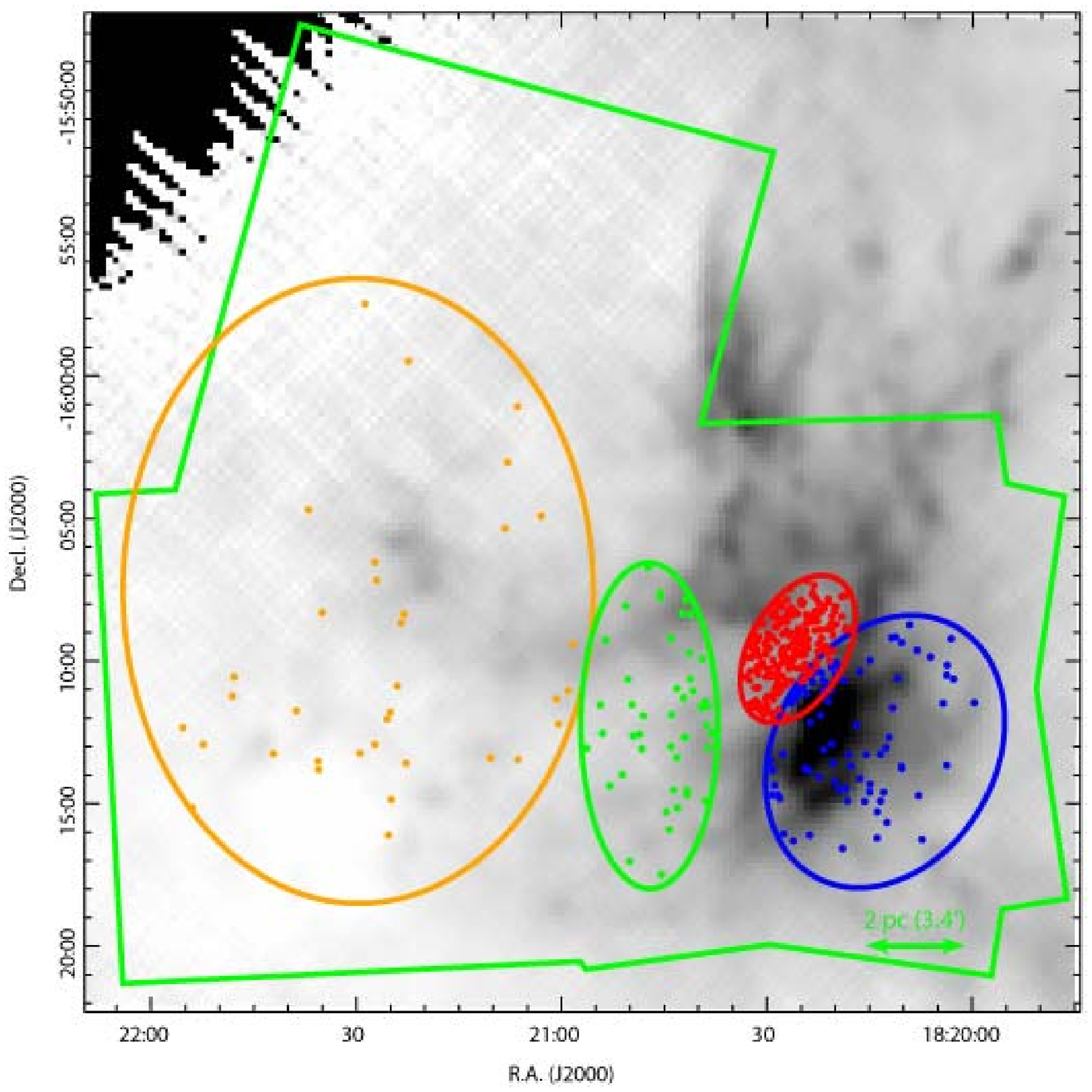} \hspace{0.0in}
  \includegraphics[angle=0.,scale=0.6]{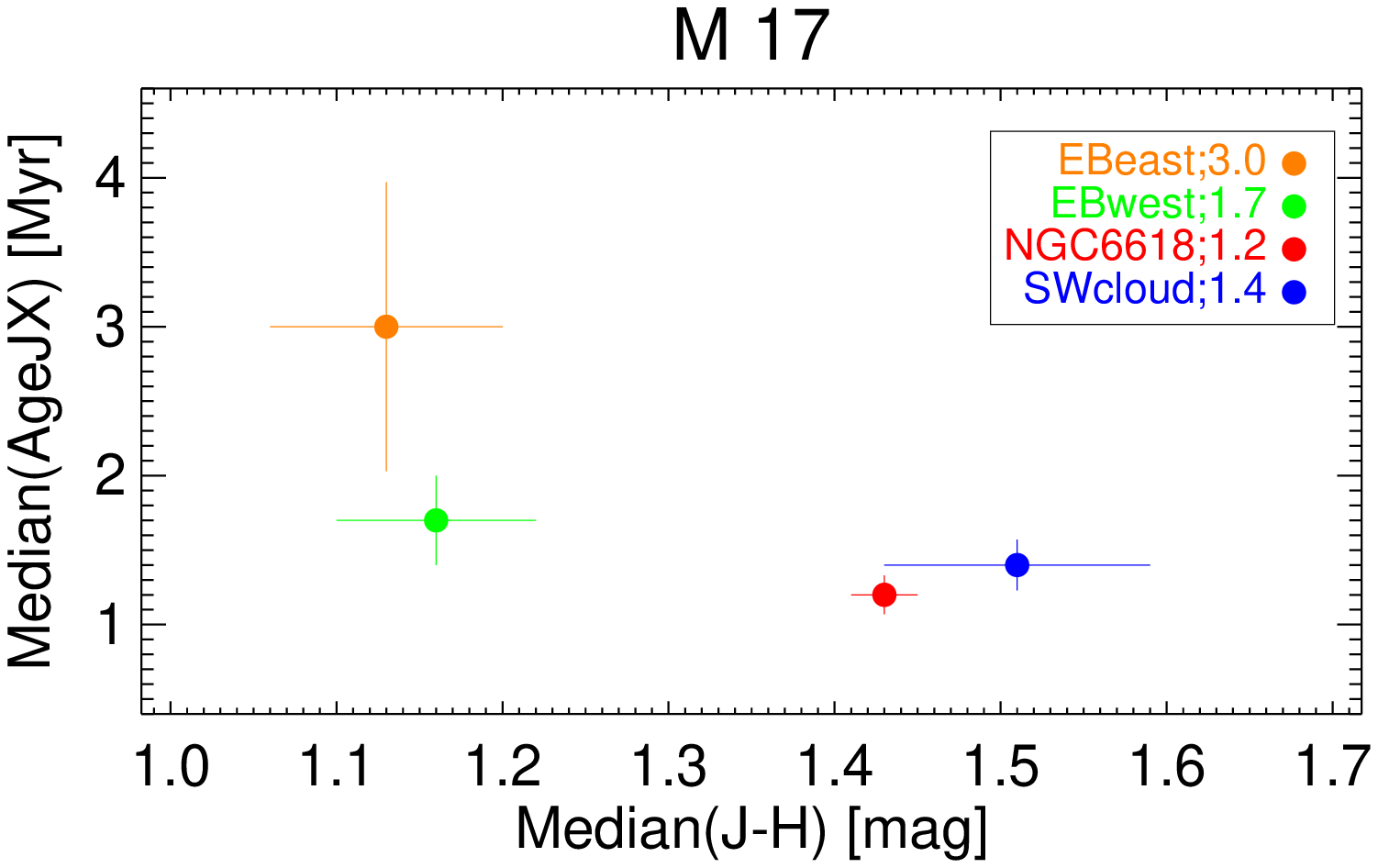} \hspace{0.0in}
\caption{\small Special age analysis of M 17. {\it Upper:} The subset of MPCM stars available for $Age_{JX}$ analysis stratified by the regions of special interest: the southwest molecular cloud (blue), the NGC~6618 cluster (red), the eastern (orange) and western (green) parts of Extended Bubble. The stars are superimposed on the 500~$\mu$m $Herschel$-SPIRE image. The {\it Chandra} field of view is outlined by the green polygon. {\it Lower:} 
The median $Age_{JX}$ estimates for the four regions of interest plotted against their respective median $(J-H)$ color indices. The legend identifies each point with a region and states the median age in Myr. \label{fig_age_gradients_m17_special}}
\end{minipage}
\end{figure}

\begin{figure}
\centering
\begin{minipage}[t]{1.0\textwidth}
  \centering
  \includegraphics[angle=0.,width=5.5in]{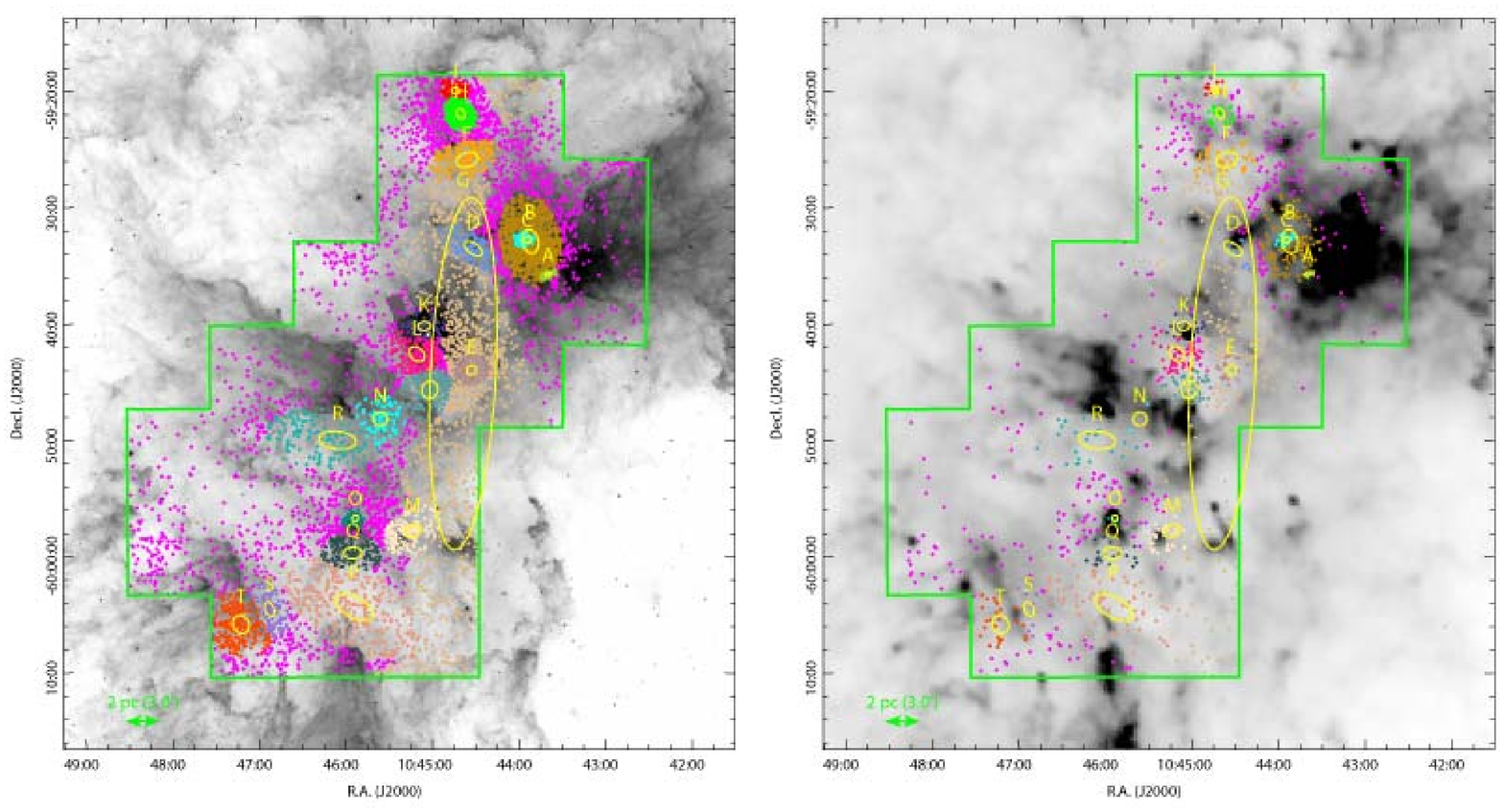} \hspace{0.0in}
  \includegraphics[angle=0.,scale=0.9]{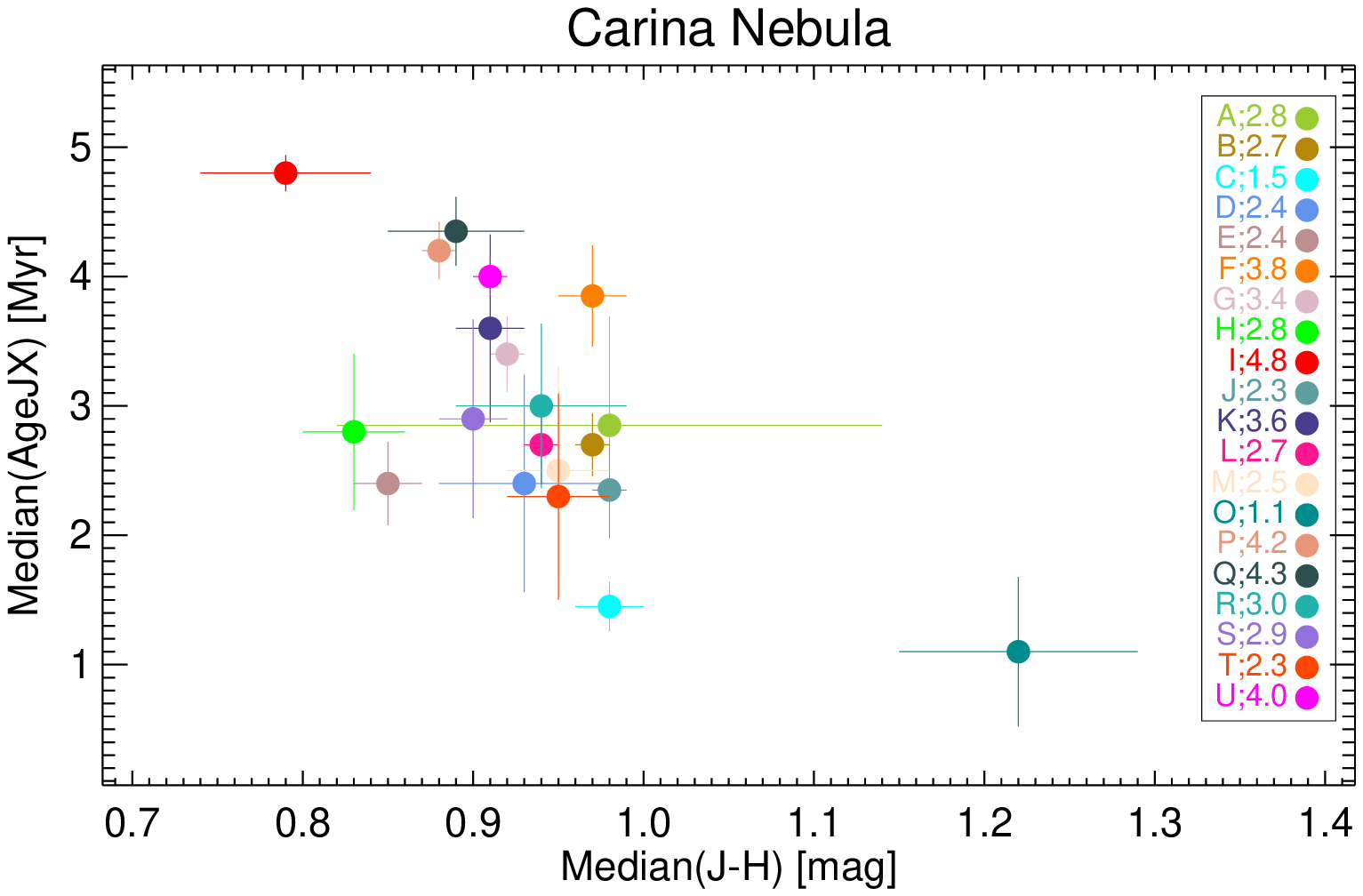} \hspace{0.0in}
\caption{$Age_{JX}$ analysis of the Carina Nebula complex. See Figure~\ref{fig_age_gradients_orion} for description. \label{fig_age_gradients_carina}}
\end{minipage}
\end{figure}

\begin{figure}
\centering
\begin{minipage}[t]{1.0\textwidth}
  \centering
  \includegraphics[angle=0.,width=5.5in]{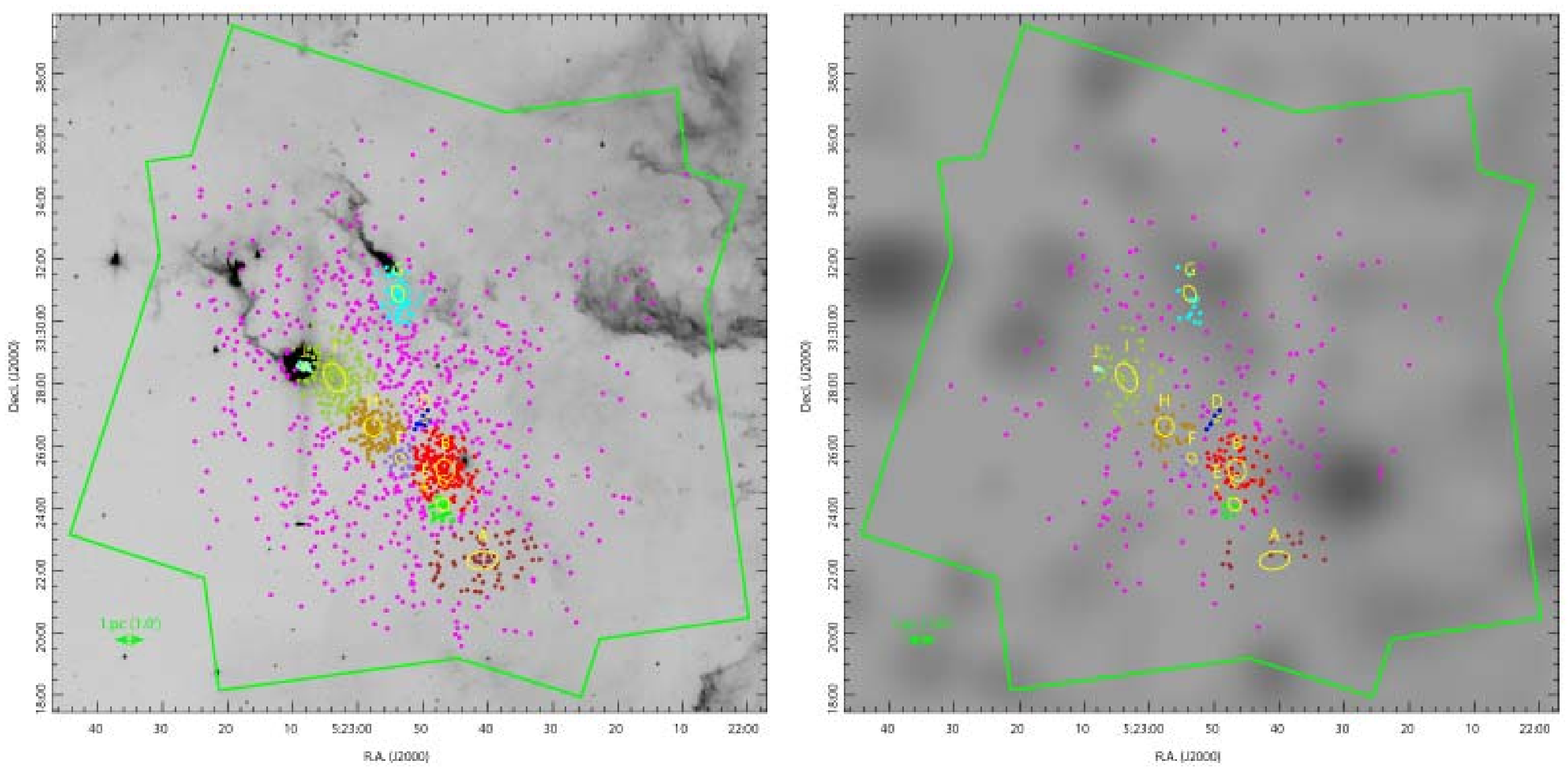} \hspace{0.0in}
  \includegraphics[angle=0.,scale=0.9]{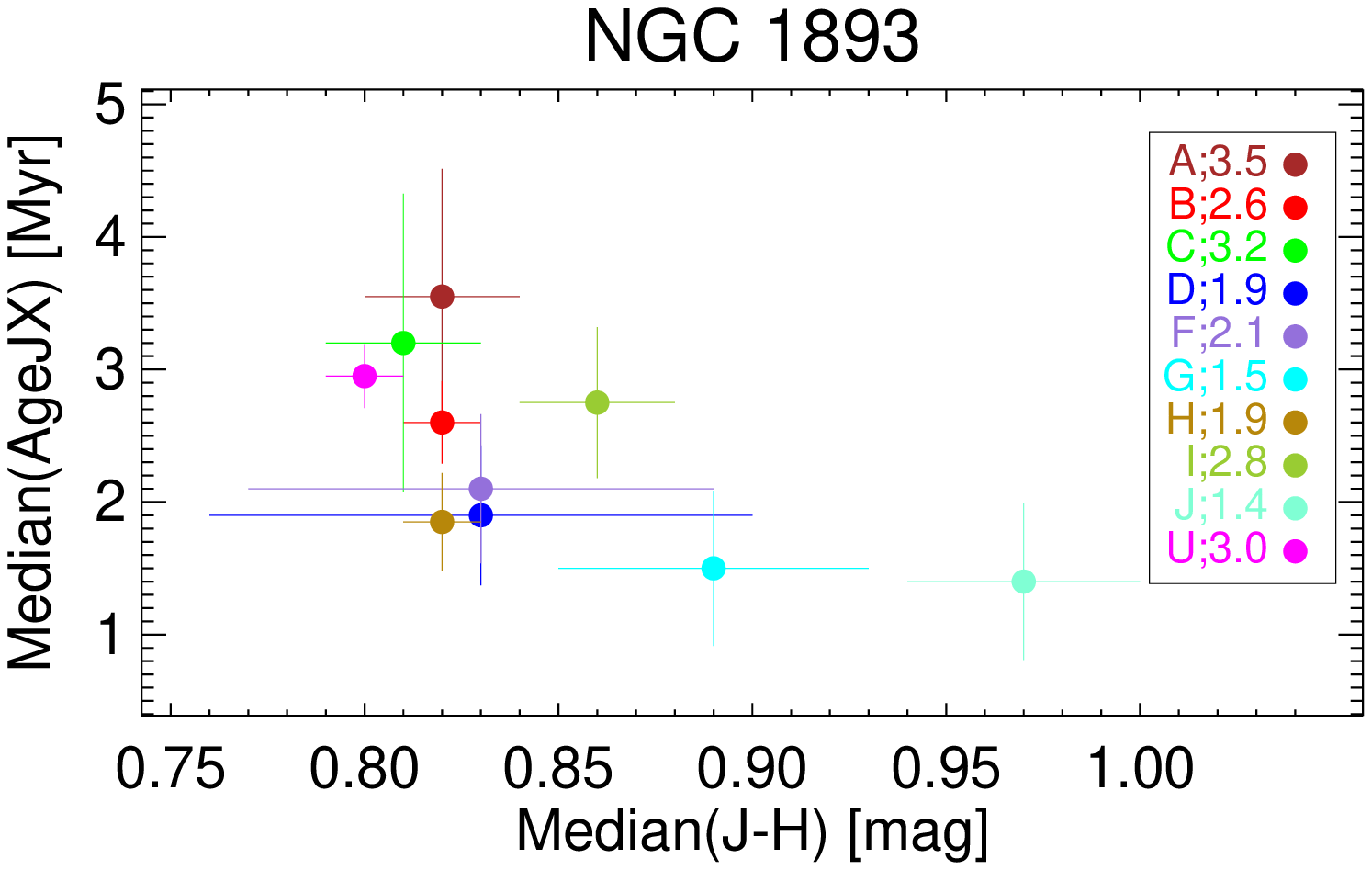} \hspace{0.0in}
\caption{$Age_{JX}$ analysis of NGC~1893. See Figure~\ref{fig_age_gradients_orion} for description. {\it Upper right:} The subset of MPCM stars available for $Age_{JX}$ analysis and the elliptical contours superposed on the $A_V$ map from \citet{Broos2013}. \label{fig_age_gradients_ngc1893}}
\end{minipage}
\end{figure}

The importance of the subcluster median ages becomes apparent when we compare them to their spatial locations and absorptions in Figures \ref{fig_age_gradients_orion} - \ref{fig_age_gradients_ngc1893}.  Here we summarize several broad and powerful results evident from the figures.   A detailed discussion of each MYStIX region is given in the Appendix.   These are the main findings of this study.  \begin{enumerate}

\item  In most MYStIX regions, we see an anti-correlation between median $Age_{JX}$ and median $J-H$ for the subclusters.  This is expected from the individual star results shown in Figure~\ref{fig_AgeJX_vs_Reddening} and validates the expected trend that younger subclusters are more embedded in molecular material than older subclusters.  

\item In all MYStIX regions, the unclustered population (shown as magenta symbols in Figures \ref{fig_age_gradients_orion} - \ref{fig_age_gradients_ngc1893}) is older than most of the subcluster populations.  This is most clearly seen in regions with a single dominant cluster: Flame Nebula, W~40, and RCW~36. This is readily interpreted as the kinematic dispersion of older generations of stars formed in the molecular cloud.  We infer that MYStIX regions have been forming stars for millions of years before the compact rich clusters we see today were formed.    

\item In some MYStIX regions, a coherent spatial gradient in subcluster ages is seen, consistent with the progression of star formation through the molecular cloud on scales of several parsecs. This is most clearly seen in the Rosette Nebula, Eagle Nebula, NGC~6634, and NGC~1893.  For a typical region extent of $\sim 10$~pc and age gradient of $\sim 2$~Myr, we infer a characteristic `speed' of 5~km~s$^{-1}$ for the propagation of star formation through a giant molecular cloud.  

\item In other cases, the subclusters have indistinguishable ages, or have distinct ages in an incoherent spatial pattern.  This is seen in the Orion Nebula, NGC~2264, Lagoon Nebula, NGC~2362, DR~21, NGC~6357, M~17, and the Carina Nebula. A few examples follow. The three clusters producing H\,{\sc{ii}} regions in NGC~6357 appear to be coeval. Cluster formation along the DR~21 filament, about 5 parsecs in length, appears roughly simultaneous. The subclusters superposed on Carina South Pillars show a broad (possibly bimodal) range of ages; this could be from contamination by the old distributed population and/or could be consistent with scenarios of triggered star formation proceeding in these clouds for millions of years \citep{Povich2011}. On occasion, a lightly absorbed cluster (or its portion) lies projected against a region harboring embedded clusters: the subcluster K in front of the IRS 1 cloud in NGC 2264, the subcluster L in front of the Rosette Molecular Cloud, and the Orion Nebula Cluster in front of OMC-1. 
 
\end{enumerate}

\section{Advantages and Limitations of $Age_{JX}$ Analysis \label{some_advantage_subsection}}

The use of $Age_{JX}$ values based on $J$ and $L_X$ measurements,  rather than other age methods such as an HRD or a CMD derived from visual band observations, is potentially advantageous for a number of reasons: \begin{enumerate}

\item $Age_{JX}$ estimates can be made for young stars over a wide range of PMS evolutionary stages and cloud absorptions.  Infrared-only studies, biased towards selection of disk-bearing YSOs, are typically restricted to the earlier stages, while optical-only studies are typically restricted to later, lightly-obscured stages.

\item Compared to optical and mid-infrared surveys where diffuse nebular emission from ionized gas and heated polycyclic aromatic hydrocarbons can be bright, X-ray and NIR are subject to less confusion from nebular emission by heated gas and dust.

\item Extinction corrections are much smaller in the $J$-band than $V$-band; recall that $A_J = 0.28 A_V$. Both X-ray and NIR surveys penetrate comparably deeper into obscuring material than optical surveys. This gives a major advantage to $Age_{JX}$ estimates in MYStIX MSFRs where absorption can vary by tens of magnitudes between subclusters and individual stars. 

\item Multi-epoch photometric surveys have shown that the NIR bands are less affected by variability compared to optical bands.  Photometric variability can arise either from changes in accretion rates or from magnetic activity.  There will be thus less scatter in estimating intrinsic photospheric luminosities from single-epoch $J$ photometry in contrast to single-epoch $V$ photometry often used in HRD/CMD studies. \citet{Herbst94} find that the photometric variability of classical T Tauri stars decreases in amplitude $>2$-fold from $V$ to $I$ bands. \citet{Frasca09} find that photometric effects of rotationally modulated spots are typically $\Delta J \simeq 0.06$~mag, several times lower than in the $V$ band.  Similarly, \citet{Goulding12} recommend the $J$ band for planetary transit surveys in magnetically active M dwarfs.  From a near-infrared study of nearby star forming regions, \citet{Scholz2012} conclude that `the variability in the J-band is less than 0.5 mag for the overwhelming majority of YSOs, i.e. a factor of $<$ 1.6 in luminosity, which is only a minor part of the observed spread in the HR diagrams'.  
 
\item Unlike optical measurements, the X-ray luminosity $L_X$ from most MSFRs is not detectably affected by accretion variations or emission from accretion shocks.  The soft X-rays from the accretion shock is generally lost due to interstellar absorption.  No causal link is seen between the sources of optical variability (often accretion related) and X-ray variability \citep{Stassun2006}. X-ray variability is equally ubiquitous on both diskless and disk-bearing stars and is mainly due to magnetic activity \citep{Wolk2005,Stelzer2007,Getman2008}. The contribution of the `very soft' X-ray emission from accretion shocks can only be detected (and modeled) in cases where there is negligible soft X-ray absorption, which is not the case for most MSFRs. However, although independent of accretion, the X-ray variability does affect the $Age_{JX}$ estimates, as detailed in \S \ref{error_propagation_section}. 

\item Cluster ages are often estimated from infrared excess properties.  In some studies, disk-bearing stars are binned into a classification scheme (Class 0, I, and II young stellar objects) that does not have an established calibration to stellar age.  In other studies, the disk fraction is measured (often with a poorly established sample of disk-free stars) and ages are estimated from a rough calibration curve \citep{Hernandez2008, Mamajek2009}. Nevertheless, the age vs. disk fraction calibration of \citet{Ybarra2013} seems consistent with the $Age_{JX}$ estimates for the highly-absorbed MPCMs (\S \ref{reddening_effect_section}).

\item Unlike some studies where different age estimation methods are applied to arbitrarily defined regions or (sub)samples, $Age_{JX}$ values are calculated for stellar samples within subclusters defined by a mathematically objective procedure in \citet{Kuhn2014c}.

\item Unlike some studies where cluster ages are the mean of age or log-age distributions, $Age_{JX}$ results reported here are calculated using median values that are robust to non-Gaussianity and unaffected by transformation of the age variable. 

\item Our homogeneous $Age_{JX}$ analysis is applied to a large number of star forming regions, whereas most previous studies applied different methods to different regions.  The $Age_{JX}$ estimates for over a hundred MYStIX subclusters are referenced to a uniform time scale, which depends on the \citet{Siess2000} models and the adopted distances to the MYStIX regions. This allows reasonably reliable comparison of star formation histories within and between MSFRs.  Our median $Age_{JX}$ estimates for MYStIX subclusters are listed in Table~\ref{tbl_cluster_ages}.  

\end{enumerate} 

\begin{figure}
\centering
\includegraphics[angle=0.,width=6.5in]{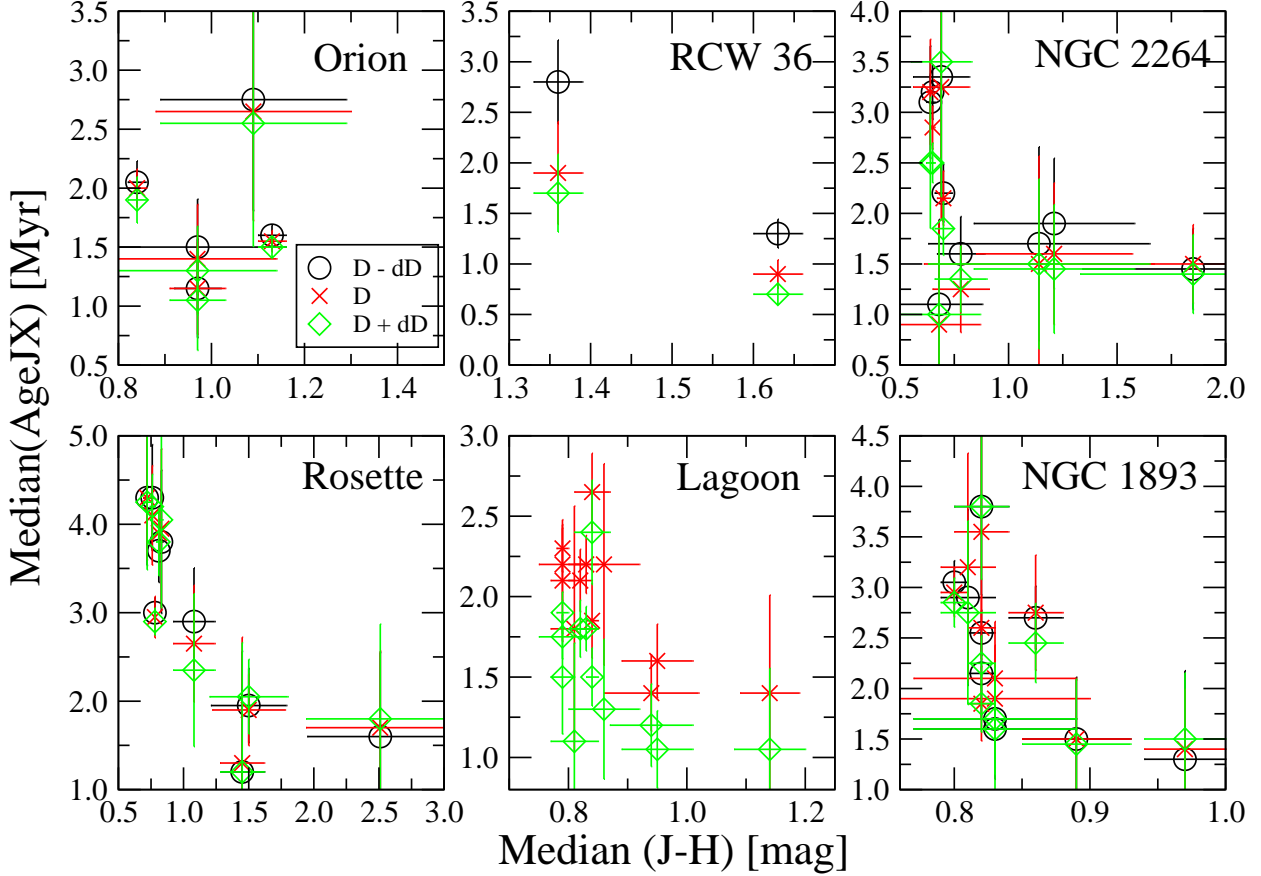}
\caption{Effect of distance uncertainties on the resulted age gradients. The inferred median $Age_{JX}$ for individual MYStIX subclusters plotted against their respective median $(J-H)$ color indices. The plots are given for the set of 6 exemplifying MYStIX regions (Orion, RCW 36, NGC 2264, Rosette, Lagoon, and NGC 1893) with distance errors reported in the literature \citep[][Table~1]{Feigelson2013}. The $Age_{JX}$ vs. $J-H$ estimates are given for the set of tree trial distances to a MYStIX region: the reported mean/median distance measurement (red $\times$; same as in Figures \ref{fig_age_gradients_orion} - \ref{fig_age_gradients_ngc1893}), the reported lower boundary on distance (black $\circ$), and the reported upper boundary on distance (green $\diamond$). \label{fig_distance_effect}}
\end{figure}

However, like most of the other age methods, the $Age_{JX}$ method is prone to some problems and limitations: \begin{enumerate}

\item Our $Age_{JX}$ estimates for individual stars are highly uncertain. The largest contribution to the error is the scatter on the $L_X-M$ diagram that causes typical $Age_{JX}$ uncertainties of $\sim 2$~Myr (\S \ref{error_propagation_section}).  However, scatter of similar amplitude is seen in optically derived ages (\S\ref{individual_ages_section}); this is the well-known problem of age spreads in young stellar clusters. We use the robust median estimator that effectively reduces both the influence of outliers and individual star age uncertainties in estimating subcluster ages.  

\item For some embedded MYStIX subclusters the inferred $Age_{JX}$ or $Age_{JH}$ estimates may overestimate true ages due to the poor sensitivity of the $Age_{JX}$ method to very young Class~0/I protostars. 

\item Our $Age_{JX}$ estimates for the individual MYStIX stars, and hence the median values for subclusters, have not been corrected for possible effects of binarity.  True ages may thus be somewhat older than we state. This deficiency is shared with most other age estimates at distances of MSFRs.  

\item Any anomalous dust extinction laws characteristic of some obscured regions of molecular clouds are ignored.

\item $Age_{JX}$ relies on the $L_X-M$ regression averaged between accreting and non-accreting stars that are known to have different X-ray luminosity functions (\S\ref{lx_mass_pms_section}). The accreting stars are on average less X-ray active. But we do not recommend use of the separate regression lines reported by \citet{Telleschi2007} for a number of reasons. First, for the typical mass range of the MYStIX $Age_{JX}$ stars, the recent published results on the presence/absence of the effect are controversial (\S \ref{lx_mass_pms_section}). Second, a clear separation between accreting and non-accreting systems is not available for most MYStIX MSFRs.  Third, the XPHOT method for estimating $L_X$ is based on the X-ray emission models already averaged between accreting and non-accreting systems \citep{Getman2010}.

\item As for any other age method, our age estimates are prone to changes as the distance estimate to a star forming region varies. Figure \ref{fig_distance_effect} exemplifies the effect of uncertain distances on the MYStIX age gradients. Following are a few notable examples of this effect. 

First, changes in distance estimates will globally shift inferred age scales with larger distances resulting in younger age scales.  Second, due to the non-uniform separation and orientation of the theoretical isochrones on the $M_J$ versus $M$ diagram (Figure \ref{fig_mj_vs_mass_ajfromccd_SMOOTH_MEDIANS_V2}) and the distance dependence of both $M_J$ and $M$ (through $L_X$) quantities, the slopes on the median($Age_{JX}$) vs. median($J-H$) plots would generally change with the changing reference distance.  Third, distance uncertainties of $\sim 5$\% \citep[][Table 1]{Feigelson2013} propagate into subcluster age differences of only a few to several percent (e.g., Orion, Rosette, and NGC 1893 in Figure \ref{fig_distance_effect}). Distance uncertainties of $\sim 30-40$\%, as seen in RCW~38 and the Lagoon Nebula,  propagate into subcluster age differences of $\sim 20-50$\%. Finally, different distances to subclusters within the same MYStIX region would likely have negligible effect on the inferred age gradients in the region. For instance, using volumetric assumptions for the nearly circular NGC~1893 and Rosette bubbles, it is reasonable to suggest that distance differences to the MYStIX subclusters in these regions of $\sim 10-20$~pc are only $0.3-2$\% of the total distance to the region and give no more than a few percent difference in inferred subcluster ages.

\end{enumerate}

\section{Conclusions} \label{conclusion_section}

Researchers have historically used a number of different methods for dating young stellar clusters: isochrone fitting on the HRD or on photometric CMD, protoplanetary disk fraction, lithium depletion, and other procedures \citep{Mamajek2009b, Soderblom2010, Jeffries2012}. The placement of PMS stars on the HRD or CMD with theoretical evolutionary isochrones is probably the most popular method in the literature for dating young ($\la 10$~Myr) clusters but the individual stellar ages are highly uncertain.   The distributions of stellar ages within rich clusters and across large-scale MSFRs often show wide age spreads; it has been unclear whether this is entirely due to poor individual age estimates or to real extended star formation histories.  

The problem of age estimation in MSFRs is particularly challenging, as reliable censuses of stellar members are hard to obtain and complex star formation histories are likely to be present.  The MYStIX project has produced rich samples in 20 OB-dominated young star forming regions at distances $<4$~kpc using sensitive X-ray, NIR, and mid-infrared photometric catalogs \citep{Feigelson2013}. \citet{Kuhn2014c} has segmented the spatial distribution of MYStIX Probable Cluster Members into dozens of distinct subclusters. Some subclusters are deeply embedded in cloud material and have mostly disk-bearing stars, while others have low obscuration and mostly disk-free stars.  The MYStIX sample thus provides a valuable opportunity to characterize star formation histories in MSFRs where optical or NIR spectroscopy is as yet unavailable.  
  
In this work, we develop a new PMS stellar age estimator, $Age_{JX}$,  based on $J$-band stellar photospheric emission and on X-ray emission from coronal magnetic flaring. Stellar masses are directly derived from absorption-corrected X-ray luminosities using the $L_X - M$ relation from the Taurus cloud. These masses are combined with $J$-band magnitudes, corrected for source extinction and distance, for comparison with PMS evolutionary models \citep{Siess2000} to estimate ages.  We provide a homogeneous set of  median $Age_{JX}$ for over a hundred subclusters identified by \citet[][Table~\ref{tbl_cluster_ages}]{Kuhn2014c}.   The method has a number of advantages over traditional methods of age estimation for young stellar clusters (\S\ref{some_advantage_subsection}).

The main conclusions of our study are as follows:

\begin{enumerate}

\item We demonstrate that the $L_X-M$ relation, rather than the $L_X/L_{bol}$ ratio,  remains slowly varying during the early PMS phase (\S \ref{lx_mass_pms_section}).

\item The $Age_{JX}$ stellar ages are anticorrelated with the source extinction $A_V$ and the $J-H$ color. At high absorptions, the $Age_{JX} - A_V$ relationship is consistent with that of \citet{Ybarra2013} that was independently derived using disk fraction techniques. The $Age_{JX}$ vs. $J-H$ trend can serve as an approximate age predictor for young, highly reddened clusters, provided that the bulk of the reddening is local to the clusters (\S \ref{reddening_effect_section}).

\item To overcome the dispersion of highly uncertain individual ages, we compute median ages (and their confidence intervals) of stellar samples within subclusters defined by the companion study of \citet{Kuhn2014c} (\S\S \ref{subregions_subclusters_section} and \ref{age_gradients_section}).  We find narrowly constrained ages for these spatially distinct structures.  Thus, we establish that (at least some) of the apparent age spread in MSFRs is real and can be attributed to clusters formed at different times.  

\item Spatial gradients in subcluster ages are often seen, consistent with astronomically reasonable patterns of star formation histories with older lightly absorbed structures on one side, younger heavily absorbed structures on the other side, and intermediate-age structures in between.  Some regions do not show coherent star formation patterns.  However, widely distributed populations nearly always have older ages than the principal clusters. The progression of star formation within a MSFR can be traced over spatial scales of several parsecs and time scales of several million years.  Patterns of star formation histories can be compared between MSFRs. We caution that uncertainties in distance to MYStIX MSFRs can cause systematic shifts in the age scales  (\S \ref{some_advantage_subsection}).

\end{enumerate} 

\acknowledgments We thank R. Jeffries (Keele University), T. Megeath (University of Toledo), and N. Da Rio (ESA) for useful comments. We also thank the anonymous referee for his time and useful comments that improved this work. The MYStIX project is supported at Penn State by NASA grant NNX09AC74G, NSF grant AST-0908038, and the {\it Chandra} ACIS Team contract SV474018 (G. Garmire \& L. Townsley, Principal Investigators), issued by the {\it Chandra} X-ray Center, which is operated by the Smithsonian Astrophysical Observatory for and on behalf of NASA under contract NAS8-03060. The Guaranteed Time Observations (GTO) included here were selected by the ACIS Instrument Principal Investigator, Gordon P. Garmire, of the Huntingdon Institute for X-ray Astronomy, LLC, which is under contract to the Smithsonian Astrophysical Observatory; Contract SV2-82024. This research made use of data products from the {\it Chandra} Data Archive and the {\it Spitzer Space Telescope}, which is operated by the Jet Propulsion Laboratory (California Institute of Technology) under a contract with NASA. The United Kingdom Infrared Telescope is operated by the Joint Astronomy Centre on behalf of the Science and Technology Facilities Council of the U.K. This work is based in part on data obtained as part of the UKIRT Infrared Deep Sky Survey and in part on data obtained in UKIRT director's discretionary time. This research used data products from the Two Micron All Sky Survey, which is a joint project of the University of Massachusetts and the Infrared Processing and Analysis Center/California Institute of Technology, funded by the National Aeronautics and Space Administration and the National Science Foundation. This research has also made use of NASA's Astrophysics Data System Bibliographic Services and SAOImage DS9 software developed by Smithsonian Astrophysical Observatory.

{\it Facilities:} \facility{CXO (ASIS)}, \facility{Spitzer (IRAC)}, \facility{CTIO:2MASS ()}, \facility{UKIRT (WFCAM)}.

\clearpage
\newpage

\section{Appendix: $Age_{JX}$ Results for Individual Massive Star Forming Regions} \label{age_gradients_section}

\subsection{Orion Nebula} \label{age_orion_subsection}

The Orion Nebula Cluster (ONC) produces the nearest ($D \sim 414$~pc) unobscured H\,{\sc{ii}} region, lying in front of the OMC~1 (Orion Molecular Cloud core \#1) molecular filament tracing the central ridge of the Orion~A cloud. For detailed description of the region see \citet[][and references therein]{Odell2001, Muench2008}.  Based on $Chandra$ X-ray \citep{Getman2005} and $Spitzer$ mid-infrared surveys \citep{Megeath2012}, over 1500 MPCMs are identified as likely members of the region \citep{Broos2013}. Of these, $43$\% have available $Age_{JX}$ estimates. The spatial distribution of the MPCMs is associated with four isothermal cluster ellipses \citep{Kuhn2014c}: the small (core radius 0.01~pc) subcluster A containing the embedded cluster around the BN-KL hot core,  subcluster B representing the small and dense Trapezium core of the ONC cluster, the rich subcluster C containing both the ONC cluster and OMC cloud members, and subcluster D extending north-south along the OMC~1 filament behind the ONC. 

Due to the unfavorable viewing angle where the ONC is superposed on the molecular cores, both the Kuhn et al. MPCM sample and the subset used here for $Age_{JX}$ analysis for the subclusters A and D are significantly contaminated by the foreground ONC stars.  A similar problem is evident in the X-ray study of the OMC 1 cores by \citet{Grosso2005}. Our inferred low absorption and relatively high age for these embedded subclusters are likely not realistic. Meanwhile, the subcluster C is likely contaminated by the OMC objects.

The sophisticated analysis of optical data by \citet{DaRio2010} for the ONC led to an age estimate of $2-3$~Myr. Further, combined with the {\it Hubble Space Telescope} photometry, these optical data of stars in ONC give an average age of $2.2$~Myr \citep{Reggiani2011}. The optical data have good coverage of the outskirts of the ONC cluster extending beyond the {\it Chandra} field but, due to the issues of absorption and high nebular emission, their sample misses many stellar members of the dense Trapezium region and the OMC~1 cloud cores.

The reliable $Age_{JX}$ result for this region is that the inner Trapezium region appears young at 1.1~Myr and the `unclustered' component appears older at 2.0~Myr (Figure~\ref{fig_age_gradients_orion}).  The apparent age gradient from the interior of the ONC to its outer region is discussed further in our companion paper \citep{Getman2013ApJL} where we do a better job of separating between ONC and OMC members.

\subsection{Flame Nebula} \label{age_flame_subsection}

The Flame Nebula is a nearby ($D \sim 414$~pc) H\,{\sc{ii}} region in the Orion B cloud associated with the rich, obscured stellar cluster NGC~2024. \citet{Meyer2008} gives a detailed description of the region. There are over 480 MPCMs in the region \citep{Broos2013} of which $21$\% have $Age_{JX}$ estimates. The spatial distribution of the MPCMs is associated with the single elliptical isothermal cluster elongated along the direction of the molecular filament \citep{Kuhn2014c}. The high median $J-H$ color for all MPCMs in the main cluster of 1.8~mag (equivalent to $A_V \sim 11$~mag) suggests that many cluster members are still embedded in the molecular filament.

The previously published age estimates of $\la 0.3-0.5-1$~Myr, based on IR spectroscopy and photometry data \citep[][and references therein]{Levine2006}, indicate that the NGC~2024 cluster is very young. Our $Age_{JX}$ analysis gives a median age of 0.8~Myr for the entire cluster, consistent with the previous estimates (Figure~\ref{fig_age_gradients_flame}).  The age of the less absorbed stellar population that is spatially distributed outside the molecular filament, in the outskirts of the {\it Chandra} field of view, is older around $\sim 1.3$~Myr.   However, we have also found a strong age gradient within the cluster such that the core is significantly older than the intermediate and outer regions.  This is presented and discussed in the companion paper \citep{Getman2013ApJL}.

\subsection{W 40} \label{age_w40_subsection}

The W 40 is a nearby ($D \sim 500$~pc) blister H\,{\sc{ii}} region at the edge of the dense molecular core DoH 279-P7, which in turn is part of the Aquila Rift molecular cloud complex.  A description of the region is given by \citet{Kuhn2010}. There are over 420 MPCMs in the region \citep{Broos2013}; $21$\% have available $Age_{JX}$ estimates. The spatial distribution of the MPCMs is modeled as a single rich isothermal sphere with core radius 0.15~pc \citep{Kuhn2014c}. The high median $(J-H)$ color of $1.9$~mag ($A_V \sim 11$~mag) is attributable to absorption both in the cloud screen and local molecular material.  

$Age_{JX}$  results are shown in Figure~\ref{fig_age_gradients_flame} and listed in Table~\ref{tbl_cluster_ages}.  The $Age_{JX} \sim 0.8$~Myr for the cluster supports the very young age $\la 1$~Myr inferred from the high fraction of $K$-band excess stars \citep{Kuhn2010}.  A strong spatial gradient in age from the young cluster to an older unclustered component ($\sim 1.5$~Myr) is seen.   

\subsection{RCW 36} \label{age_rcw36_subsection}

RCW 36 is a nearby ($D \sim 700$~pc)  H\,{\sc{ii}} region associated with one of the numerous molecular clumps of the giant molecular filament Vela-C \citep{Pettersson2008,Ellerbroek2013}. There are over 380 MPCMs in the region \citep{Broos2013}; $22$\% have $Age_{JX}$ estimates. According to \citet{Kuhn2014c} the spatial distribution of the MPCMs is characterized by a core-halo cluster structure with the core (halo) ellipses elongated perpendicular (parallel) to  the molecular filament. We find here that the stars in the cluster core have a redder median $J-H$ color of 2.2~mag ($A_V \sim 14$~mag) than the stars in the cluster halo ($J-H = 1.7$~mag, $A_V \sim 10$~mag), suggesting that the core is more embedded in the cloud.  

There are insufficient number of 2MASS-$JHK_s$ stars with reliable photometry for $Age_{JX}$ analysis of the cluster core (B).  The inferred age of the cluster halo (A) is 0.9~Myr, and the dispersed unclustered stars are older with $Age_{JX} \sim  1.9$~Myr. From a spectroscopic study of several dozen members, \citet{Ellerbroek2013} find the most common age of members is around 1.1~Myr with a tail to several Myr. They do not find a spatial gradient in ages, but their sample generally lies within the core radius of our component A, omitting many dispersed stars in our unclustered component.  

\subsection{NGC 2264} \label{age_ngc2264_subsection}

NGC~2264 ($D \sim 913$~pc) is a composite star forming region with both embedded and revealed clustered components and a filamentary molecular structure within the Mon~OB1 molecular cloud complex \citep{Dahm2008, Feigelson2013}. There are nearly 1200 MPCMs in the region \citep{Broos2013}; $25$\% have $Age_{JX}$ estimates. The spatial mixture model of \citet{Kuhn2014c} has 13 subclusters: seven in the lightly absorbed northern region (A, B, C, D, and H around the Fox Fur Nebula; E and F near/around the O-type binary S Mon), and six in the mostly embedded southern region (G, I, and J projected against the molecular core in the IRS~1 region; K, L, and M in the IRS~2 region). 

The stellar population of NGC~2264 has been studied for decades.  Recent published age estimates for the stars in the complex, mainly based on optical data, range from 1.5 to 3~Myr \citep{Dahm2007,Mayne2007,Baxter2009}. The $Age_{JX}$ estimates are generally similar.  Our principal new result is an age gradient between the richest northern subclusters (D and E; $Age_{JX} \sim 3.2$~Myr) and the southern subclusters (G, I, J, and M; $\sim 1.5$~Myr). This is consistent with the $Herschel$ satellite image of the cold dust where the northern subclusters are associated with more dispersed molecular material.  Both the rich southern subcluster K ($Age_{JX} \sim 2.2$~Myr) and the dispersed stars in the region ($Age_{JX} \sim 2.8$~Myr) exhibit intermediate ages.

\subsection{Rosette Nebula} \label{age_rosette_subsection}

The Rosette Nebula ($D \sim 1.33$~kpc) is an H\,{\sc{ii}} bubble region ionized by the unobscured rich cluster NGC~2244.  Star formation in the region has been extensively studied  \citep{RomanZunigaLada2008,Wang2010,Ybarra2013}. The {\it Chandra} mosaic encloses NGC~2244 and an extended portion of the filamentary Rosette Molecular Cloud (RMC) to the southeast. There are over 1700 MPCMs in the region \citep{Broos2013}; $19$\% have available $Age_{JX}$ estimates. The spatial analysis of \citet{Kuhn2014c} identified 15 subclusters: 8 northwestern subclusters inside the ionized bubble (A, B, C, D, E, F, G, and H associated with the NGC~2244 and NGC~2237 clusters), 3 subclusters at the northwestern end of the molecular cloud (I, J, and K), and 4 subclusters that lie projected against the densest parts of the southeastern cloud (L, M, N, and O).   As expected, we find that the clusters inside the bubble are much less absorbed with median $(J-H) \sim 0.6-0.8$~mag (or $A_V \la 1.5$~mag) than the clusters associated with cloud cores with median $(J-H) \sim 1-2$~mag (or  $A_V \sim  4-15$~mag).

\citet{Ybarra2013} estimates ages of $\sim 2-3$~Myr for NGC~2244 and $\sim 1$~Myr for the stars in the cloud clusters based on their infrared disk fractions.  Our $Age_{JX}$ values are consistent with these estimates. The main result of our analysis is that we detect an age gradient:  the distributed stellar population and the sparse subclusters B, C, F on the periphery of NGC~2244 have ages $\ga 4$~Myr; NGC~2244 itself (E) has age $\sim 3$~Myr; and the molecular cloud subclusters (J, M, N, and O) have ages  $\sim 1-2$~Myr. Notice that for subclusters M and O, the $Age_{JH}$ values are preferred over the $Age_{JX}$ estimates due to the poor sensitivity of the $Age_{JX}$ sample to heavily embedded members.  We also find that the rich cluster L is superposed in front of the Rosette Molecular Cloud with less absorbed stars and an older ($\ga 2.7$~Myr) inferred age. 

\subsection{Lagoon Nebula} \label{age_lagoon_subsection}

The Lagoon Nebula ($D \sim 1.3$~kpc) is defined by a large H\,{\sc{ii}} bubble ionized by the cluster NGC~6530. The {\it Chandra} field encloses the southern portion of the bubble with the main cluster, a complex interface between the bubble and molecular cloud with numerous bright-rimmed components, and the bright Hourglass Nebula region to the west of the main cluster. Descriptions of the star forming complex are given by  \citet{Tothill2008} and \citet{KumarAnandarao2010}. The region is very rich in X-ray sources with over 2000 MPCM stars \citep{Broos2013}; $31$\% have $Age_{JX}$ estimates.  The spatial analysis of \citet{Kuhn2014c} identified 11 subclusters: three associated with the Hourglass region (B, C, and D), seven associated with NGC~6530 (E, F, G, H, I, J, and K), and one to the northwest of the Hourglass (A). Except for the B and K with median $(J-H) \sim 1.1$~mag ($A_V \sim 4$~mag), all are lightly absorbed with $(J-H) \sim 0.8-0.9$~mag ($A_V<3$~mag).  

The age analysis of \citet{Prisinzano2012} based on optical data indicated a younger $<2$~Myr stellar population in the southeastern portion of the field and an older ($>2$~Myr) stellar populations in the northern and southwestern parts of the field. Our age results are somewhat different.  We find younger ($\la 2$~Myr) star concentrations in clusters projected against the molecular cloud including the  K subcluster near the M8~E cloud, E subcluster near the South Eastern Bright Rim cloud, and B and C subclusters near the HW clump \citep[these molecular structures are shown in][]{Tothill2008}. The subclusters north of the cloud (A, F, I, and J) and G and H subclusters near the Central Ridge clump are found to be older with ages $2.0-2.6$~Myr. The large-scale unclustered component has median age of 2.2~Myr.

\subsection{NGC~2362} \label{age_ngc2362_subsection}

With no molecular material left in its immediate vicinity, the NGC~2362 cluster ($D \sim 1.48$~kpc) is perhaps the oldest MYStIX target. There are nearly 500 MPCMs \citep{Broos2013} of which $24$\% have $Age_{JX}$ estimates. For detailed description of the region see \citet{Dahm2008b}. The spatial analysis of \citet{Kuhn2014c} identifies two subclusters; the principal cluster B harbors the most massive star $\tau $~CMa and most of the cluster members; and the small subcluster A represents a concentration of stars to the north. Both components exhibit low absorption,  $(J-H) \simeq 0.6$~mag (or $A_V \la 1$~mag).

The previous age estimate  based on optical data is $\sim 3.5-5$~Myr \citep{Dahm2005}. We find a large-scale age gradient:  the stars in the primary cluster are significantly younger ($\sim 2.9$~Myr) than the stars of the distributed population ($\sim 3.8$~Myr).  There is also a  systematic difference between the Dahm et al. ages and those derived here, which is entirely due to different choice of pre-main sequence evolutionary models, \citet{Baraffe1998} in place of \citet{Siess2000}.

\subsection{DR~21} \label{age_dr21_subsection}

DR~21 ($D \sim 1.5$~kpc) is a star forming region associated with a long and dense molecular filament in the giant Cygnus~X star forming complex \citep{Kumar2007,ReipurthSchneider2008}. There are nearly 1000 MPCMs, both embedded in the DR~21 filament and distributed within the Cygnus-X complex; $14$\% have $Age_{JX}$ estimates. The spatial analysis of \citet{Kuhn2014c} identifies nine clusters, all positioned and elongated along the molecular filament. With median $J-H \simeq 2-3$~mag (corresponding to $A_V \simeq 13-22$~mag),  these are among the most heavily absorbed MYStIX structures.

\citet{Kumar2007} find an age of $\sim 3$~Myr for the NIR stellar population in the region, but suggest a younger age of $\la 1$~Myr for the clustered mid-infrared population. The results of our age analysis are consistent with these values.  We find a strong age difference between the older $\sim 2.5$~Myr unclustered stellar population uniformly distributed outside the molecular filament, and the younger $\la 1$~Myr stellar population grouped in the small clusters embedded in the molecular cores along the filament. $Age_{JX}$ estimates are not available for some of the heavily embedded clusters (B, C, G, and H), but the alternative $Age_{JH}$ estimates indicate that these are all very young ($\la 1$~Myr).  The age gradient may be due either to the dispersion of older stars from the DR~21 clouds, or to an age difference between the embedded DR~21 stellar population and the large-scale Cygnus X population of pre-main sequence stars arising from many past star forming regions in the complex.

\subsection{NGC~6334} \label{age_ngc6334_subsection}

NGC~6334 ($D \sim 1.7$~kpc) is a massive star forming region associated with a long and dense molecular filament part of a larger cloud that also produced the NGC~6357 complex \citep{PersiTapia2008,Feigelson2009,Russeil2010}. The MYStIX field has nearly 1700 MPCMs \citep{Broos2013}, but only $7$\% have $Age_{JX}$ estimates due to the shallow X-ray data that inhibit derivation of reliable stellar $L_X$ estimates.  The spatial analysis of \citet{Kuhn2014c} identifies 14 stellar structures:  six clusters lie projected against the molecular filament (D, E, J, K, L, and M), one subcluster is associated with a molecular patch northeast of the main filament (N), and seven clusters lie superposed around the filament (A, B, C, F, G, H, and I). This last group has lower obscuration, $(J-H) \simeq 1.0-1.6$~mag ($A_V \simeq 3-9$~mag), than the clusters embedded in the densest parts of the filament (J, K, L, and N) with $(J-H) \simeq 1.9-3.0$~mag ($A_V \simeq 12-22$~mag).

\citet{PersiTapia2008} assume an age of $\la 1$~Myr.  Our $Age_{JX}$ analysis gives an age  $\sim 1.1$~Myr for the combined heavily absorbed subclusters L and J, significantly younger than the dispersed stellar population with median age $\sim 1.9$~Myr. The  alternative $Age_{JH}$ procedure indicates that the heavily absorbed subclusters D, E, K, L, and N are very young with ages $0.6-0.9$~Myr.  The remaining subclusters have insufficient information for age analysis.

\subsection{NGC~6357} \label{age_ngc6357_subsection}

NGC~6357 ($D \sim 1.7$~kpc) is a massive star forming region divided into three major cluster complexes that are associated with the three H\,{\sc{ii}} bubbles, CS~59, CS~61, and CS~63 \citep{Wang2007,Russeil2010,Fang2012}. The stellar population is rich with over 2200 MPCMs \citep{Broos2013}; $13$\% have available $Age_{JX}$ estimates. The spatial analysis of \citet{Kuhn2014c} identifies 6 subclusters:  A, associated with the rich Pismis~24 cluster, and B lie in the CS~61 bubble; C, D, and E in the CS~63 bubble; and F in the CS~59 bubble. All subclusters are subject to similar absorption with median $J-H \simeq 1.3-1.4$~mag ($A_V \simeq 6-7$~mag).

Based on analysis of optical properties of Pismis~24, \citet{Fang2012} find an age of $1$~Myr.  Our $Age_{JX}$ value for subcluster A is close at $1.4$~Myr.  We find no promising age gradients in the full MYStIX region. Despite the diverse morphology of the stellar and the molecular components in the region, most of the MYStIX stars, both in the clusters and those uniformly distributed across the bubbles and the bubble edges, have similar ages of $1.0-1.5$~Myr. It seems that the star formation proceeded nearly simultaneously across this region. 

\subsection{Eagle Nebula} \label{age_eagle_subsection}

The Eagle Nebula ($D \sim 1.75$~kpc) is an H\,{\sc{ii}} region ionized by the rich cluster NGC~6611. The giant molecular cloud surrounding the cluster is composed of numerous dense molecular clumps and bright-rimmed components including the famous Pillars of Creation, Spire Pillar, and Bright Rimmed Cloud 30  \citep{Indebetouw2007,Oliveira2008,Hill2012}. The MYStIX region has over 2500 MPCM stars \citep{Broos2013} of which $19$\% have $Age_{JX}$ estimates. The spatial analysis of \citet{Kuhn2014c} identifies twelve subclusters: three associated with the NGC~6611 cluster (A, B, and C); two possibly associated with the Spire Pillar (D and K); four related to the M16-N cloud (E, G, H, and I); two at the northern edge of the {\it Chandra} field with no obvious cloud associations (J and F); and one at the southeastern edge of the field projected against the M16-E cloud (L). The NGC~6611 components have the lowest obscuration with $(J-H) \la 1.0$~mag ($A_V \la 3$~mag), while the small clusters associated with dense molecular patches (E, G, H, I, and L) are subject to the highest absorption with $(J-H) \sim 1.6-2.6$~mag ($A_V \simeq 9-18$~mag).  The remaining subclusters (D, K, J, and F) are subject to the intermediate absorption.

Based on optical and IR data, \citet{Guarcello2010b} report an age gradient in Eagle Nebula: with the youngest stars toward the north ($\sim 0.3-0.8$~Myr), intermediate-age stars in the NGC~6611 cluster ($0.8-1.3$~Myr), and the oldest stars to the southeast ($\sim 2.6$~Myr). The result of our age analysis qualitatively agrees with these findings: we find the northern structures E, G, H, and I associated with cores of the M16-N cloud to be much younger ($\simeq 0.7-1.3$~Myr) than the stars in the subclusters A, B, C, and D that compose the principal ionizing cluster NGC~6611 ($\simeq 1.5-2.5$~Myr).  However, our individual inferred age values for these cluster structures are systematically older than those found by Guarcello et al. We obtain different results in the southeast region.  Instead of the oldest stars, we find intermediate-age compact subcluster K near the base of the molecular Spire pillar ($Age_{JX} \sim 1.9$~Myr) and very young subcluster L near a core in the M16-E cloud ($Age_{JH} \sim 1.0$~Myr).

\subsection{M 17} \label{age_m17_subsection}

M~17 ($D \sim 2.0$~kpc), one of the brightest H\,{\sc{ii}} regions in the sky, is ionized by the massive cluster NGC~6618. The H\,{\sc{ii}} region lies at the southwestern edge of the enormous extended bubble M17~EB, and it interacts energetically with the surrounding giant molecular cloud producing photodissociation regions at the edges of the two massive cloud components, M17 SW and M17 North \citep{Broos2007,ChiniHoffmeister2008,Povich2009}. There are nearly 2400 MPCMs in the MYStIX field \citep{Broos2013}; $21$\% have $Age_{JX}$ estimates.  The spatial analysis of \citet{Kuhn2014c} identifies fifteen subclusters: three associated with dense portions of the M17 SW cloud (A, B, and F); three projected against the edge of the M17 SW cloud (C, G, and H); one near the edge of the M17 North cloud (M); and eight constituting parts of the rich but clumpy NGC~6618 cluster (D, E, I, J, K, L, N, and O). Most of the structures that lie projected against cloud components are highly absorbed (A, B, C, F, and M have median $(J-H) \simeq 1.7-2.5$~mag or $A_V \simeq 10-17$~mag), whereas most of the structures associated with NGC~6618 are less absorbed ($(J-H) \simeq 1.3-1.5$~mag or $A_V \simeq 6-8$~mag).  Note that Kuhn et al. performed the cluster analysis using only the {\it Chandra} exposure centered on NGC~6618 (Figure \ref{fig_age_gradients_m17}), while the $Age_{JX}$ analysis here also treats outer {\it Chandra} exposures (Figure \ref{fig_age_gradients_m17_special}).  We consider these stars to be unclustered.  

Based on optical and IR data, \citet{Hoffmeister2008} estimate an age of $\sim 0.5$~Myr for the youngest generation of stars in the region.  We find homogeneous subcluster ages within the NGC~6618 and M17~SW regions at $Age_{JX} \simeq 1-1.4$~Myr.   We find older ($\sim 3$~Myr) stars in the eastern {\it Chandra} field that covers part of the extended bubble M17~EB,  and intermediate-age stars that lie projected in between NGC~6618 and M17~EB with $Age_{JX} \sim 1.7$~Myr. A large-scale east-west age gradient is thus present in this region but no evident small-scale temporal structure.  Within the M17 SW cloud, our $Age_{JX}$ analysis is likely not sensitive to the embedded population associated with the UC1 and Kleinman-Wright Object ultra-compact H\,{\sc{ii}} regions, but instead samples an older ($\sim 1.4$~Myr) less absorbed and distributed sub-population. Our alternative $Age_{JH}$ analysis indicates a younger age of $\la 1$~Myr for the subclusters in the cloud (A, B, C, F, and M).

\subsection{Carina Nebula} \label{age_carina_subsection}

The Carina Nebula is one of the richest star-forming complexes in the Galaxy's spiral arms, a starburst `cluster of clusters' with star formation over an extended time and region.  It was extensively studied at X-ray and infrared wavebands in the {\it Chandra Carina Complex Project} \citep{Townsley2011}.  The region harbors a chain of very rich stellar clusters $-$ Tr~15, Tr~14, and Tr~16 from north to south $-$, several secondary clusters, and active star formation in the South Pillars region. Tr~15 appears to be oldest, missing its massive O stars (earlier than O9) probably to supernova explosions \citep{Wang2011}. Tr~16 is also old with post-main sequence massive stars including the famous luminous blue variable $\eta$~Car which is expected to go supernova soon. Tr~14, on the other hand, lies near a dense molecular cloud and appears to be significantly younger.  There are over 7300 MPCMs in the MYStIX study region \citep{Broos2013}; $16$\% have $Age_{JX}$ estimates. 

The spatial analysis of \citet{Kuhn2014c} identifies twenty subclusters: three subclusters associated with Tr~15 (I, H, and F); three associated with Tr~14 (A, B, and C); one for Collinder 232 (D);  four associated with Tr~16 (E, J, K, and L); and the remaining sub-clusters (M, N, O, P, Q, R, S, and T) associated with the South Pillars. The very elongated ellipse G covers heterogeneous structures in the Carina Complex and should not be viewed as a distinct structure. Most of the structures are lightly absorbed with medians  $(J-H) \simeq 0.8-1.0$~mag ($A_V \simeq 1-3$~mag).  The principal exception is  embedded Treasure Chest cluster (O) with $(J-H) \simeq 1.2$~mag ($A_V \ga 5$~mag).

Based on near-IR data, \citet{Preibisch2011} give age estimates of (from north to south) $5-8$~Myr for Tr~15, $\la 3$~Myr for Tr~14 and Collinder~232, $3-4$~Myr for Tr~16, and $<1$~Myr for the Treasure Chest. For the major Trumpler clusters, our $Age_{JX}$ estimates show similar results when averaged over the rich clusters, although our ages are systematically younger.  But we find heterogeneity in structure and age within these rich clusters.  In Tr~15, the central cluster H with the massive stars has age $\sim 2.8$~Myr while the remaining two secondary subclusters, I and F, and the stellar population in the outskirts of the region are substantially older ($\ga 4.5$ Myr).  Tr~14 is youngest but with a younger core component at 1.5~Myr and an older halo component at 2.7~Myr; this result is qualitatively consistent with the core-halo age gradient reported by \citet{Ascenso2007}. The Tr~16 region is diverse. The subcluster K harboring $\eta$~Car is oldest at $3.6 \pm 0.7$~Myr, consistent with the theoretical minimum age of $\sim 3$~Myr for the supergiant $\eta$~Car \citep{SmithBrooks2008}. The remaining cluster structures in Tr~16 are younger ($\sim 2.5$~Myr) and lie projected closer to the denser parts of the molecular cloud. 

The South Pillars region harbors stellar populations of different ages with a wide range of ages from 1 to 5~Myr.  Some portions of the South Pillars region show a bimodal age distribution suggesting superposed younger and older populations.  This supports the scenario presented by \citet{Povich2011} and \citet{Feigelson2011} that these clouds have been (and will continue to) produce stars over an extended period of triggered star formation.   The Treasure Chest is the only localized distinct cluster that is young with $Age_{JX} \simeq 1.1$ Myr. 

\subsection{NGC~1893} \label{age_ngc1893_subsection}

The most distant MYStIX target, NGC~1893 at $D \simeq 3.6$~kpc, is a giant H\,{\sc{ii}} region surrounded by  $\sim 1.5^\circ$ diameter incomplete ring of molecular cloud. The MYStIX  field encompasses the central part of the region where most cloud material has been evacuated \citep{Caramazza2008,Prisinzano2011,Pandey2013}. There are about 1300 MPCMs \citep{Broos2013}; $27$\% have $Age_{JX}$ estimates by virtue of an unusually long {\it Chandra} exposure.  The spatial analysis of \citet{Kuhn2014c} divides the central cluster into a linear chain of ten subclusters. The subclusters G and J are located at the heads of giant molecular pillars and exhibit mild absorption with $(J-H) \simeq 1.0$~mag ($A_V \simeq 3$~mag).  All the other subclusters are lightly absorbed with $A_V \simeq 1$~mag.

According to the optical analyses of \citet{Prisinzano2011} and \citet{Pandey2013}, the  majority of young stars in the region have ages between $\la 1$ and 5~Myr with median age around 1.5~Myr. Our age analysis shows that the distributed stellar population is older, with median $Age_{JX} \simeq 3$~Myr, than the majority of the clustered population. A dramatic spatial gradient in subcluster ages is seen with the older structures ($2.6-3.5$~Myr; A, B, and C) lying at the southwestern end of the cluster chain and the younger subclusters ($1.4-2.1$~Myr; D, F, G, H, and J) lying at the northeastern end of the chain. The northeastern structure I is an exception from this gradient; however, the age histogram for its constitute stars appears bimodal suggesting superposition of younger and older stellar populations.

\clearpage
\newpage

%% FIGURES======================================================================
\clearpage

\clearpage

\clearpage

\clearpage

\clearpage

\clearpage
\newpage

\clearpage
\newpage

\clearpage
\newpage

\clearpage
\newpage

\clearpage
\newpage

\clearpage
\newpage

\clearpage
\newpage

\clearpage
\newpage

\clearpage
\newpage

\clearpage
\newpage

\clearpage
\newpage

\clearpage
\newpage

\clearpage
\newpage

\clearpage
\newpage

\clearpage
\newpage

\clearpage
\newpage

\clearpage
\newpage

%%TABLES========================================================================

\begin{deluxetable}{ccccccccccc}
\centering \tabletypesize{\footnotesize} \rotate \tablewidth{0pt} \tablecolumns{11}

\tablecaption{Individual Stellar Age Estimates \label{tbl_individual_ages}}

\tablehead{\colhead{SF} & \colhead{Source} & \colhead{R.A.} & \colhead{Decl.} & \colhead{$J$} & \colhead{$H$} & \colhead{$K_s$} & \colhead{$L_X$} & \colhead{$A_J$} & \colhead{$Age_{JX}$} & \colhead{Mem}\\

\colhead{Region} & \colhead{Name} & \multicolumn{2}{c}{(J2000 deg)} & \colhead{(mag)} & \colhead{(mag)} & \colhead{(mag)} & \colhead{(10$^{30}$~erg s$^{-1}$)} & \colhead{(mag)} & \colhead{(Myr)} & \colhead{}\\

\colhead{(1)} & \colhead{(2)} & \colhead{(3)} & \colhead{(4)} & \colhead{(5)} & \colhead{(6)}  & \colhead{(7)}  & \colhead{(8)}  & \colhead{(9)}  & \colhead{(10)}  & \colhead{(11)}}

\startdata
NGC 1893 & 052242.31+332608.3 &    80.676310 &    33.435659 & $ 15.54\pm 0.01 $ & $ 14.75\pm 0.01$  & $ 14.49\pm 0.01 $ & $    2.11\pm   0.97 $ &    0.3 & 1.2 & U\\
NGC 1893 & 052242.38+332422.0 &    80.676610 &    33.406116 & $ 17.15\pm 0.01 $ & $ 16.27\pm 0.01$  & $ 15.96\pm 0.01 $ & $    1.62\pm   0.78 $ &    0.5 & 4.6 & B\\
NGC 1893 & 052242.55+333256.6 &    80.677320 &    33.549070 & $ 15.62\pm 0.01 $ & $ 14.94\pm 0.01$  & $ 14.75\pm 0.01 $ & $    0.41\pm   0.18 $ &    0.0 & 0.7 & U\\
NGC 1893 & 052242.63+332449.4 &    80.677660 &    33.413723 & $ 15.88\pm 0.01 $ & $ 14.91\pm 0.01$  & $ 14.51\pm 0.01 $ & $    2.98\pm   1.95 $ &    0.8 & 1.0 & B\\
NGC 1893 & 052242.67+332454.9 &    80.677820 &    33.415276 & $ 17.36\pm 0.01 $ & $ 16.40\pm 0.01$  & $ 16.00\pm 0.01 $ & $    1.66\pm   1.01 $ &    0.8 & 3.5 & B\\
NGC 1893 & 052242.86+332445.6 &    80.678620 &    33.412678 & $ 16.50\pm 0.01 $ & $ 15.55\pm 0.01$  & $ 15.15\pm 0.01 $ & $    2.60\pm   1.25 $ &    0.8 & 2.6 & B\\
NGC 1893 & 052243.04+332632.0 &    80.679364 &    33.442225 & $ 16.34\pm 0.01 $ & $ 15.57\pm 0.01$  & $ 15.28\pm 0.01 $ & $    3.05\pm   1.40 $ &    0.3 & 4.1 & U\\
NGC 1893 & 052243.25+332011.0 &    80.680220 &    33.336400 & $ 16.03\pm 0.01 $ & $ 15.30\pm 0.01$  & $ 14.95\pm 0.01 $ & $    1.69\pm   1.12 $ &    0.3 & 1.9 & U\\
NGC 1893 & 052243.25+332735.4 &    80.680249 &    33.459840 & $ 17.19\pm 0.01 $ & $ 16.35\pm 0.01$  & $ 16.08\pm 0.01 $ & $    1.50\pm   0.86 $ &    0.4 & 4.6 & U\\
NGC 1893 & 052243.31+332641.1a &    80.680470 &    33.444770 & $ 17.42\pm 0.01 $ & $ 16.58\pm 0.01$  & $ 16.15\pm 0.02 $ & $    0.34\pm   0.21 $ &    0.6 & 1.7 & U\\
\enddata

\tablecomments{This table is available in its entirety (5525 MYStIX stars) in the machine-readable form in the on-line journal. A portion is shown here for guidance regarding its form and content. Column 1: Name of a star forming region. Column 2: MYStIX source's IAU designation. Columns 3 and 4: Right ascension and declination for epoch J2000.0 in degrees. Columns 5-7: NIR $JHK_s$ magnitudes from \citet{Broos2013}. Column 8: Intrinsic X-ray luminosity in the $(0.5-8)$~keV band from \citet{Broos2013}. The statistical and systematic errors on $L_X$ are summed in quadrature. Column 9: Estimate of the source extinction in the $J$-band. Column 10: $Age_{JX}$ estimate. Individual age estimates of 5~Myr indicate that ages run into our truncation limit of 5~Myr. Column 11: Cluster membership from \citet{Kuhn2014c}. Unclustered or ambiguous stars are denoted as 'U'. All $Age_{JX}$ members of the W~3, W~4, and NGC~3576 regions are denoted as 'U', since these regions were omitted from the analyses of Kuhn et al.}
\end{deluxetable}

\begin{deluxetable}{cccccccccccc}
\centering \tabletypesize{\footnotesize} \rotate \tablewidth{0pt} \tablecolumns{12}

\tablecaption{Age Estimates for the MYStIX Clusters \label{tbl_cluster_ages}}

\tablehead{\colhead{SF} & \colhead{MYStIX} & \colhead{R.A.} & \colhead{Decl.} & \colhead{$R\arcmin$} & \colhead{$R_{pc}$} & \colhead{$N_{tot}$} & \colhead{$median$} & \colhead{$N_{AgeJX}$} & \colhead{$median$} & \colhead{$median$} & \colhead{Age${\rm_{JH}}$}\\

\colhead{} & \colhead{} & \colhead{} & \colhead{} & \colhead{} & \colhead{} & \colhead{} & \colhead{$(J-H)_{tot}$} & \colhead{} & \colhead{$(J-H)_{AgeJX}$} & \colhead{$(Age_{JX})$} & \colhead{}\\

\colhead{Region} & \colhead{Cluster} & \multicolumn{2}{c}{(J2000 deg)} & \colhead{($4R_c\arcmin$)} & \colhead{($4R_c$~pc)} & \colhead{} & \colhead{(mag)} & \colhead{} & \colhead{(mag)} & \colhead{(Myr)} & \colhead{(Myr)}\\

\colhead{(1)} & \colhead{(2)} & \colhead{(3)} & \colhead{(4)} & \colhead{(5)} & \colhead{(6)}  & \colhead{(7)}  & \colhead{(8)}  & \colhead{(9)}  & \colhead{(10)}  & \colhead{(11)} & \colhead{(12)}}

\startdata
Orion & A &    83.811003 &    -5.375278 &    0.4 &    0.0 &    25 &    1.0 &     9 & $ 1.0\pm 0.2$ & $ 1.4\pm 0.5$ & \nodata\\
Orion & B &    83.815418 &    -5.389725 &    1.6 &    0.2 &    41 &    0.9 &    20 & $ 1.0\pm 0.1$ & $ 1.1\pm 0.4$ & \nodata\\
Orion & C &    83.819538 &    -5.376180 &    7.7 &    0.9 &  1211 &    1.1 &   522 & $ 1.1\pm 0.0$ & $ 1.5\pm 0.1$ & \nodata\\
Orion & D &    83.824266 &    -5.276333 &    4.4 &    0.5 &    30 &    1.2 &    12 & $ 1.1\pm 0.2$ & $ 2.7\pm 0.9$ & \nodata\\
Orion & U &   \nodata &   \nodata &  \nodata &  \nodata &   217 &    0.8 &    94 & $ 0.8\pm 0.0$ & $ 2.0\pm 0.2$ & \nodata\\
Flame & A &    85.427087 &    -1.903796 &    4.2 &    0.5 &   343 &    1.8 &    56 & $ 1.7\pm 0.1$ & $ 0.8\pm 0.2$ &   1.1:\\
Flame & U &   \nodata &   \nodata &  \nodata &  \nodata &   141 &    1.3 &    47 & $ 1.3\pm 0.1$ & $ 1.3\pm 0.3$ & \nodata\\
W 40 & A &   277.861454 &    -2.094043 &    4.5 &    0.6 &   235 &    2.1 &    68 & $ 1.9\pm 0.1$ & $ 0.8\pm 0.1$ &   0.9:\\
W 40 & U &   \nodata &   \nodata &  \nodata &  \nodata &   191 &    1.9 &    34 & $ 1.6\pm 0.1$ & $ 1.5\pm 0.2$ &   1.0:\\
RCW 36 & A &   134.863096 &   -43.755983 &    3.2 &    0.6 &   227 &    1.7 &    40 & $ 1.6\pm 0.0$ & $ 0.9\pm 0.1$ &   1.1:\\
RCW 36 & B &   134.863598 &   -43.757322 &    0.4 &    0.1 &    32 &    2.2 & \nodata & \nodata & \nodata &   0.9:\\
RCW 36 & U &   \nodata &   \nodata &  \nodata &  \nodata &    49 &    1.4 &    43 & $ 1.4\pm 0.0$ & $ 1.9\pm 0.5$ & \nodata\\
NGC 2264 & A &   100.131244 &     9.831153 &    1.3 &    0.3 &    17 &    0.9 &     5 & $ 0.7\pm 0.2$ & $ 0.9\pm 1.0$ & \nodata\\
NGC 2264 & B &   100.154592 &     9.791891 &    0.4 &    0.1 &     9 &    0.6 & \nodata & \nodata & \nodata & \nodata\\
NGC 2264 & C &   100.168359 &     9.851570 &    0.4 &    0.1 &     6 &  \nodata & \nodata & \nodata & \nodata & \nodata\\
NGC 2264 & D &   100.191014 &     9.817594 &    1.8 &    0.5 &    21 &    0.6 &     4 & $ 0.7\pm 0.1$ & $ 3.2\pm 1.3$ & \nodata\\
NGC 2264 & E &   100.246348 &     9.872885 &    3.5 &    0.9 &    86 &    0.6 &    23 & $ 0.6\pm 0.0$ & $ 3.2\pm 0.5$ & \nodata\\
NGC 2264 & F &   100.246812 &     9.899766 &    0.8 &    0.2 &    19 &    0.6 & \nodata & \nodata & \nodata & \nodata\\
NGC 2264 & G &   100.247385 &     9.603203 &    1.3 &    0.3 &    33 &    2.2 &    10 & $ 1.9\pm 0.5$ & $ 1.5\pm 0.4$ &   0.9:\\
NGC 2264 & H &   100.258800 &     9.812112 &    2.6 &    0.7 &    30 &    0.6 & \nodata & \nodata & \nodata & \nodata\\
NGC 2264 & I &   100.268675 &     9.599235 &    1.2 &    0.3 &    55 &    1.1 &     9 & $ 1.1\pm 0.5$ & $ 1.5\pm 1.1$ & \nodata\\
NGC 2264 & J &   100.276199 &     9.569136 &    2.2 &    0.6 &   105 &    1.2 &    20 & $ 1.2\pm 0.4$ & $ 1.6\pm 0.7$ & \nodata\\
NGC 2264 & K &   100.284295 &     9.498011 &    2.7 &    0.7 &   104 &    0.7 &    34 & $ 0.7\pm 0.0$ & $ 2.2\pm 0.3$ & \nodata\\
NGC 2264 & L &   100.303280 &     9.486087 &    0.4 &    0.1 &    13 &    1.8 & \nodata & \nodata & \nodata &   1.0:\\
NGC 2264 & M &   100.312268 &     9.444966 &    1.2 &    0.3 &    27 &    0.8 &    10 & $ 0.8\pm 0.1$ & $ 1.2\pm 0.4$ & \nodata\\
NGC 2264 & U &   \nodata &   \nodata &  \nodata &  \nodata &   648 &    0.6 &   176 & $ 0.6\pm 0.0$ & $ 2.8\pm 0.2$ & \nodata\\
Rosette & A &    97.737964 &     4.965760 &   10.7 &    4.1 &    49 &    0.8 & \nodata & \nodata & \nodata & \nodata\\
Rosette & B &    97.835327 &     4.837189 &    1.6 &    0.6 &     8 &    0.7 &     4 & $ 0.7\pm 0.0$ & $ 4.3\pm 0.7$ & \nodata\\
Rosette & C &    97.883154 &     4.849333 &    1.9 &    0.7 &    21 &    0.8 &    11 & $ 0.8\pm 0.0$ & $ 4.1\pm 0.6$ & \nodata\\
Rosette & D &    97.980824 &     4.944177 &    0.9 &    0.4 &    22 &    0.6 & \nodata & \nodata & \nodata & \nodata\\
Rosette & E &    97.996974 &     4.913893 &    9.1 &    3.5 &   637 &    0.8 &   202 & $ 0.8\pm 0.0$ & $ 3.0\pm 0.2$ & \nodata\\
Rosette & F &    98.022888 &     4.805445 &    1.5 &    0.6 &     9 &    0.8 &     4 & $ 0.8\pm 0.0$ & $ 4.0\pm 0.9$ & \nodata\\
Rosette & G &    98.193852 &     4.759768 &    1.6 &    0.6 &     3 &  \nodata & \nodata & \nodata & \nodata & \nodata\\
Rosette & H &    98.279829 &     4.782528 &    4.8 &    1.9 &    44 &    0.8 & \nodata & \nodata & \nodata & \nodata\\
Rosette & I &    98.292418 &     4.517794 &    1.9 &    0.7 &    15 &    1.0 & \nodata & \nodata & \nodata & \nodata\\
Rosette & J &    98.312868 &     4.585495 &    1.3 &    0.5 &    11 &    2.2 & \nodata & \nodata & \nodata &   0.9:\\
Rosette & K &    98.333639 &     4.616929 &    0.9 &    0.4 &     5 &  \nodata & \nodata & \nodata & \nodata & \nodata\\
Rosette & L &    98.544391 &     4.418399 &    8.9 &    3.5 &   265 &    1.1 &    24 & $ 1.1\pm 0.2$ & $ 2.7\pm 0.7$ & \nodata\\
Rosette & M &    98.553634 &     4.318479 &    4.6 &    1.8 &    75 &    2.2 &     4 & $ 1.5\pm 0.3$ & $ 1.9\pm 0.4$ &   0.9:\\
Rosette & N &    98.630872 &     4.319126 &    1.8 &    0.7 &    36 &    1.6 &     9 & $ 1.5\pm 0.2$ & $ 1.3\pm 1.4$ &   1.2:\\
Rosette & O &    98.654683 &     4.217308 &    1.4 &    0.6 &    20 &    1.8 &     3 & $ 2.5\pm 0.6$ & $ 1.7\pm 0.9$ &   1.1:\\
Rosette & U &   \nodata &   \nodata &  \nodata &  \nodata &   510 &    0.8 &    65 & $ 0.8\pm 0.0$ & $ 3.9\pm 0.2$ & \nodata\\
Lagoon & A &   270.849088 &   -24.255214 &    2.7 &    1.0 &    39 &    0.9 &    20 & $ 0.8\pm 0.0$ & $ 2.2\pm 0.2$ & \nodata\\
Lagoon & B &   270.917151 &   -24.377850 &    0.6 &    0.2 &    74 &    1.2 &    20 & $ 1.1\pm 0.1$ & $ 1.4\pm 0.6$ & \nodata\\
Lagoon & C &   270.943049 &   -24.367034 &    2.0 &    0.7 &    81 &    0.9 &    24 & $ 0.9\pm 0.1$ & $ 1.6\pm 0.2$ & \nodata\\
Lagoon & D &   270.963550 &   -24.352091 &    0.8 &    0.3 &    23 &    0.8 &     9 & $ 0.8\pm 0.0$ & $ 1.8\pm 0.8$ & \nodata\\
Lagoon & E &   271.031512 &   -24.431435 &    5.1 &    1.9 &   139 &    0.8 &    48 & $ 0.8\pm 0.0$ & $ 1.9\pm 0.2$ & \nodata\\
Lagoon & F &   271.055521 &   -24.307465 &    9.9 &    3.7 &   500 &    0.8 &   191 & $ 0.8\pm 0.0$ & $ 2.3\pm 0.1$ & \nodata\\
Lagoon & G &   271.083619 &   -24.380807 &    0.7 &    0.3 &    33 &    0.8 &    11 & $ 0.9\pm 0.1$ & $ 2.2\pm 0.6$ & \nodata\\
Lagoon & H &   271.097216 &   -24.353499 &    2.5 &    0.9 &   143 &    0.8 &    37 & $ 0.8\pm 0.0$ & $ 2.1\pm 0.4$ & \nodata\\
Lagoon & I &   271.117721 &   -24.379582 &    3.5 &    1.3 &   229 &    0.8 &    64 & $ 0.8\pm 0.0$ & $ 2.1\pm 0.2$ & \nodata\\
Lagoon & J &   271.164951 &   -24.388953 &    2.6 &    1.0 &    76 &    0.9 &    22 & $ 0.8\pm 0.0$ & $ 2.7\pm 0.2$ & \nodata\\
Lagoon & K &   271.210306 &   -24.438561 &    3.8 &    1.4 &   127 &    1.0 &    19 & $ 0.9\pm 0.1$ & $ 1.4\pm 0.2$ & \nodata\\
Lagoon & U &   \nodata &   \nodata &  \nodata &  \nodata &   592 &    0.8 &   171 & $ 0.8\pm 0.0$ & $ 2.2\pm 0.2$ & \nodata\\
NGC 2362 & A &   109.657400 &   -24.899454 &    1.6 &    0.7 &    44 &    0.6 &    11 & $ 0.7\pm 0.0$ & $ 3.2\pm 0.6$ & \nodata\\
NGC 2362 & B &   109.678847 &   -24.962153 &    3.8 &    1.6 &   240 &    0.6 &    51 & $ 0.7\pm 0.0$ & $ 2.9\pm 0.2$ & \nodata\\
NGC 2362 & U &   \nodata &   \nodata &  \nodata &  \nodata &   207 &    0.6 &    57 & $ 0.7\pm 0.0$ & $ 3.8\pm 0.1$ & \nodata\\
DR 21 & A &   309.715318 &    42.314510 &    1.2 &    0.5 &    19 &    2.1 &     3 & $ 1.7\pm 0.7$ & $ 0.6\pm 1.1$ &   0.9:\\
DR 21 & B &   309.740409 &    42.297373 &    0.5 &    0.2 &     9 &    2.0 & \nodata & \nodata & \nodata &   1.0:\\
DR 21 & C &   309.750805 &    42.313176 &    0.5 &    0.2 &    22 &    3.0 & \nodata & \nodata & \nodata &   0.6:\\
DR 21 & D &   309.751734 &    42.329353 &    1.1 &    0.5 &    96 &    2.9 &     7 & $ 2.2\pm 0.5$ & $ 0.7\pm 0.5$ &   0.6:\\
DR 21 & E &   309.752011 &    42.377500 &    1.6 &    0.7 &   109 &    2.5 &     8 & $ 2.0\pm 0.4$ & $ 1.0\pm 0.8$ &   0.8:\\
DR 21 & F &   309.753634 &    42.411888 &    0.4 &    0.2 &    15 &  \nodata & \nodata & \nodata & \nodata & \nodata\\
DR 21 & G &   309.765774 &    42.280760 &    1.1 &    0.5 &    13 &    2.7 & \nodata & \nodata & \nodata &   0.7:\\
DR 21 & H &   309.766340 &    42.426072 &    0.8 &    0.3 &    31 &    3.0 & \nodata & \nodata & \nodata &   0.6:\\
DR 21 & I &   309.772383 &    42.354995 &    1.0 &    0.4 &    24 &    3.0 &     4 & $ 2.8\pm 0.4$ & $ 1.1\pm 1.1$ &   0.6:\\
DR 21 & U &   \nodata &   \nodata &  \nodata &  \nodata &   641 &    1.6 &   110 & $ 1.3\pm 0.1$ & $ 2.5\pm 0.3$ &   1.3:\\
NGC 6334 & A &   259.991145 &   -35.900501 &    1.0 &    0.5 &    38 &    1.4 & \nodata & \nodata & \nodata & \nodata\\
NGC 6334 & B &   259.993594 &   -35.939946 &    3.0 &    1.5 &   129 &    1.4 &     6 & $ 1.1\pm 0.2$ & $ 2.3\pm 0.4$ & \nodata\\
NGC 6334 & C &   260.011243 &   -35.973044 &    0.5 &    0.2 &    17 &    1.1 & \nodata & \nodata & \nodata & \nodata\\
NGC 6334 & D &   260.061955 &   -35.912041 &    0.6 &    0.3 &    18 &    1.9 & \nodata & \nodata & \nodata &   1.0:\\
NGC 6334 & E &   260.079016 &   -35.916462 &    1.9 &    1.0 &   100 &    1.9 & \nodata & \nodata & \nodata &   1.0:\\
NGC 6334 & F &   260.096742 &   -35.950008 &    1.7 &    0.8 &    39 &    1.3 & \nodata & \nodata & \nodata & \nodata\\
NGC 6334 & G &   260.104977 &   -35.734512 &    0.9 &    0.4 &    25 &    1.6 & \nodata & \nodata & \nodata &   1.2:\\
NGC 6334 & H &   260.130189 &   -35.903772 &    1.2 &    0.6 &    40 &    1.4 &     4 & $ 1.2\pm 0.1$ & $ 1.6\pm 0.4$ & \nodata\\
NGC 6334 & I &   260.146879 &   -35.988912 &    0.9 &    0.5 &    12 &    1.0 & \nodata & \nodata & \nodata & \nodata\\
NGC 6334 & J &   260.163313 &   -35.824230 &    2.6 &    1.3 &   225 &    2.2 &    14 & $ 2.0\pm 0.3$ & $ 1.5\pm 0.4$ &   0.9:\\
NGC 6334 & K &   260.201620 &   -35.716380 &    0.9 &    0.5 &    19 &    3.0 & \nodata & \nodata & \nodata &   0.6:\\
NGC 6334 & L &   260.226540 &   -35.761934 &    2.0 &    1.0 &   122 &    2.7 &     5 & $ 2.7\pm 0.5$ & $ 0.7\pm 0.3$ &   0.7:\\
NGC 6334 & M &   260.238922 &   -35.661256 &    1.4 &    0.7 &    19 &  \nodata & \nodata & \nodata & \nodata & \nodata\\
NGC 6334 & N &   260.385219 &   -35.674128 &    2.3 &    1.1 &    15 &    2.2 & \nodata & \nodata & \nodata &   0.9:\\
NGC 6334 & U &   \nodata &   \nodata &  \nodata &  \nodata &   845 &    1.4 &    75 & $ 1.2\pm 0.0$ & $ 1.9\pm 0.2$ & \nodata\\
NGC 6357 & A &   261.181999 &   -34.201943 &    2.0 &    1.0 &   282 &    1.3 &    29 & $ 1.2\pm 0.0$ & $ 1.4\pm 0.1$ & \nodata\\
NGC 6357 & B &   261.194671 &   -34.256275 &    3.9 &    1.9 &   239 &    1.3 &    23 & $ 1.3\pm 0.0$ & $ 1.4\pm 0.2$ & \nodata\\
NGC 6357 & C &   261.388830 &   -34.412054 &    2.2 &    1.1 &   229 &    1.3 &    27 & $ 1.3\pm 0.0$ & $ 1.2\pm 0.3$ & \nodata\\
NGC 6357 & D &   261.392877 &   -34.385974 &    0.4 &    0.2 &    49 &    1.3 &     8 & $ 1.2\pm 0.1$ & $ 1.1\pm 0.6$ & \nodata\\
NGC 6357 & E &   261.449624 &   -34.453379 &    5.0 &    2.5 &   126 &    1.3 &    14 & $ 1.4\pm 0.1$ & $ 1.4\pm 0.4$ & \nodata\\
NGC 6357 & F &   261.508987 &   -34.278275 &    2.2 &    1.1 &   299 &    1.4 &    59 & $ 1.4\pm 0.0$ & $ 1.5\pm 0.2$ & \nodata\\
NGC 6357 & U &   \nodata &   \nodata &  \nodata &  \nodata &  1011 &    1.4 &   138 & $ 1.3\pm 0.0$ & $ 1.5\pm 0.2$ & \nodata\\
Eagle & A &   274.666334 &   -13.794696 &    0.9 &    0.5 &    56 &    0.9 &     6 & $ 0.9\pm 0.0$ & $ 2.4\pm 1.0$ & \nodata\\
Eagle & B &   274.675703 &   -13.784286 &    5.3 &    2.7 &  1117 &    1.0 &   253 & $ 1.0\pm 0.0$ & $ 2.1\pm 0.1$ & \nodata\\
Eagle & C &   274.720055 &   -13.778617 &    1.4 &    0.7 &    39 &    1.0 &    10 & $ 1.0\pm 0.1$ & $ 1.7\pm 0.3$ & \nodata\\
Eagle & D &   274.738796 &   -13.756268 &    9.0 &    4.6 &   542 &    1.0 &   127 & $ 1.0\pm 0.0$ & $ 2.5\pm 0.2$ & \nodata\\
Eagle & E &   274.782841 &   -13.607665 &    0.6 &    0.3 &    23 &    2.4 &     3 & $ 1.6\pm 0.2$ & $ 1.0\pm 1.3$ &   0.8:\\
Eagle & F &   274.803157 &   -13.434045 &    3.3 &    1.7 &    55 &    1.6 & \nodata & \nodata & \nodata &   1.2:\\
Eagle & G &   274.804783 &   -13.562508 &    1.2 &    0.6 &    25 &    2.6 & \nodata & \nodata & \nodata &   0.7:\\
Eagle & H &   274.812350 &   -13.656935 &    0.8 &    0.4 &    13 &    1.6 & \nodata & \nodata & \nodata &   1.3:\\
Eagle & I &   274.829961 &   -13.609316 &    2.3 &    1.2 &    44 &    1.7 &     5 & $ 1.7\pm 0.1$ & $ 0.8\pm 0.4$ &   1.2:\\
Eagle & J &   274.873687 &   -13.385557 &    1.9 &    1.0 &    15 &    1.5 & \nodata & \nodata & \nodata &   1.4:\\
Eagle & K &   274.877955 &   -13.758213 &    0.5 &    0.2 &    14 &    1.4 &     5 & $ 1.3\pm 0.1$ & $ 1.9\pm 0.8$ & \nodata\\
Eagle & L &   275.011498 &   -13.807294 &    0.2 &    0.1 &     5 &    2.0 & \nodata & \nodata & \nodata &   1.0:\\
Eagle & U &   \nodata &   \nodata &  \nodata &  \nodata &   626 &    1.0 &    80 & $ 1.0\pm 0.1$ & $ 2.2\pm 0.2$ & \nodata\\
M 17 & A &   275.075084 &   -16.236297 &    0.5 &    0.3 &    13 &    2.0 & \nodata & \nodata & \nodata &   1.0:\\
M 17 & B &   275.081195 &   -16.224450 &    0.2 &    0.1 &     7 &    2.5 & \nodata & \nodata & \nodata &   0.8:\\
M 17 & C &   275.089865 &   -16.176439 &    0.9 &    0.6 &    77 &    1.7 &    18 & $ 1.6\pm 0.1$ & $ 1.4\pm 0.5$ &   1.1:\\
M 17 & D &   275.093603 &   -16.139853 &    1.9 &    1.1 &   257 &    1.5 &    75 & $ 1.5\pm 0.0$ & $ 1.1\pm 0.2$ &   1.4:\\
M 17 & E &   275.093839 &   -16.164482 &    0.6 &    0.3 &    27 &    1.5 &     5 & $ 1.2\pm 0.2$ & $ 2.4\pm 1.2$ &   1.4:\\
M 17 & F &   275.095044 &   -16.206489 &    0.3 &    0.2 &    15 &    1.8 & \nodata & \nodata & \nodata &   1.0:\\
M 17 & G &   275.104021 &   -16.192919 &    0.3 &    0.1 &    17 &    1.6 & \nodata & \nodata & \nodata &   1.3:\\
M 17 & H &   275.106514 &   -16.188004 &    0.8 &    0.4 &    55 &    1.3 &     9 & $ 1.2\pm 0.2$ & $ 1.0\pm 0.9$ & \nodata\\
M 17 & I &   275.108478 &   -16.159423 &    1.2 &    0.7 &    83 &    1.4 &    40 & $ 1.4\pm 0.0$ & $ 1.4\pm 0.3$ & \nodata\\
M 17 & J &   275.115791 &   -16.165053 &    0.1 &    0.0 &     2 &  \nodata & \nodata & \nodata & \nodata & \nodata\\
M 17 & K &   275.118917 &   -16.182870 &    1.2 &    0.7 &    98 &    1.4 &    19 & $ 1.4\pm 0.1$ & $ 1.0\pm 0.3$ & \nodata\\
M 17 & L &   275.124664 &   -16.179366 &    0.9 &    0.5 &   142 &    1.8 &    30 & $ 1.6\pm 0.1$ & $ 1.2\pm 0.5$ &   1.0:\\
M 17 & M &   275.128330 &   -16.054331 &    1.2 &    0.7 &    27 &    1.8 & \nodata & \nodata & \nodata &   1.0:\\
M 17 & N &   275.130814 &   -16.160819 &    0.9 &    0.5 &    71 &    1.3 &    23 & $ 1.3\pm 0.1$ & $ 1.6\pm 0.5$ & \nodata\\
M 17 & O &   275.131351 &   -16.189476 &    0.5 &    0.3 &    48 &    1.4 &    11 & $ 1.3\pm 0.1$ & $ 0.7\pm 0.3$ & \nodata\\
M 17 & U &   \nodata &   \nodata &  \nodata &  \nodata &  1031 &    1.4 &   263 & $ 1.3\pm 0.0$ & $ 1.6\pm 0.2$ & \nodata\\
Carina & A &   160.924382 &   -59.596843 &    0.8 &    0.6 &    42 &    1.0 &     6 & $ 1.0\pm 0.2$ & $ 2.8\pm 0.8$ & \nodata\\
Carina & B &   160.975828 &   -59.550664 &    3.4 &    2.3 &   962 &    1.0 &   111 & $ 1.0\pm 0.0$ & $ 2.7\pm 0.2$ & \nodata\\
Carina & C &   160.985040 &   -59.548341 &    1.2 &    0.8 &   294 &    0.9 &    30 & $ 1.0\pm 0.0$ & $ 1.5\pm 0.2$ & \nodata\\
Carina & D &   161.137282 &   -59.561635 &    2.7 &    1.8 &   119 &    0.9 &    15 & $ 0.9\pm 0.1$ & $ 2.4\pm 0.8$ & \nodata\\
Carina & E &   161.141734 &   -59.735484 &    1.7 &    1.1 &   129 &    0.8 &    28 & $ 0.9\pm 0.0$ & $ 2.4\pm 0.3$ & \nodata\\
Carina & F &   161.155625 &   -59.434109 &    3.1 &    2.1 &   169 &    0.9 &    42 & $ 1.0\pm 0.0$ & $ 3.8\pm 0.4$ & \nodata\\
Carina & G &   161.165355 &   -59.740477 &   36.0 &   24.1 &  1122 &    0.9 &   202 & $ 0.9\pm 0.0$ & $ 3.4\pm 0.3$ & \nodata\\
Carina & H &   161.174039 &   -59.367985 &    1.6 &    1.0 &   242 &    0.8 &    24 & $ 0.8\pm 0.0$ & $ 2.8\pm 0.6$ & \nodata\\
Carina & I &   161.188758 &   -59.335179 &    1.2 &    0.8 &    91 &    0.8 &     8 & $ 0.8\pm 0.1$ & $ 4.8\pm 0.1$ & \nodata\\
Carina & J &   161.260036 &   -59.763797 &    2.9 &    1.9 &   231 &    1.0 &    48 & $ 1.0\pm 0.0$ & $ 2.3\pm 0.4$ & \nodata\\
Carina & K &   161.275744 &   -59.672471 &    1.8 &    1.2 &    86 &    0.9 &    19 & $ 0.9\pm 0.0$ & $ 3.6\pm 0.7$ & \nodata\\
Carina & L &   161.296066 &   -59.712681 &    2.5 &    1.7 &   230 &    0.9 &    58 & $ 0.9\pm 0.0$ & $ 2.7\pm 0.3$ & \nodata\\
Carina & M &   161.307215 &   -59.966023 &    2.4 &    1.6 &   169 &    0.9 &    34 & $ 0.9\pm 0.0$ & $ 2.5\pm 0.8$ & \nodata\\
Carina & N &   161.400506 &   -59.805549 &    2.3 &    1.5 &    85 &    1.4 & \nodata & \nodata & \nodata & \nodata\\
Carina & O &   161.472697 &   -59.948001 &    1.1 &    0.7 &    73 &    1.1 &     7 & $ 1.2\pm 0.1$ & $ 1.1\pm 0.6$ & \nodata\\
Carina & P &   161.476497 &   -60.075608 &    5.3 &    3.5 &   305 &    0.9 &    74 & $ 0.9\pm 0.0$ & $ 4.2\pm 0.2$ & \nodata\\
Carina & Q &   161.479838 &   -59.997525 &    2.5 &    1.7 &   155 &    0.9 &    20 & $ 0.9\pm 0.0$ & $ 4.3\pm 0.3$ & \nodata\\
Carina & R &   161.522459 &   -59.835953 &    4.7 &    3.1 &   136 &    1.0 &    21 & $ 0.9\pm 0.1$ & $ 3.0\pm 0.6$ & \nodata\\
Carina & S &   161.719518 &   -60.077775 &    2.1 &    1.4 &   108 &    0.9 &    13 & $ 0.9\pm 0.0$ & $ 2.9\pm 0.8$ & \nodata\\
Carina & T &   161.801941 &   -60.099413 &    3.0 &    2.0 &   233 &    0.9 &    25 & $ 0.9\pm 0.0$ & $ 2.3\pm 0.8$ & \nodata\\
Carina & U &   \nodata &   \nodata &  \nodata &  \nodata &  2353 &    0.9 &   354 & $ 0.9\pm 0.0$ & $ 4.0\pm 0.2$ & \nodata\\
NGC 1893 & A &    80.669746 &    33.372321 &    1.6 &    1.6 &    68 &    0.8 &    16 & $ 0.8\pm 0.0$ & $ 3.5\pm 1.0$ & \nodata\\
NGC 1893 & B &    80.693946 &    33.420652 &    1.3 &    1.3 &   180 &    0.8 &    54 & $ 0.8\pm 0.0$ & $ 2.6\pm 0.3$ & \nodata\\
NGC 1893 & C &    80.695468 &    33.402485 &    0.8 &    0.8 &    23 &    0.8 &     8 & $ 0.8\pm 0.0$ & $ 3.2\pm 1.1$ & \nodata\\
NGC 1893 & D &    80.706555 &    33.447851 &    0.4 &    0.4 &    14 &    0.8 &     7 & $ 0.8\pm 0.1$ & $ 1.9\pm 0.5$ & \nodata\\
NGC 1893 & E &    80.706743 &    33.411659 &    0.1 &    0.1 &     5 &    0.8 & \nodata & \nodata & \nodata & \nodata\\
NGC 1893 & F &    80.722654 &    33.426924 &    0.6 &    0.6 &    30 &    0.8 &     7 & $ 0.8\pm 0.1$ & $ 2.1\pm 0.6$ & \nodata\\
NGC 1893 & G &    80.724232 &    33.515267 &    0.9 &    1.0 &    55 &    0.9 &    14 & $ 0.9\pm 0.0$ & $ 1.5\pm 0.6$ & \nodata\\
NGC 1893 & H &    80.740159 &    33.443803 &    1.2 &    1.2 &   111 &    0.8 &    32 & $ 0.8\pm 0.0$ & $ 1.9\pm 0.4$ & \nodata\\
NGC 1893 & I &    80.764739 &    33.470349 &    1.6 &    1.7 &   148 &    0.8 &    46 & $ 0.9\pm 0.0$ & $ 2.8\pm 0.6$ & \nodata\\
NGC 1893 & J &    80.784690 &    33.476036 &    0.2 &    0.2 &    17 &    1.0 &     3 & $ 1.0\pm 0.0$ & $ 1.4\pm 0.6$ & \nodata\\
NGC 1893 & U &   \nodata &   \nodata &  \nodata &  \nodata &   624 &    0.8 &   166 & $ 0.8\pm 0.0$ & $ 3.0\pm 0.2$ & \nodata\\
\enddata

\tablecomments{Column 1: Name of a star forming region. Column 2: MYStIX cluster designation. Columns 3 and 4: Coordinates of the center of a MYStIX cluster. Right ascension and declination for epoch J2000.0 in degrees. Columns 5-6: Spatial extent of a MYStIX cluster; this is $4\times$ cluster core radius in arc-minutes and parsecs, respectively. Columns 7-8: Total number of MYStIX stars assigned as cluster members and the median of their $J-H$ color \citep{Kuhn2014c}. Columns 9-10: Number of $Age_{JX}$ stars assigned as cluster members and the median of their $J-H$ color.  Column 11:  Cluster $median(Age_{JX})$ estimate. Column 12: For the heavily reddened MYStIX clusters, with $(J-H)_{tot} > 1.5$~mag, independent age estimates are given  based on the calibration curve of Figure \ref{fig_AgeJX_vs_Reddening}. The `:' indicates that these estimates could be quite uncertain.}
\end{deluxetable}

\end{document}